\documentclass[12pt]{article}

\usepackage{amsmath}
\usepackage{enumerate}
\usepackage{url} 
\usepackage{lmodern}
\usepackage{natbib}
\bibpunct{(}{)}{;}{a}{}{,}

\usepackage[unicode=true, pdfusetitle, bookmarks=true, bookmarksnumbered=false, bookmarksopen=false, breaklinks=false, pdfborder={0 0 0}, pdfborderstyle={}, backref=false, colorlinks=true]{hyperref}
\hypersetup{linkcolor=blue, citecolor=blue}

\usepackage{tabularx} 
\usepackage{amssymb}
\usepackage{multirow}
\usepackage{amsthm}
\usepackage{commath}
\usepackage{stmaryrd}
\usepackage{subfigure}
\usepackage{bbm}
\usepackage[T1]{fontenc}
\usepackage[utf8]{inputenc}
\usepackage[ruled, lined, linesnumbered]{algorithm2e}
\usepackage{cases}
\usepackage[toc,page]{appendix}
\usepackage{filecontents,url}
\usepackage{tikz}
\usepackage{color}

\usepackage{graphicx,psfrag,epsf}

\newcommand{\say}[1]{\emph{#1}}
\usetikzlibrary{decorations.pathmorphing,topaths}
\usetikzlibrary{fit}
\usetikzlibrary{positioning,automata,patterns}
\usetikzlibrary{arrows,shapes}

\newcommand{\B}[1]{\mathbf{#1}} 
\newcommand{\pr}[1]{\left(#1\right)}
\newcommand{\br}[1]{\left[#1\right]} 
\newcommand{\bbr}[1]{\left\{#1\right\}}
\newcommand{\nr}[1]{\left\|#1\right\|} 
\newcommand{\joint}[1]{\pi_{\rho}#1} 
\newcommand{\marginal}[1]{\pi_{\rho}#1}

\newtheorem{property}{Property}
\newtheorem{proposition}{Proposition}
\newtheorem{theorem}{Theorem}
\newtheorem{remark}{Remark}

\newtheorem{definition}{Definition}
\newtheorem{corollary}{Corollary}

\newtheorem{exampleinn}{Example}
\newtheorem*{example*}{Example \DO}
\newenvironment{example}
  {\xdef\DO{\number\numexpr\value{exampleinn}+1\relax\ (continued)}%
   \exampleinn}
  {\endexampleinn}

\makeatletter
\def\bstctlcite{\@ifnextchar[{\@bstctlcite}{\@bstctlcite[@auxout]}}
\def\@bstctlcite[#1]#2{\@bsphack
  \@for\@citeb:=#2\do{%
    \edef\@citeb{\expandafter\@firstofone\@citeb}%
    \if@filesw\immediate\write\csname #1\endcsname{\string\citation{\@citeb}}\fi}%
  \@esphack}
\makeatother

\makeatletter
\newcommand{\removelatexerror}{\let\@latex@error\@gobble}
\makeatother

\makeatletter
\newcommand{\thickhline}{%
    \noalign {\ifnum 0=`}\fi \hrule height 1pt
    \futurelet \reserved@a \@xhline
}

\newcommand{\blind}{0}

\graphicspath{{./images/}} 

\begin{document}

\def\spacingset#1{\renewcommand{\baselinestretch}%
{#1}\small\normalsize} \spacingset{1}

\if0\blind
{
  \title{\bf Asymptotically exact data augmentation:\\ models, properties and algorithms}
  \author{Maxime Vono and Nicolas Dobigeon\hspace{.2cm}\\
    Univ. of Toulouse, IRIT/INP-ENSEEIHT, Toulouse, France\\
    and \\
    Pierre Chainais \\
    Univ. of Lille, Centrale Lille, UMR CNRS 9189 - CRIStAL, Lille, France
}
  \maketitle
} \fi

\if1\blind
{
  \bigskip
  \bigskip
  \bigskip
  \begin{center}
    {\LARGE\bf Asymptotically exact data augmentation:\\ models, properties and algorithms}
\end{center}
  \medskip
} \fi

\bigskip
\begin{abstract}
Data augmentation, by the introduction of auxiliary variables, has become an ubiquitous technique to improve convergence properties, simplify the implementation or reduce the computational time of inference methods such as Markov chain Monte Carlo ones.
Nonetheless, introducing appropriate auxiliary variables while preserving the initial target probability distribution and offering a computationally efficient inference cannot be conducted in a systematic way.
To deal with such issues, this paper studies a unified framework, coined asymptotically exact data augmentation (AXDA), which encompasses both well-established and more recent approximate augmented models.
In a broader perspective, this paper shows that AXDA models can benefit from interesting statistical properties and yield efficient inference algorithms. 
In non-asymptotic settings, the quality of the proposed approximation is assessed with several theoretical results.
The latter are illustrated on standard statistical problems.
Supplementary materials including computer code for this paper are available online.
\end{abstract}

\noindent%
\vfill

\newpage
\spacingset{1.45} 
\section{Introduction}
\label{sec:introduction}

Starting at least from the 1960s with the seminal paper of \cite{Hartley1958} on the expectation-maximization (EM) algorithm, introducing auxiliary variables has been a widely adopted strategy to derive iterative algorithms able to deal with possibly complicated inference problems.
Indeed, either by coming from statistical physics \citep{Swendsen1987} or by the broad statistical community \citep{Dempster1977}, auxiliary (also called latent) variables have been used to improve \citep{Duane1987,Edwards1988,Marnissi2018} and/or simplify \citep{TannerWong1987,Doucet2002} inference methods, such as maximum likelihood (ML) estimation or simulation-based ones.
Insightful reviews of these methods were conducted by \cite{Besag1993,Dyk2001,Tanner2010}.
Among many others, slice sampling and half-quadratic (HQ) methods are archetypal instances of such auxiliary variable-based methods.
These methods, by introducing auxiliary variables, appear to be an interesting alternative when sampling cannot be performed directly from a target distribution $\pi$.
Nonetheless, the superiority of simulation-based algorithms based on data augmentation (DA) over classical Markov chain Monte Carlo (MCMC) methods without DA is not obvious as pointed out by \cite{Polson1996,Damien1999}. DA methods have been found to be slower than single-site update approaches in some cases \citep{Hurn1997} and some improvements have been derived to cope with these problems such as partial decoupling \citep{Higdon1998} or the introduction of a working parameter \citep{Meng1997}.
Moreover, DA techniques are often used on a case-by-case basis \citep{Geman1992,Albert1993,Geman1995,Polson2013} and could not be applied in general scenarios due to the absence of exact DA schemes yielding an efficient inference and low computation costs.

Similarly to approximate Bayesian computation (ABC) methods to circumvent intractable likelihoods \citep{Beaumont2002,Sisson2018}, these limitations can be tackled by considering approximate DA schemes that become exact asymptotically.
For instance, inspired from the variable splitting technique used in the alternating direction method of multipliers (ADMM) \citep{Boyd2011}, \cite{Vono2019} and \cite{Rendell2018} recently and independently proposed a novel and broad Bayesian inference framework that can circumvent limitations of exact DA approaches.
By introducing a collection of instrumental (also called \say{splitting}) variables, the aforementioned authors considered the inference from an approximate probability distribution which can be simpler, more efficient and distributed over multiple computational workers (e.g., machines or kernels).

This paper aims at deeply investigating a broad framework coined asymptotically exact data augmentation (AXDA) which encompasses previously proposed special instances such as approximate models used in \cite{Vono2019,Rendell2018}, among others.
More precisely, Section \ref{sec:asymptotic_data_augmentation} details how such models can be built in a quasi-systematic and simple way which is highly appreciable compared to the case-by-case search of computationally efficient DA schemes.
In Section \ref{sec:related_work}, we revisit some already-proposed special instances of AXDA models in order to show the potential benefits of AXDA on specific examples and to exhibit interesting properties which can be generally inherited by AXDA approaches.
In Section \ref{sec:properties}, we assess quantitatively the bias of AXDA models with non-asymptotic theoretical results by considering Wasserstein and total variation distances.
Then, Section \ref{subsec:numerical_illustrations} illustrates the previous theoretical results and the benefits of the proposed methodology on several statistical problems.
In order to facilitate the use of AXDA, we eventually point out that the supplementary material involves a dedicated section (Section 5) presenting how such models can be instantiated to perform efficient inference through classical simulation-based, variational Bayes (VB), optimization or expectation-maximization (EM) methods.
The proofs are also given in the supplementary material, see Section 1.

\section{Asymptotically exact data augmentation}
\label{sec:asymptotic_data_augmentation}

This section introduces AXDA schemes that aim to circumvent exact DA main issue: \textit{the art} \citep{Dyk2001} of finding the exact DA associated to a statistical model and its inference limitations. For sake of simplicity, with little abuse, we shall use the same notations for a probability distribution and its associated probability density function (pdf).

\subsection{Motivations}
\label{subsec:motivations}

In this paper, we are interested in performing the inference of a variable of interest $\boldsymbol{\theta} \in \Theta \subseteq \mathbb{R}^d$, where $\Theta$ is a closed convex set and $\mathrm{dim}(\Theta) = d$, by relying on a probability distribution with density $\pi$ writing
\begin{align}
	\pi(\boldsymbol{\theta}) \propto \exp\left(-f(\boldsymbol{\theta})\right), \quad \text{or} \quad \pi(\B{y}|\boldsymbol{\theta}) \propto \exp\left(-f(y;\boldsymbol{\theta})\right), \label{eq:target_density}
\end{align}
where the potential $f$ taking values in $\mathbb{R}$ is such that $\pi$ defines a proper, bounded and continuous probability distribution.
For sake of generality, note that $\pi$ in \eqref{eq:target_density} shall describe various quantities.
First, with a little abuse of notations, $\pi(\boldsymbol{\theta})$ may simply refer to a pdf associated to the random variable $\boldsymbol{\theta}$, e.g., its prior distribution $\pi(\boldsymbol{\theta})$ or its posterior distribution $\pi(\boldsymbol{\theta}) \triangleq \pi(\boldsymbol{\theta}|\B{y})$ when referring to a set of observations denoted by $\B{y}$.
Depending on the problem, we also allow $\pi$ to stand for a likelihood function $\pi(\B{y}|\boldsymbol{\theta})$.
We will work under this convention and write explicitly the form of $\pi$ when required.
For sake of simplicity and clarity, only the case corresponding to $\pi(\boldsymbol{\theta})$ will be detailed in this section.
The application of the proposed methodology to $\pi(\B{y}|\boldsymbol{\theta})$ is very similar and can be retrieved by a straightforward derivation.

We consider situations where direct inference from \eqref{eq:target_density} is difficult because intractable or computationally prohibitive. 
To overcome these issues, an option is to rely on exact DA which introduces some auxiliary variables stacked into a vector $\B{z} \in \mathcal{Z} \subseteq \mathbb{R}^k$ and defines a new density, simpler to handle, such that
\begin{align}
  \int_{\mathcal{Z}}\pi(\boldsymbol{\theta},\B{z})\mathrm{d}\B{z} = \pi(\boldsymbol{\theta}).  \label{eq:data_aug}
\end{align}
Much research has been devoted to these models in order to simplify an inference task or to improve the convergence properties of direct inference approaches (e.g., slice sampling and HQ methods introduced in Section \ref{sec:introduction}).
Nonetheless, these approaches have several limitations.
Indeed, finding a convenient form for the augmented density in order to satisfy \eqref{eq:data_aug} while leading to efficient algorithms generally requires some knowledge and can even be impossible in some cases \citep{Geman1995}.
For instance, the mixture representation of a binomial likelihood function based on the Polya-Gamma distribution has been used to derive a promising Gibbs sampler for logistic regression problems \citep{Polson2013}.
Nonetheless, even if this algorithm has been proved to be uniformly ergodic by \cite{Choi2013}, the corresponding ergodicity constant depends exponentially on the number of observations $n$ and on the dimension of the regression coefficients vector $d$.

To tackle these limitations, we propose to relax the constraint \eqref{eq:data_aug} and consider an {\em approximate} DA model.
This will permit the choice of an augmented density with more flexibility, fix the issues associated to the initial model and make inference more efficient in some cases.
To this purpose, Section \ref{subsec:model} presents the so-called AXDA framework which embeds approximate DA models controlled by a positive scalar parameter $\rho$.
These models become asymptotically exact when $\rho$ tends towards 0.
Of course, some assumptions will be required on the approximate augmented density to guarantee a good approximation.
The quality of this approximation will be assessed in Section \ref{sec:properties} with non-asymptotic theoretical results.

\subsection{Model}
\label{subsec:model}

Instead of searching for an exact data augmentation scheme \eqref{eq:data_aug}, some auxiliary variables $\B{z}$ can be introduced in order to define an approximate but asymptotically exact probability distribution.
One possibility is to introduce an augmented distribution depending on a parameter $\rho > 0$ and such that the associated marginal density defined by
\begin{align}
  \pi_{\rho}(\boldsymbol{\theta}) = \int_{\mathcal{Z}}\pi_{\rho}(\boldsymbol{\theta},\B{z})\mathrm{d}\B{z}, \label{eq:marginal_p_rho}
\end{align}
satisfies the following property.
\begin{property}
	\label{assumption:phi_rho}
	For all $\boldsymbol{\theta} \in \Theta$,
	$\lim_{\rho \rightarrow 0} \marginal(\boldsymbol{\theta}) = \pi(\boldsymbol{\theta})$ .
\end{property}

\noindent By applying Scheff\'e's lemma \citep{Scheffe1947}, this property yields the convergence in total variation, that is $\nr{\pi_{\rho}-\pi}_{\mathrm{TV}} \rightarrow 0$ as $\rho \rightarrow 0$.
A natural question is: how to choose the augmented density in \eqref{eq:marginal_p_rho} such that Property \ref{assumption:phi_rho} is met?
In this paper, we assume that $\mathcal{Z} = \Theta$ and investigate AXDA schemes associated to an initial density \eqref{eq:target_density} and defined by the approximate augmented density
\begin{align}
	\joint(\boldsymbol{\theta},\B{z}) = \pi(\B{z})\kappa_{\rho}(\B{z},\boldsymbol{\theta}), \label{eq:split_density} 
\end{align}
where $\kappa_{\rho}$ is such that \eqref{eq:split_density} defines a proper density. 

\begin{remark}
  When $\pi$ stands for a product of $J$ densities, that is $\pi \propto \prod_{j=1}^J \pi_j$, the proposed approximate model can naturally be generalized to $\joint(\boldsymbol{\theta},\B{z}_{1:J}) \propto \prod_{j=1}^J\pi_j(\B{z}_j)\kappa_{\rho}(\B{z}_j,\boldsymbol{\theta})$.
  Such a generalization will for instance be considered in Sections \ref{subsec:tractable_inference} and \ref{subsec:distributed_inference}.
\end{remark}

The introduction of the proposed model \eqref{eq:split_density} is aimed at avoiding a case-by-case search of an appropriate augmented approach.
Although there might exist other marginal densities $\pi_{\rho}$ satisfying Property \ref{assumption:phi_rho}, we restrict our analysis to models where $\kappa_{\rho}(\cdot,\boldsymbol{\theta})$ weakly converges towards the Dirac measure at $\boldsymbol{\theta}$ as $\rho \rightarrow 0$ \citep{Aguirregabiria2002}. 
This is a sufficient condition to satisfy Property \ref{assumption:phi_rho}.
In the sequel, we will call AXDA any approach based on \eqref{eq:split_density} and satisfying these properties.

A natural choice for $\kappa_{\rho}$ is to consider a standard kernel $K$ \citep{WandJones1995}. 
Based on the latter, we  define for all $\B{z},\boldsymbol{\theta} \in \Theta$, $\kappa_{\rho}(\B{z},\boldsymbol{\theta}) \propto_{\B{z}} \rho^{-d}K(\rho^{-1}(\boldsymbol{\theta}-\B{z}))$ \citep{Dang2012}.
Beyond standard kernels but motivated by the same idea of measuring the discrepancy between $\B{z}$ and $\boldsymbol{\theta}$, one can also build on divergence functions widely used in the optimization literature to define a potentially asymmetric density $\kappa_{\rho}$ such that for all $\B{z},\boldsymbol{\theta} \in \Theta$, $\kappa_{\rho}(\B{z},\boldsymbol{\theta}) \propto_{\B{z}} \exp(-\rho^{-1}\phi(\B{z},\boldsymbol{\theta}))$ where $\phi$ is a strictly convex function w.r.t. $\B{z}$ admitting a unique minimizer $\B{z}^* = \boldsymbol{\theta}$ \citep{Ben-Tal2001,Krichene2015,Fellows2019}.
Specific instances of such potentials are Bregman divergences such as the logistic loss and the Kullback-Leibler divergence, see Definition \ref{definition:Bregman}.
\begin{definition}[Bregman divergence]
  \label{definition:Bregman}
  Let $\psi$ a continuously-differentiable and strictly convex function defined on a closed convex set.
  The Bregman divergence associated to $\psi$ is defined by
  \begin{equation}
    d_{\psi}(\B{z},\boldsymbol{\theta}) = \psi(\B{z}) - \psi(\boldsymbol{\theta})  - \nabla \psi(\boldsymbol{\theta})^T(\B{z} - \boldsymbol{\theta}).
  \end{equation} 
\end{definition}
Additional details associated to standard kernels and Bregman divergences are given in Section 2 in the supplementary material.

\section{Benefits of AXDA by revisiting existing models}
\label{sec:related_work}

Before providing theoretical guarantees for AXDA models, this section proposes to review some important state-of-the-art works from the AXDA perspective described in Section \ref{sec:asymptotic_data_augmentation}.
We do not pretend to give new insights about these approaches.
We rather use them to illustrate potential benefits that can be gained by resorting to the proposed framework.
For sake of clarity, these benefits are directly highlighted in the title of the following sections before being discussed in the latter.

\subsection{Tractable posterior inference}
\label{subsec:tractable_inference}

This first section illustrates how an AXDA approach can alleviate the intractability of an initial posterior distribution $\pi$ and significantly aid in the computations.

To this purpose, we consider the case where the posterior distribution $\pi$ is intractable. 
Such a model for instance appears when $\pi$ involves a constraint on some set \citep{Liechty2009}, admits a non-standard potential function such as the total variation norm \citep{Chambolle2010,Pereyra2016B,Vono2019} or yields complicated conditional posterior distributions \citep{Holmes2003}.
To simplify the inference, the aforementioned authors have considered special instances of AXDA by relying on an additional level involving latent variables $\B{z}$, leading a hierarchical Bayesian model.
In these cases, AXDA has been invoked in order to move a difficulty to the conditional posterior of $\B{z}$ where it can be dealt with more easily by using standard inference algorithms, see Section 5 in the supplementary material for more details.
The following example, derived from \cite{Holmes2003}, illustrates this idea.

\begin{example}
Let $\B{y} \in \mathbb{R}^n$ be a set of observations and $\B{X} = (\B{x}_1,\hdots,\B{x}_n)^T \in \mathbb{R}^{n \times d}$ a design matrix filled with covariates.
We consider a generalized non-linear model which writes
\begin{align}
  &y_i | \boldsymbol{\theta} \sim p(y_i \ | \ g^{-1}(h(\B{x}_i,\boldsymbol{\theta})),\sigma^2), \quad \forall i \in [n], \\
  &\boldsymbol{\theta} \sim \mathcal{N}(\boldsymbol{\theta} \ | \ \B{0}_d,\nu^2\B{I}_d),
\end{align}
where $p$ belongs to the exponential family and has mean $g^{-1}(h(\B{x}_i,\boldsymbol{\theta}))$ and variance $\sigma^2$ where $g$ is a link function.
As in classical regression problems, we are interested in infering the regression coefficients $\boldsymbol{\theta} \in \mathbb{R}^d$.
In the sequel, we set the non-parametric model $h$ to be 
\begin{equation}
  h(\B{x}_i,\boldsymbol{\theta}) = \sum_{j=1}^k\theta_j B(\B{x}_i,\B{k}_j),
\end{equation}
where $B(\B{x}_i,\B{k}_j)$ is a non-linear function of $\B{x}_i$ (e.g., regression splines) and $\B{k}_j$ is the knot location of the $j$-th basis.
The difficulty here is the non-linearity of $h$ which, combined with the non-Gaussian likelihood, rules out the use of efficient simulation schemes to sample from the posterior $\pi(\boldsymbol{\theta}|\B{y})$.
In order to mitigate this issue, \cite{Holmes2003} proposed to rely on an additional level which boils down to consider the approximate model \eqref{eq:split_density}.
More specifically, the aforementioned authors treated the non-linear predictor $h$ as a Gaussian random latent variable which leads to the approximate model
\begin{align}
  &y_i | z_i \sim p(y_i \ | \ g^{-1}(z_i),\sigma^2), \quad \forall i \in [n], \\
  &z_i | \boldsymbol{\theta} \sim \mathcal{N}(z_i \ | \ h(\B{x}_i,\boldsymbol{\theta}),\rho^2), \quad \forall i \in [n], \label{eq:splines_kappa_rho} \\
  &\boldsymbol{\theta} \sim \mathcal{N}(\boldsymbol{\theta} \ | \ \B{0}_d,\nu^2\B{I}_d).
\end{align}
Here, AXDA has been applied only to the likelihood function with $\kappa_{\rho}$ chosen as the univariate normal distribution \eqref{eq:splines_kappa_rho} leading to a smoothed likelihood function.
The main advantage of relying on such a model is that the posterior conditional distribution $\pi_{\rho}(\boldsymbol{\theta} | \B{z}, \B{X})$, with $\B{z} = [z_1,\hdots,z_n]^T$, is now a multivariate normal distribution.
In addition, by moving the difficulty induced by $h$ to the conditional posterior of $z_i$, we are now dealing with a generalized linear model where standard techniques can be applied \citep{Albert1993,Polson2013}.
\end{example}

Beyond the widely-used Gaussian choice for $\kappa_{\rho}$ \citep{Holmes2003,Liechty2009,barbos2017clone,Vono2019}, more general AXDA approaches can be built by taking inspiration from these works. 
To this purpose, we recommand to choose $\kappa_{\rho}$ w.r.t. the prior and likelihood at stake.
For instance, when a Poisson likelihood function and a complex prior distribution on its intensity $\boldsymbol{\theta}$ are considered, one option for $\phi$ (see Section \ref{subsec:model}) would be an Itakura-Saito divergence since it preserves the positivity constraint on $\boldsymbol{\theta}$ and yields the well-known Gamma-Poisson model \citep{Canny2004}.

\subsection{Distributed inference}
\label{subsec:distributed_inference}

When data are stored on multiple machines and/or one is interested in respecting their privacy, this section illustrates how AXDA can be resorted to perform distributed computations.

Let consider observed data $\{y_i,\B{x}_i\}_{i=1}^n$, where $\B{x}_i$ stands for the covariates associated to observation $y_i$, which are distributed among $B$ nodes within a cluster.
By adopting a prior $\nu(\boldsymbol{\theta})$ and by assuming that the likelihood can be factorized w.r.t. the $B$ nodes, the posterior distribution of the variable of interest $\boldsymbol{\theta}$ writes
\begin{align}
\pi(\boldsymbol{\theta}|\B{y},\B{X}) \propto \nu(\boldsymbol{\theta})\prod_{b=1}^B\prod_{i \in \text{node } b} \exp\pr{-f_i(y_i;h(\B{x}_i,\boldsymbol{\theta}))}. \label{eq:distributed_MCMC}
\end{align}

Such models classically appear in statistical machine learning when generalized linear models (GLMs) \citep{DobsonBarnett2008} are considered.
In these cases, $h(\B{x}_i,\boldsymbol{\theta}) = \B{x}_i^T\boldsymbol{\theta}$.
Due to the distributed environment, sampling efficiently from \eqref{eq:distributed_MCMC} is challenging and a lot of ``divide-and-conquer'' approaches have been proposed in the past few years to cope with this issue \citep{Wang2013,Scott2016}.
These methods launch independent Markov chains on each node $b$ and then combine the outputs of these local chains to obtain an approximation of the posterior of interest \eqref{eq:distributed_MCMC}.
Nonetheless, the averaging schemes used to combine the local chains might lead to poor approximations when $\pi$ is high-dimensional and non-Gaussian.
Instead, considering a special instance of AXDA circumvents the previously mentioned drawbacks by introducing local auxiliary variables on each node such that
\begin{align}
  \joint(\boldsymbol{\theta},\B{z} |\B{y},\B{X}) \propto \nu(\boldsymbol{\theta})\prod_{b=1}^B\prod_{i \in \text{node } b} \exp\pr{-f_i(y_i;z_i)}\kappa_{\rho}(z_i,h(\B{x}_i,\boldsymbol{\theta})).
\end{align}
The posterior distribution of the auxiliary variables conditionally to $\boldsymbol{\theta}$ only depends on the data available at a given node. 
Based on this nice property, the joint posterior can be sampled efficiently with a Gibbs sampler, see \cite{Rendell2018} for a comprehensive review.
We emphasize that the benefits described in this section for Monte Carlo sampling also hold when one wants to use other types of algorithms (e.g., expectation-maximization or variational Bayes), see Section 5 in the supplementary material.

\subsection{Robust inference}
\label{subsec:robust_Bayesian_model}

By noting that classical robust hierarchical models fall into the proposed framework, this section shows that AXDA is also a relevant strategy to perform robust inference by coping with model misspecification by modeling additional sources of uncertainty.

\begin{figure}
\centering
\subfigure[Initial model]
{\begin{tikzpicture}
  \node[circle,draw=gray,inner sep=2mm] (x) at (2,0) {$\boldsymbol{\theta}$};
  \node[circle,draw=gray,inner sep=2mm] (y) at (4,0) {$\B{y}_i$};
  \node (dim) at (4.7,-0.5) {\small $n$};
  \node[draw,fit=(y) (dim)] {};
  \draw[->] (x) -- (y);
\end{tikzpicture}}
\hspace{1cm}
\subfigure[Localized hierarchical model]
{\begin{tikzpicture}
  \node[circle,draw=gray,inner sep=2mm] (theta) at (0,0) {$\boldsymbol{\theta}$};
  \node[circle,draw=gray,inner sep=2mm] (x) at (2,0) {$\B{z}_i$};
  \node[circle,draw=gray,inner sep=2mm] (y) at (4,0) {$\B{y}_i$};
  \node (dim) at (4.7,-0.5) {\small $n$};
  \node[draw,fit=(y) (x) (dim)] {};
  \draw[->] (theta) -- (x);
  \draw[->] (x) -- (y);
\end{tikzpicture}}
\caption{Concept of localization. Comparison between the initial (left) and the localized hierarchical Bayesian (right) models with $n$ the number of observations $y_i$.}
\label{fig:localized_model}
\end{figure}
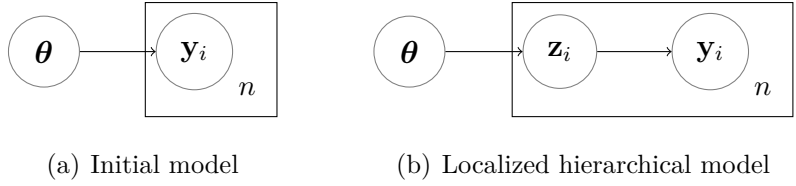

Considering a well-chosen \textit{demarginalization} procedure is known to yield robustness properties in some cases \citep{Robert2004}.
Some approaches took advantage of this idea in order to build robust hierarchical Bayesian models w.r.t. possible outliers in the data.
For instance, such models can be built by allowing each observation to be randomly drawn from a local statistical model, as described in the recent review of \cite{Wang2018}.
This ``localization'' idea is illustrated in Figure \ref{fig:localized_model}.
Many of these models can be viewed as particular instances of AXDA.
Indeed, assume that $n$ data points $y_i$ are independently and identically distributed (i.i.d.) defining the likelihood function 
\begin{align}
  \pi(\B{y}|\boldsymbol{\theta}) \propto \prod_{i=1}^n\pi(y_i|\boldsymbol{\theta}), \label{eq:likelihood_prod}
\end{align}
where $\boldsymbol{\theta} \in \Theta$ is a common parameter.
Applying AXDA as described in Section \ref{sec:asymptotic_data_augmentation} by introducing $n$ $d$-dimensional auxiliary variables stacked into the vector $\B{z}_{1:n}$ leads to the augmented likelihood 
\begin{align}
  \joint{(\B{y},\B{z}_{1:n}|\boldsymbol{\theta})} \propto \prod_{i=1}^n\pi(y_i|\B{z}_i)\kappa_{\rho}(\B{z}_i,\boldsymbol{\theta}). \label{eq:likelihood_prod_split}
\end{align}
The statistical model defined by \eqref{eq:likelihood_prod_split} implies a hierarchical Bayesian model similar to the localized one depicted on Figure \ref{fig:localized_model}(b) and corresponds in general to an approximation of the initial one, see Example \ref{example:robust_model2}.
\begin{example}
  \label{example:robust_model2}
    \textbf{Robust logistic regression --} Assume that for all $i \in [n]$, $\pi(y_i|\boldsymbol{\theta}) = \mathcal{B}\pr{\sigma\pr{\B{x}_i^T\boldsymbol{\theta}}}$, where $\mathcal{B}$ stands for the Bernoulli distribution, $\sigma$ for the sigmoid function, $\B{x}=\left[\B{x}_1,\ldots,\B{x}_n\right]$ for the transpose of the design matrix and $\boldsymbol{\theta}$ for the regression coefficients vector to infer.
    Then as proposed by \cite{Wang2018}, one can robustify the inference by assuming that each observation $y_i$ is drawn from a local and independent model $\mathcal{B}\pr{\sigma\pr{\B{x}_i^T\B{z}_i}}$ associated to an auxiliary parameter $\B{z}_i \sim \mathcal{N}(\boldsymbol{\theta},\rho^2\B{I}_{d})$.
    In this case, $\kappa_{\rho}(\B{z},\boldsymbol{\theta}) \propto \prod_{i=1}^n\mathcal{N}(\B{z}_i \ | \ \boldsymbol{\theta},\rho^2\B{I}_{d})$.
\end{example}

Beyond the convenient Gaussian prior $\kappa_{\rho}$, its choice can be motivated by robust loss functions.
In the statistical machine learning literature, the absolute or Huber losses are of common use \citep{She2011}.
In Bayesian linear inverse problems considered in the signal processing community, it is classical to approximate a complicated forward physical model in order to yield tractable computations.
If the latter can be written as $\B{y} = h(\boldsymbol{\theta}) + \boldsymbol{\epsilon}$, with $\boldsymbol{\epsilon} \sim \pi(\boldsymbol{\epsilon})$, then introducing a latent variable $\B{z} \sim \kappa_{\rho}(\B{z},h(\boldsymbol{\theta}))$ such that $\B{y} = \B{z} + \boldsymbol{\epsilon}$ allows to take into consideration the model approximation.
In those cases, one can set $\kappa_{\rho}$ to be the distribution of the modeling error which could be adjusted thanks to some expertise.

\subsection{Inheriting sophisticated inference schemes from ABC}
\label{subsec:ABC}

Finally, this section shows that AXDA models, by sharing strong connections with ABC, might inherit sophisticated algorithms to sample from \eqref{eq:split_density}.

ABC stands for a family of methods that permit to cope with intractable likelihoods by sampling from the latter instead of evaluating them.
In a nutshell, if one's goal is to infer a parameter $\boldsymbol{\theta}$ based on a posterior of interest, the simplest ABC rejection sampler is as follows.
At iteration $t$, draw a candidate $\boldsymbol{\theta}^{(t)}$ from the prior, generate pseudo-observations $\B{z}$ from the likelihood given this candidate and accept $\boldsymbol{\theta}^{(t)}$ if $\B{z} = \B{y}$ where $\B{y}$ is the observations vector.
Many more sophisticated ABC samplers have been derived.
We refer the interested reader to the recent review by \cite{Sisson2018handbook} for more information about ABC methods.

Among a huge literature on ABC (also called likelihood-free) methods, \textit{noisy ABC} approaches proposed and motivated by \cite{Fearnhead2012} and \cite{Wilkinson2013} are strongly related to AXDA.
Indeed, only comparing the underlying models, AXDA with observation splitting is equivalent to noisy ABC.
To see this, let $\pi(\B{y}|\boldsymbol{\theta})$ stand for an intractable likelihood. Noisy ABC replaces the exact inference based on $\pi$ by considering the pseudo-likelihood with density
\begin{align}
  \marginal(\B{y}|\boldsymbol{\theta}) \triangleq \int_{\Theta} \joint(\B{y},\B{z}|\boldsymbol{\theta})\mathrm{d}\B{z} = \int_{\Theta}\pi(\B{z}|\boldsymbol{\theta})\kappa_{\rho}(\B{z},\B{y})\mathrm{d}\B{z}. \label{eq:ABC_split_likelihood}
\end{align}
This density has exactly the same formulation as the one defined in \eqref{eq:split_density} except that noisy ABC splits the observations $\B{y}$ instead of the parameter of interest $\boldsymbol{\theta}$.
Capitalizing on this equivalence property, also pointed out by \cite{Rendell2018}, one can derive efficient algorithms for AXDA from the ABC framework.
For instance, \cite{Rendell2018} recently built on the works of \cite{Beaumont2002,DelMoral2012} in the ABC context to propose a bias correction approach and a sequential Monte Carlo (SMC) algorithm avoiding the tuning of the tolerance parameter $\rho$.
Obviously, many other inspirations from ABC can be considered, such as the parallel tempering approach of \cite{Baragatti2013} among others, to make the inference from an AXDA model more flexible and efficient.

\section{Theoretical guarantees}
\label{sec:properties}

By building on existing approaches, Section \ref{sec:related_work} showed that AXDA can be used in quite general and different settings depending on ones motivations.
In order to further promote the use of such approximate augmented models, this section goes beyond the empirical bias analysis performed by previous works and provides quantitative bounds on the error between the initial and the approximate model.
More precisely, for a fixed tolerance parameter $\rho > 0$, non-asymptotic results on the error associated to densities, potentials and credibility regions are derived.
We will assume all along this section that $\Theta = \mathbb{R}^d$.
The proofs of the results of this section can be found in Section 1 of the supplementary material.

\subsection{Results for standard kernels}
\label{subsec:theory_kernels}

In this section, we consider the case $\kappa_{\rho}(\B{z},\boldsymbol{\theta}) \propto \rho^{-d}K(\rho^{-1}(\boldsymbol{\theta} - \B{z}))$ where $K$ is a kernel, see Section \ref{subsec:model}. 
Under this model, $\pi_{\rho}$ stands for the convolution of $\pi$ and $\kappa_{\rho}$ and the following results hold.
\begin{proposition}
\label{prop:differentiability_prior}
  Let $\pi \in L^1$.
  The marginal with density $\marginal$ in \eqref{eq:marginal_p_rho} has the following properties.
  \begin{enumerate}
    \item[i)] Let $\pi$ stand for a pdf associated to the random variable $\boldsymbol{\theta}$ and $\mathbb{E}_{\kappa_{\rho}}(X) = 0$. 
    Then, the expectation and variance under $\marginal$ are given by
    \begin{align}
      \mathbb{E}_{\marginal\text{}}(\theta) &= \mathbb{E}_{\pi}(\theta) \\
      \mathrm{var}_{\marginal\text{}}(\theta) &= \mathrm{var}_{\pi}(\theta) + \mathrm{var}_{\kappa_{\rho}}(\theta).
    \end{align}
    \item[ii)] $\mathrm{supp}(\pi_{\rho}) \subseteq S$ where $S$ is the closure of $\{\B{x} + \B{z} ; \B{x} \in \mathrm{supp}(\pi), \B{z} \in \mathrm{supp}(\kappa_{\rho}) \}$. The notation $\mathrm{supp(h)} = \{\B{x} \in \mathcal{X} \ | \ h(\B{x}) \neq 0 \}$ refers to the support of a function $h: \mathcal{X} \rightarrow \mathbb{R}$.
    \item[iii)] If both $\pi$ and $\kappa_{\rho}$ are log-concave, then $\marginal$ is log-concave.
    \item[iv)] If $\kappa_{\rho} \in \mathcal{C}^{\infty}(\mathbb{R}^d)$ and $|\partial^k\kappa_{\rho}|$ is bounded for all $k \geq 0$, then $\marginal$ is infinitely differentiable w.r.t. $\boldsymbol{\theta}$.
  \end{enumerate}
\end{proposition}

Proposition \ref{prop:differentiability_prior} permits to draw several conclusions about the inference based on $\joint$.
Firstly, the infinite differentiability of $\marginal$ (Property \emph{iv)}) implies that it stands for a smooth approximation of $\pi$, see Figure \ref{fig:prior_potential} in Section \ref{example:LASSO}.
Secondly, Property \emph{i)} of Proposition \ref{prop:differentiability_prior} is reassuring regarding the inference task.
Indeed, if $\pi$ stands for a prior distribution, then considering the approximation $\marginal$ simply corresponds to a more diffuse prior knowledge around the same expected value, see Section \ref{example:LASSO}.
Thus, more weight will be given to the likelihood if a posterior distribution is derived with this prior.
On the other hand, if $\pi$ stands for a likelihood, then considering the approximation $\marginal$ yields the opposite behavior: the likelihood becomes less informative w.r.t. the prior.
This idea is directly related to robust hierarchical Bayesian models discussed in Section \ref{subsec:robust_Bayesian_model}.

We now provide quantitative bounds on the approximation implied by considering the marginal $\pi_{\rho}$ instead of $\pi$.
For $p \geq 1$, we define the $p$-Wasserstein distance between $\pi$ and $\pi_{\rho}$ by 
  \begin{align}
    W_p(\pi,\pi_{\rho}) = \pr{\min_{\mu}\bbr{\int_{\mathbb{R}^d}\int_{\mathbb{R}^d}\nr{\boldsymbol{\theta}-\B{z}}_2^p\mathrm{d}\mu(\B{z},\boldsymbol{\theta}); \mu \in \Gamma(\pi_{\rho},\pi)}}^{1/p},
    \label{eq:wasserstein}
  \end{align}
where $\Gamma(\pi_{\rho},\pi)$ is the set of probability distributions $\mu(\boldsymbol{\theta},\B{z})$ with marginals $\pi_{\rho}$ and $\pi$ w.r.t. $\boldsymbol{\theta}$ and $\B{z}$, respectively.
Under mild assumptions on the kernel $K$, Proposition \ref{prop:bound_kernel} gives a simple and practical upper bound on \eqref{eq:wasserstein}.
\begin{proposition}
  \label{prop:bound_kernel}
  Assume that $\pi_{\rho}$ in \eqref{eq:marginal_p_rho} stands for a pdf associated to the variable $\boldsymbol{\theta}$.
  Let $p \geq 1$ such that $m_p \triangleq \pr{\displaystyle\int_{\mathbb{R}^d}\nr{\B{u}}_2^pK(\B{u})\mathrm{d}\B{u}}^{1/p} < \infty$.
  Then, we have
  \begin{align}
    W_p(\pi,\pi_{\rho}) \leq \rho m_p. \label{eq:kernel_upp_bound}
  \end{align}
\end{proposition}
Note that \eqref{eq:kernel_upp_bound} holds without assuming additional assumptions on the initial density $\pi$ such as infinite differentiability.
If the latter is assumed w.r.t. the parameter of interest $\boldsymbol{\theta}$, then one can estimate the bias $\pi - \pi_{\rho}$ with a Taylor expansion of $\pi$ similarly to bias analysis in ABC, see \cite{Sisson2018}. 
Table \ref{table:kernels_bounds} gives closed-form expressions of $m_2$ for the multivariate generalizations of standard kernels.
One can see that the constant $m_2$ has the same dependence w.r.t. the dimension $d$ for the considered standard kernels $K$.
Hence, in high-dimensional scenarios, the approximation quality will be more affected by an inappropriate value for the tolerance parameter $\rho$ rather than by the choice of $K$.
In Section \ref{subsec:numerical_illustrations}, we illustrate Proposition \ref{prop:bound_kernel} with numerical experiments.
\begin{table}
\caption{Closed-form expressions of $m_2$ appearing in \eqref{eq:kernel_upp_bound} for multivariate generalizations of standard kernels where $d$ denotes the dimension.}
\begin{center}
\begin{tabular}{cccccccc}
& Gaussian & Cauchy & Laplace & Dirichlet & Uniform & Triangular & Epanechnikov\\
\hline
$m_2$ & $\sqrt{d}$ & - & $\sqrt{2d}$ & - & $\sqrt{d/3}$ & $\sqrt{d/6}$ & $\sqrt{d/5}$ \\
\end{tabular}
\end{center}
\label{table:kernels_bounds}
\end{table} 

\subsection{Pointwise bias for Bregman divergences}

In complement to Section \ref{subsec:theory_kernels} where $\kappa_{\rho}$ was built using kernels, we now analyze the bias induced by considering $\pi_{\rho}$ when $\kappa_{\rho}$ is derived from a Bregman divergence $d_{\psi}$ (see Definition \ref{definition:Bregman}), that is
\begin{equation}
	\kappa_{\rho}(\B{z},\boldsymbol{\theta}) \propto \exp\pr{-\frac{d_{\psi}(\B{z},\boldsymbol{\theta})}{\rho}}.
\end{equation}
Under regularity assumptions on both $\pi$ and $\kappa_{\rho}$, Proposition \ref{proposition:Bregman_bias} shows the dependence of the pointwise bias $\pi_{\rho} - \pi$ w.r.t. to the tolerance parameter $\rho$ when the latter is sufficiently small.
\begin{proposition}
	\label{proposition:Bregman_bias}
  Assume that $\pi$ is analytic and twice differentiable on $\mathbb{R}^d$ and so does $d_{\psi}$ w.r.t. its first argument.
  Let $\boldsymbol{\theta} \in \mathbb{R}^d$ such that both $\B{H}_{\pi}(\boldsymbol{\theta})$ and $\B{H}_{d_{\psi}}(\boldsymbol{\theta})^{-1}$ exist and are continuous, where $\B{H}_{\pi}(\boldsymbol{\theta})$ is the Hessian matrix of $\pi$ and $\B{H}_{d_{\psi}}(\boldsymbol{\theta}) \triangleq \frac{\partial^2d_{\psi}(\B{z},\boldsymbol{\theta})}{\partial \B{z}^2}\Bigr|_{\substack{\B{z}=\boldsymbol{\theta}}}$ is the Hessian matrix associated to $d_{\psi}(\cdot,\boldsymbol{\theta})$. 
  Then, if
  \begin{itemize}
      \item $\nr{\B{H}_{\pi}} \leq C < \infty$
      \item $\nr{\B{H}_{d_{\psi}}} \geq c > 0$,
    \end{itemize}
	it follows that
	  \begin{equation}
    \pi_{\rho}(\boldsymbol{\theta}) - \pi(\boldsymbol{\theta}) = \mathcal{O}(\sqrt{\rho}). 
  \end{equation}
  In addition, if we have $\displaystyle\int_{\mathbb{R}^d}\B{u}\kappa_{\rho}(\boldsymbol{\theta}-\sqrt{\rho}\B{u},\boldsymbol{\theta})\mathrm{d}\B{u} = \B{0}_d$, then
  \begin{equation}
    \pi_{\rho}(\boldsymbol{\theta}) - \pi(\boldsymbol{\theta}) = \frac{\rho}{2} \mathrm{Trace}\pr{\B{H}_{\pi}(\boldsymbol{\theta})\B{H}_{d_{\psi}}(\boldsymbol{\theta})^{-1}} + o(\rho). \label{eq:kernel_dep}
  \end{equation}

\end{proposition}
Note that when $\psi(\B{z}) = \nr{\B{z}}_2^2/2$, $\kappa_{\rho}$ stands for a Gaussian smoothing kernel, see Section \ref{subsec:theory_kernels}.
In that case, we have the sanity check that the dependence w.r.t. $\rho$ of the bias between $\pi$ and $\pi_{\rho}$ in \eqref{eq:kernel_dep} is the same as the one derived by \cite{Sisson2018} when interpreting $\kappa_{\rho}$ as a kernel.

\subsection{A detailed non-asymptotic analysis for Gaussian smoothing}
\label{subsec:theory_gaussian}

The previous sections gave quantitative approximation results for a large class of densities $\kappa_{\rho}$ built either via a kernel or a Bregman divergence.
In this section, we provide complementary results by restricting our analysis on the case 
\begin{equation}
  \kappa_{\rho}(\B{z},\boldsymbol{\theta}) =  \mathcal{N}(\B{z}|\boldsymbol{\theta},\rho^2\B{I}_d).
\end{equation}
This particular yet convenient assumption will allow to complement and sharpen results of Section \ref{subsec:theory_kernels} by deriving quantitative bounds which take into account the regularity properties of $f$.
Furthermore, these bounds can be extended to a sum of potential functions $ f = \sum_i f_i$ and used to assess the bias associated to both log-densities and credibility regions.
This analysis is also motivated by the fact that the Gaussian smoothing case has been widely advocated in the literature since it generally leads to simple inference steps \citep{Holmes2003,Giovannelli2008,Liechty2009,Dumbgen2009}, and can be related to both the ADMM in optimization \citep{Boyd2011,Vono2019} and the approximation involved in proximal MCMC methods \citep{Pereyra2016B,Durmus2018,Salim2019}.
Unfortunately, a straigthforward generalization of the proof techniques used in the sequel does not give informative upper bounds for smoothing associated to other Bregman divergences.

\subsubsection{Assumptions}

To derive non-asymptotic bounds between quantities related to $\joint$ defined in \eqref{eq:marginal_p_rho} and $\pi$ in \eqref{eq:target_density}, some complementary assumptions on $f = -\log \pi$ will be required. They are detailed hereafter.
For simplicity and with a little abuse of notations, we also denote here by $f(\boldsymbol{\theta})$ the potential associated to \eqref{eq:target_density} when $\pi(\B{y}|\boldsymbol{\theta})$ stands for a likelihood.
  \begin{enumerate}
    \item[($A_1$)] $f$ is $L_f$-Lipschitz w.r.t. $\nr{\cdot}_2$, that is $\exists \ L_f \geq 0$ such that for all $\boldsymbol{\theta},\boldsymbol{\eta} \in \mathbb{R}^d$, $|f(\boldsymbol{\theta})-f(\boldsymbol{\eta})| \leq L_f \nr{\boldsymbol{\theta}-\boldsymbol{\eta}}_2$.
    When $\pi$ is a likelihood, it is further assumed that $L_f$ is independent of $\B{y}$.
    \item[($A_2$)] $f$ is continuously differentiable and has an $M_f$-Lipschitz continuous gradient w.r.t. $\nr{\cdot}_2$, that is $\exists \ M_f \geq 0$ such that for all $\boldsymbol{\theta},\boldsymbol{\eta} \in \mathbb{R}^d$, $\nr{\nabla f(\boldsymbol{\theta})-\nabla f(\boldsymbol{\eta})}_2 \leq M_f \nr{\boldsymbol{\theta}-\boldsymbol{\eta}}_2$.
    \item[($A_3$)] $f$ is convex, that is for every $\alpha \in [0,1]$, $\boldsymbol{\theta},\boldsymbol{\eta} \in \mathbb{R}^d$, $f(\alpha \boldsymbol{\theta} + (1 - \alpha)\boldsymbol{\eta}) \leq \alpha f(\boldsymbol{\theta}) + (1 - \alpha)f(\boldsymbol{\eta})$.
    \item[($A_4$)] $\mathsf{M}_f = \displaystyle\int_{\mathbb{R}^d} \nr{\nabla f(\boldsymbol{\theta})}_2^2\pi(\boldsymbol{\theta})\mathrm{d}\boldsymbol{\theta} < \infty$.
  \end{enumerate}

Assumptions $(A_1)$, $(A_2)$ and $(A_3)$ on the potential $f$ stand for standard regularity assumptions in the optimization literature and cover a large class of functions $f$ \citep{Beck2009,Bolte2014}.
In the broad statistical community, $(A_1)$ has been used by \cite{Durmus2018} to derive non-asymptotic bounds on the total variation distance between probability distributions while $(A_2)$ stands for a sufficient condition to have a strong solution to the overdamped Langevin stochastic differential equation \citep{Durmus2017}.

Under the previous assumptions (not used all at once), non-asymptotic upper bounds on the total variation distance between $\marginal$ and $\pi$ are derived in Section \ref{subsec:control_TV}.
Then, Sections \ref{subsec:bounds_potentials} and \ref{subsec:bounds_credibility_regions} take advantage of this bound to state theoretical properties on the potential functions and credibility regions.

\subsubsection{Non-asymptotic bounds on the total variation distance}
\label{subsec:control_TV}

In this section, we make additional regularity assumptions on the potential $f$ in order to show quantitative results depending explicitly on regularity constants associated to $f$.
Two different cases will be considered, namely Lipschitz potentials, and differentiable, gradient-Lipschitz and convex ones.

\textbf{Lipschitz potential --} When the potential function $f$ is assumed to be Lipschitz continuous but not necessarily continuously differentiable, the following result holds.

\begin{theorem}
\label{theorem:controlTVnorm}
Let a potential function $f$ satisfy $(A_1)$.
Then,
\begin{align}
  \nr{\marginal-\pi}_{\mathrm{TV}} \leq 1 - \Delta_d(\rho) \label{eq:controlTVnorm},
\end{align}
where
\begin{align}
\Delta_d(\rho) = \dfrac{D_{-d}(L_f\rho)}{D_{-d}(-L_f\rho)}.
\end{align}
The function $D_{-d}$ is a parabolic cylinder function defined for all $d>0$ and $z \in \mathbb{R}$ by
\begin{align}
D_{-d}(z) = \dfrac{\exp(-z^2/4)}{\Gamma(d)}\int_0^{+\infty}e^{-xz - x^2/2}x^{d-1}\mathrm{d}x.
\end{align}
\end{theorem}

As expected from Property \ref{assumption:phi_rho}, note that this bound tends towards zero when $\rho \rightarrow 0$.
Additionally, this bound depends on few quantities that can be computed, bounded or approximated in real applications: the dimension of the problem $d$, the Lipschitz constant $L_f$ associated to the regularized potential $f$ and the tolerance parameter $\rho$.
In the limiting case $\rho\rightarrow0$, the following equivalent function for the upper bound derived in \eqref{eq:controlTVnorm} holds.
\begin{corollary}
\label{corollary:DLTVnorm}
  In the limiting case $\rho\rightarrow0$, we have:
  \begin{align}
    \nr{\pi_{\rho}-\pi}_{\mathrm{TV}} \leq \rho L_f\dfrac{2\sqrt{2}\Gamma\pr{\dfrac{d+1}{2}}}{\Gamma\pr{\dfrac{d}{2}}} + o(\rho), \label{eq:upperBound_equivalent}
  \end{align}
  where for all $z > 0$ as $\Gamma(z) = \displaystyle\int_0^{+\infty}x^{z-1}e^{-x}\mathrm{d}x$.
\end{corollary}

Under some regularity conditions (here Lipschitz continuity) on the potential function $f$, Proposition \ref{corollary:DLTVnorm} states that $\nr{\marginal-\pi}_{\mathrm{TV}}$ grows at most linearly w.r.t. the parameter $\rho$ and w.r.t. $L_{f}$ when $\rho$ is sufficiently small.
Moreover, using Stirling-like approximations when $d$ is large in the equivalence relation \eqref{eq:upperBound_equivalent} may give a mild dependence on the dimensionality of the problem in $\mathcal{O}\pr{L_fd^{1/2}}$.
Potential functions verifying the hypothesis of Theorem \ref{theorem:controlTVnorm} are common in machine learning and signal/image processing problems, see Section 3 in the online supplementary material.
As an archetypal example, the sparsity promoting potential function defined for all $\boldsymbol{\theta} \in \mathbb{R}^d$ by $f(\boldsymbol{\theta}) = \tau\nr{\boldsymbol{\theta}}_1$ with $\tau > 0$ is Lipschitz continuous with Lipschitz constant $L_f = \tau \sqrt{d}$ and satisfies Theorem \ref{theorem:controlTVnorm} and Proposition \ref{corollary:DLTVnorm}.
In this case, the dependence of \eqref{eq:upperBound_equivalent} is linear w.r.t. $d$ when $d$ is large and $\rho$ is small.
Note also that continuously differentiable functions on a compact set are Lipschitz continuous.\vspace{0.5cm}

\noindent\textbf{Convex and gradient-Lipschitz potential --} We now show a complementary result by assuming $f$ to be convex and continuously differentiable with a Lipschitz-continuous gradient.

\begin{theorem}
\label{theorem:controlTVnorm_2}
Let a potential function $f$ satisfy $(A_2)$, $(A_3)$ and $(A_4)$.
Then, when $\pi$ stands for a pdf associated to $\boldsymbol{\theta}$, we have:
\begin{align}
  \nr{\marginal-\pi}_{\mathrm{TV}} \leq 1 - \dfrac{1}{(1 + 2\rho^2M_f)^{d/2}}\pr{1 - \dfrac{\rho^4M_f\mathsf{M}_f}{1+2\rho^2M_f}} \label{eq:controlTVnorm_2}.
\end{align}
\end{theorem}
In the limiting case $\rho\rightarrow0$, the upper bound in \eqref{eq:controlTVnorm_2} has a simpler expression as shown hereafter.
\begin{corollary}
\label{corollary:DLTVnorm_2}
  In the limiting case $\rho\rightarrow0$, we have:
  \begin{align}
    \nr{\marginal-\pi}_{\mathrm{TV}} \leq \rho^2dM_f + o(\rho^2). \label{eq:upperBound_equivalent_2}
  \end{align}
\end{corollary}
Note that the dependences w.r.t. both $\rho$ and $d$ in Corollary \ref{corollary:DLTVnorm} and \ref{corollary:DLTVnorm_2} are similar to the ones found by \cite{Nesterov2017} for optimization purposes. 

\begin{figure}
\centering
\mbox{{\includegraphics[scale=0.50]{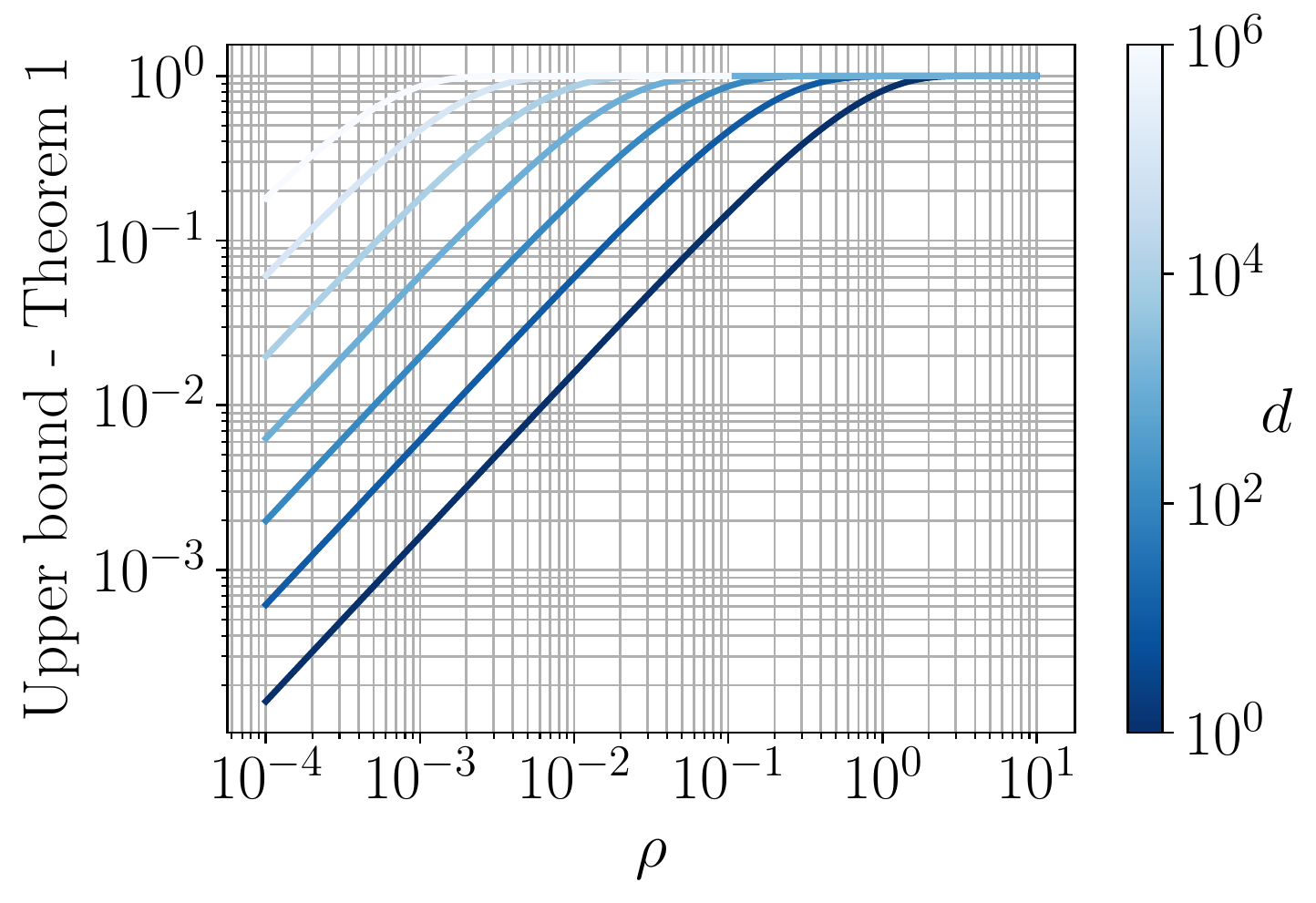}}}
  \mbox{{\includegraphics[scale=0.50]{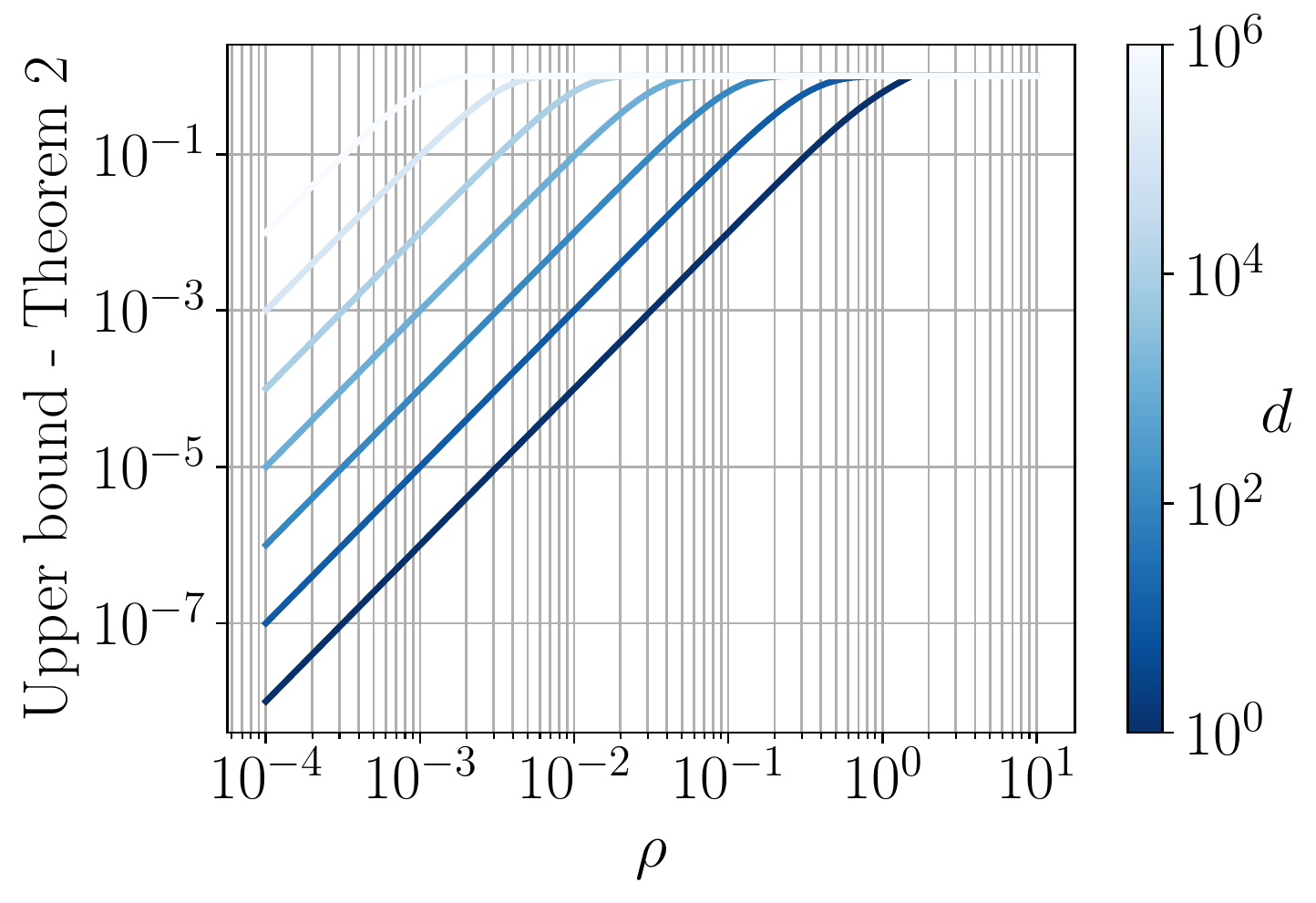}}}
\caption{Behavior of the quantitative bounds shown in Theorems \ref{theorem:controlTVnorm} and \ref{theorem:controlTVnorm_2} w.r.t. $\rho$ in log-log scale for a set of dimensions $d$. The other quantities appearing in the bounds have been set to 1.}
\label{fig:TVnorm}
\end{figure}

Figure \ref{fig:TVnorm} gives the behavior of the upper bounds in \eqref{eq:controlTVnorm} and \eqref{eq:controlTVnorm_2} w.r.t. the dimensionality $d$ of the problem ranging from $1$ to $10^6$ and as a function of $\rho$ in log-log scale.
The linear (resp. quadratic) relation between this upper bound and $\rho$ shown in \eqref{eq:upperBound_equivalent} (resp. \eqref{eq:upperBound_equivalent_2}) is clearly observed for small values of $\rho$.
Nonetheless, these upper bounds are not a silver bullet.
Indeed, as expected, for a fixed value of the parameter $\rho$, the approximation error increases as the dimension $d$ grows.
Thus, these bounds suffer from the curse of dimensionality and become non-informative in high-dimension if $\rho$ is not sufficiently small.

Theorem \ref{theorem:controlTVnorm} is easily extended to the case where the initial density $\pi$ is expressed as a product of several terms.
If $\pi$ stands for the pdf associated to the variable $\boldsymbol{\theta}$, this boils down to considering
\begin{align}
  \pi(\boldsymbol{\theta}) \propto \prod_{j=1}^J \pi_j(\boldsymbol{\theta}) \propto \exp\pr{-\sum_{j=1}^J f_j(\boldsymbol{\theta})}, \label{eq:potential_sum}
\end{align}
where for all $j \in [J]$, $f_j: \mathbb{R}^d \rightarrow \mathbb{R}$, and a natural generalization of AXDA when applied to each $\pi_j$, which writes
\begin{align}
  \joint(\boldsymbol{\theta},\B{z}_{1:J}) \propto \prod_{j=1}^J\pi_j(\B{z}_j)\kappa_{\rho_j}(\B{z}_j;\boldsymbol{\theta}) \propto \exp\pr{-\sum_{j=1}^J f_j(\B{z}_j) + \dfrac{1}{2\rho_j^2}\nr{\B{z}_j-\boldsymbol{\theta}}_2^2}.
\end{align}
Under this product form, we have the following corollary.
\begin{corollary}
\label{theorem:controlTVnorm_generalization}
For all $j \in [J]$, let $f_j$ satisfy $(A_1)$.
Then,
\begin{align}
  \nr{\marginal-\pi}_{\mathrm{TV}} \leq 1 - \prod_{j=1}^J \Delta_d^{(j)}(\rho_j) \label{eq:controlTVnorm_generalization},
\end{align}
where $\Delta_d^{(j)}(\rho_j) = D_{-d}(L_{f_j}\rho_j)/D_{-d}(-L_{f_j}\rho_j)$.
\end{corollary}
Unfortunately, Theorem \ref{theorem:controlTVnorm_2} cannot be extended to the multiple splitting scenario.
We are nevertheless confident that quantitative bounds can be found with different proof techniques but this task goes beyond the scope of this paper.

\subsubsection{Uniform bounds on potentials}
\label{subsec:bounds_potentials}

From an optimization point of view, it is quite common to consider potential functions associated to densities. 
For such applications, we give hereafter a quantitative uniform bound on the difference between the potential functions associated to $\pi$ and $\pi_{\rho}$.
Similarly to the definition of the potential function $f$ in \eqref{eq:target_density}, we define the potential function $f_{\rho}$ associated to the approximate marginal $\pi_{\rho}$ in \eqref{eq:marginal_p_rho}, for all $\boldsymbol{\theta} \in \mathbb{R}^d$, by
\begin{align}
  f_{\rho}(\boldsymbol{\theta}) = -\log \int_{\mathbb{R}^d}\exp\pr{-f(\B{z})}\kappa_{\rho}(\B{z},\boldsymbol{\theta})\mathrm{d}\B{z}.
\end{align}
By considering a Gaussian smoothing kernel $\kappa_{\rho}$, the potential $f_{\rho}$ becomes
\begin{align}
  f_{\rho}(\boldsymbol{\theta}) = -\log \int_{\mathbb{R}^d}\exp\pr{-f(\B{z})-\dfrac{1}{2\rho^2}\nr{\B{z}-\boldsymbol{\theta}}_2^2}\mathrm{d}\B{z} + \frac{d}{2}\log(2\pi\rho^2). \label{eq:potential}
\end{align}
Note that  $f_{\rho}(\boldsymbol{\theta})$ appears as a regularized version of $ f(\boldsymbol{\theta})$. 

\begin{proposition}
\label{prop:control_prior}
  Let $f$ satisfy $(A_1)$.
Then, for all $\boldsymbol{\theta} \in \mathbb{R}^d$,
\begin{align}
  L_{\rho} \leq f_{\rho}(\boldsymbol{\theta}) - f(\boldsymbol{\theta}) \leq U_{\rho}, \label{eq:bound_prior}
\end{align}
with
\begin{align}
&L_{\rho} = \log N_{\rho} - \log D_{-d}(-L_f\rho), \\
&U_{\rho} = \log N_{\rho} - \log D_{-d}(L_f\rho),
\end{align}
and
\begin{align}
N_{\rho} = \dfrac{2^{d/2-1}\Gamma\pr{d/2}}{\Gamma(d)\exp\pr{L_f^2\rho^2/4}}. \label{eq:M_rho}
\end{align}
\end{proposition}

It is easily observed that these bounds are informative in the limiting case $\rho \rightarrow 0$ since they both tend towards zero.

\subsubsection{Uniform bounds on credibility regions}
\label{subsec:bounds_credibility_regions}

When $\pi$ stands for the density associated to a posterior distribution, one advantage of Bayesian analysis is its ability to derive the underlying probability distribution of the variable of interest $\boldsymbol{\theta}$ and thereby to provide credibility information under this distribution.
This uncertainty information is particularly relevant and essential for real-world applications.
Since the marginal $\marginal$ stands for an approximation of the original target distribution $\pi$, it is important to control the credibility regions under $\marginal$ w.r.t. those drawn under $\pi$.
The control in total variation distance given by Theorem \ref{theorem:controlTVnorm} is already a good indication.
However, it is possible to quantify more precisely the difference between the credible regions \citep{Robert94} with confidence level ($1-\alpha$) under $\marginal$ and $\pi$, as stated below.

\begin{proposition}
\label{prop:credibility_intervals_2}
  Let $\pi$ be a posterior distribution associated to $\boldsymbol{\theta}$ and $f$ such that $(A_1)$ is verified.
  Let $\mathcal{C}_{\alpha}^{\rho}$ an arbitrary $(1-\alpha)$-credibility region under $\marginal$, that is $\mathbb{P}_{\marginal\text{}}\pr{\boldsymbol{\theta} \in \mathcal{C}_{\alpha}^{\rho}} = 1 - \alpha$ with $\alpha \in (0,1)$.
  Then,
  \begin{align}
    (1-\alpha)\dfrac{N_{\rho}}{D_{-d}(-L_f\rho)} \leq \int_{\mathcal{C}_{\alpha}^{\rho}}\pi(\boldsymbol{\theta})\mathrm{d}\boldsymbol{\theta}\leq \min\pr{1,(1-\alpha)\dfrac{N_{\rho}}{D_{-d}(L_f\rho)}}, \label{eq:credibility_intervals_control_2}
  \end{align}
  where $N_{\rho}$ is defined in \eqref{eq:M_rho}.
\end{proposition}

Proposition \ref{prop:credibility_intervals_2} states that the coverage of $\pi$ under $\mathcal{C}_{\alpha}^{\rho}$ can be determined for a fixed value of $\rho$.
Thus, it is even possible to obtain a theoretical comprehensive description of $\mathcal{C}_{\alpha}^{\rho}$ w.r.t. the initial target density $\pi$ before conducting an AXDA-based inference.
The bounds in \eqref{eq:credibility_intervals_control_2} permit to choose a parameter $\rho$ in order to ensure a prescribed coverage property.
The behavior of these bounds w.r.t. $\rho$ is the same as in Section \ref{subsec:control_TV}, i.e., linear behavior w.r.t. $\rho$ when this parameter is sufficiently small.

\section{Numerical illustrations}
\label{subsec:numerical_illustrations}

This section illustrates the quantitative results shown in Sections \ref{subsec:theory_kernels} and \ref{subsec:theory_gaussian} on three different examples which classically appear in statistical signal processing and machine learning.
As shown in Table \ref{table:kernels_bounds}, the bias induced by considering $\pi_{\rho}$ is mostly driven by the value of the tolerance parameter $\rho$ rather than by the choice of $\kappa_{\rho}$.
Hence, for simplicity, most of the numerical illustrations hereafter consider the case where $\kappa_{\rho}$ is a Gaussian smoothing kernel.
Additional illustrations can be found in the online supplementary material.

\subsection{Multivariate Gaussian example}

\begin{figure}
\centering
  \mbox{{\includegraphics[scale=0.5]{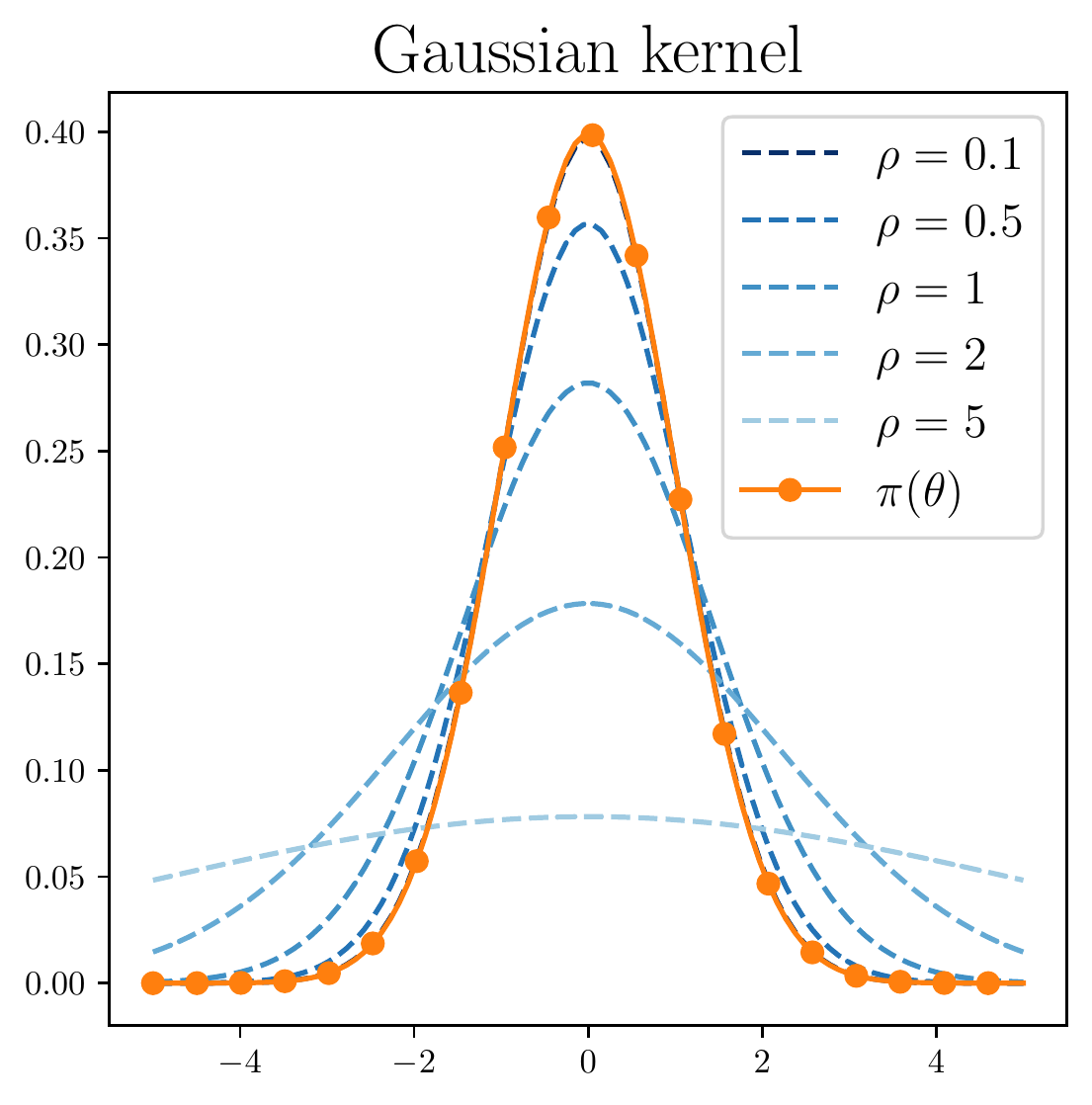}}}
  \mbox{{\includegraphics[scale=0.5]{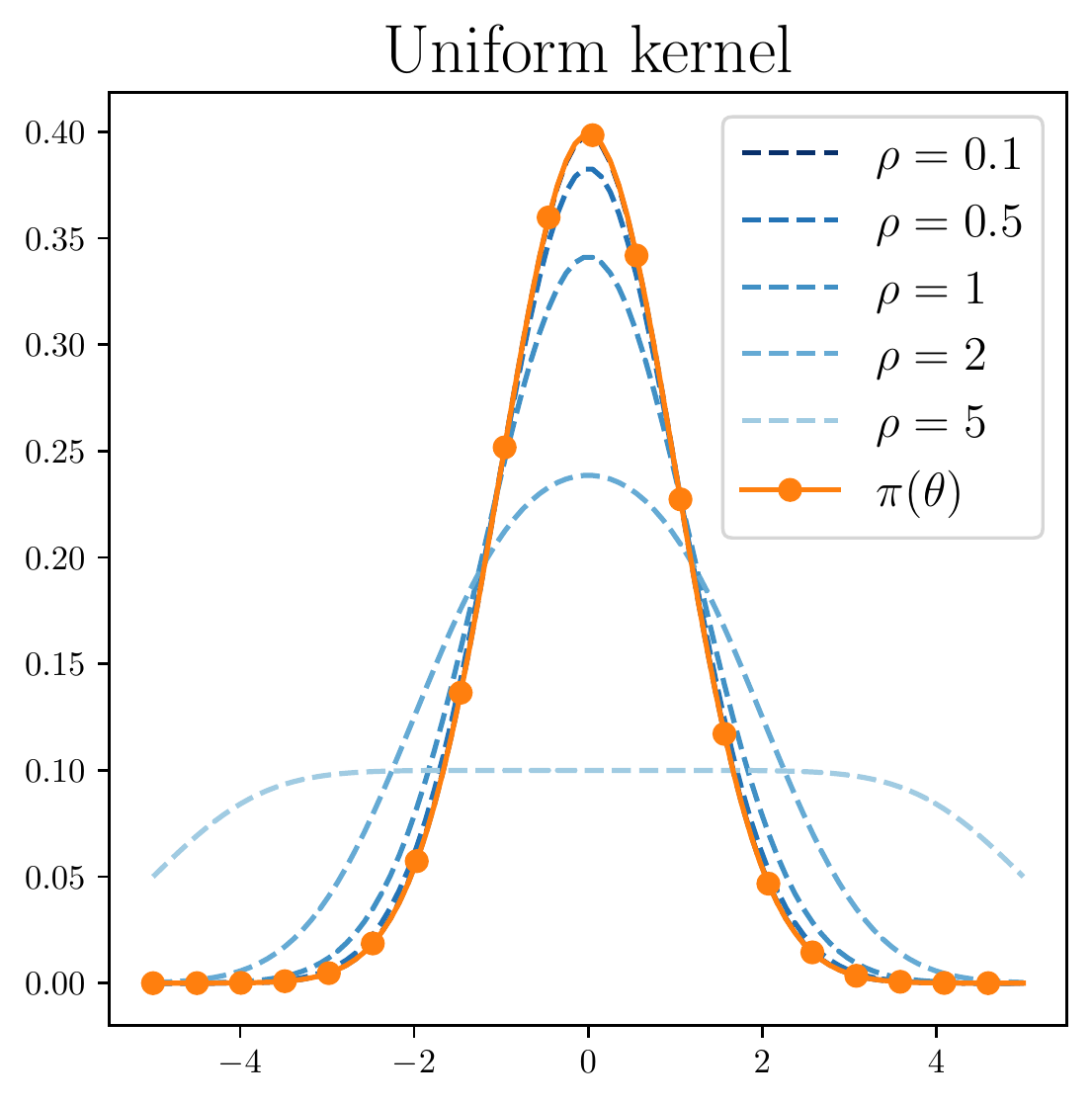}}}
  \caption{Bias between $\pi_{\rho}$ and $\pi$ in the case $\Theta = \mathbb{R}$, $\pi = \mathcal{N}(\mu,\sigma^2)$ with $\mu = 0$ and $\sigma = 1$.
  (left) $\pi_{\rho}$ is built with a Gaussian kernel $\mathcal{N}(0,\rho^2)$ and (right) with a uniform kernel on $[-\rho,\rho]$.
  Note that the curves associated to $\pi$ and $\pi_{\rho}$ for $\rho = 0.1$ are overlapping.}
  \label{fig:gaussian_example_1}
\end{figure}

We start by performing a sanity check with the simple case where $\pi$ stands for a multivariate Gaussian density that is 
\begin{equation}
	\pi(\boldsymbol{\theta}) = \mathcal{N}(\boldsymbol{\theta}|\boldsymbol{\mu},\B{\Sigma}),
\end{equation}
where $\B{\Sigma}$ is assumed to be positive definite.
If $\kappa_{\rho}(\cdot,\boldsymbol{\theta})$ is taken to be Gaussian density with mean $\boldsymbol{\theta}$ and covariance matrix $\rho^2\B{I}_d$, then one can show that
\begin{equation}
	\pi_{\rho}(\boldsymbol{\theta}) = \mathcal{N}(\boldsymbol{\theta}|\boldsymbol{\mu},\B{\Sigma} + \rho^2\B{I}_d).
\end{equation}
In particular, let consider the univariate setting, that is $\Theta = \mathbb{R}$, $\B{\Sigma} = \sigma^2$.
In this case, the variance under $\pi_{\rho}$ is $\sigma^2 + \rho^2$ and simply corresponds to the variance under $\pi$ inflated by a factor $\rho^2$.
Therefore, the approximation will be reasonable if $\rho^2/\sigma^2$ is sufficiently small, see Figure \ref{fig:gaussian_example_1}.
In this Figure, we also show the approximation induced by considering a uniform kernel instead of a Gaussian one.
The smoothing via the uniform kernel performs slightly better than Gaussian smoothing due to its lower variance.
In both cases, the approximation is reasonable for small $\rho$ although $\pi_{\rho}$, built with a uniform kernel, no longer belongs to the Gaussian family.

In order to illustrate the proposed upper bounds on both 2-Wasserstein and total variation distances, we consider a covariance matrix $\B{\Sigma}$ which stands for a squared exponential matrix commonly used in applications involving Gaussian processes \citep{Higdon2007} and which writes
\begin{equation}
  \Sigma_{ij} = 2 \exp\pr{-\dfrac{(s_i-s_j)^2}{2 a^2}} + 10^{-6}\delta_{ij}, \forall i,j \in [d] \label{eq:simu1_ex1}
\end{equation}
where $a = 1.5$, $s_{i,i\in[d]}$ are regularly spaced scalars on $[-3,3]$ and $\delta_{ij} = 1$ if $i=j$ and zero otherwise.

\begin{figure}[h]
\centering
  \mbox{{\includegraphics[scale=0.33]{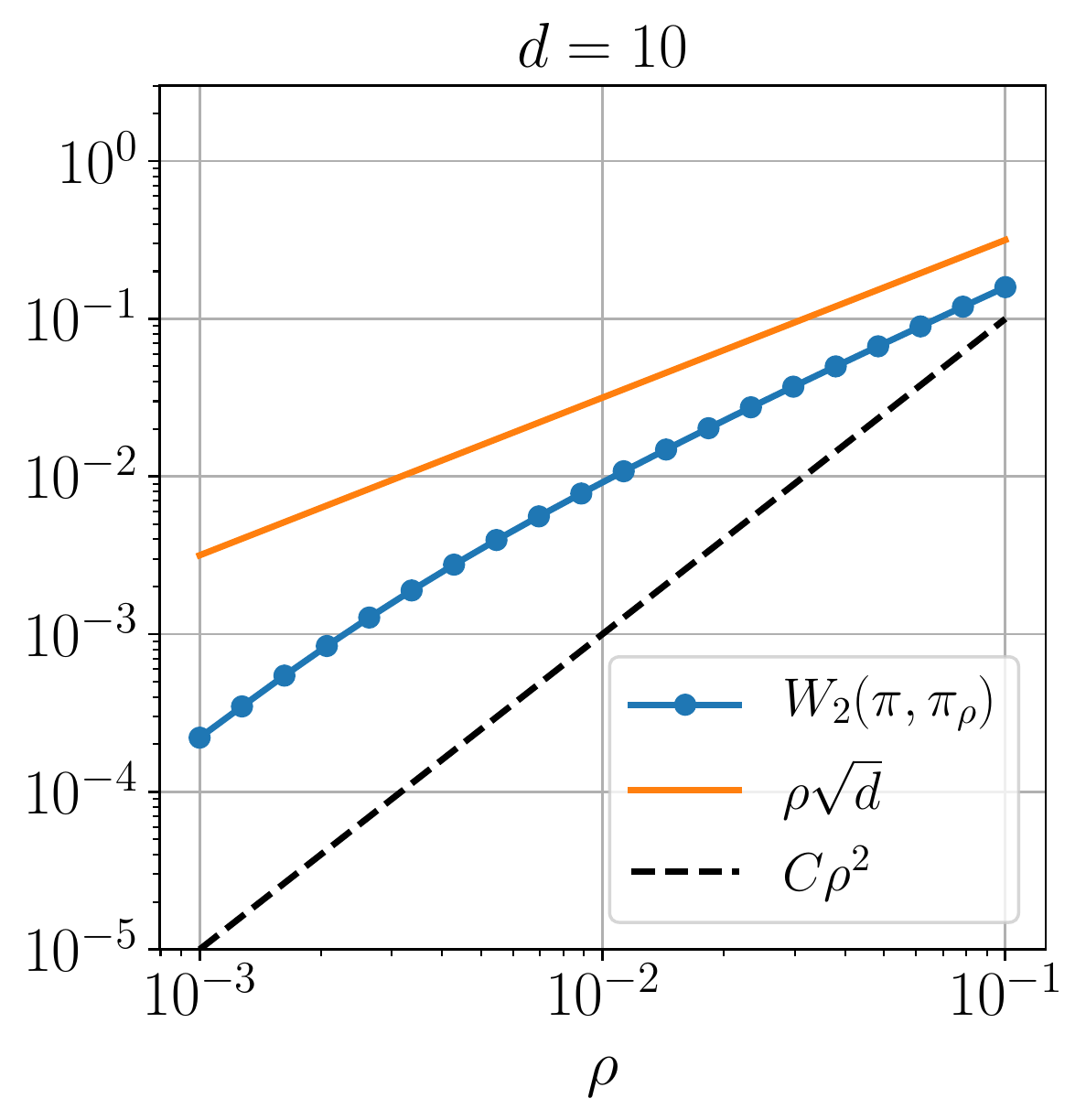}}}
  \mbox{{\includegraphics[scale=0.33]{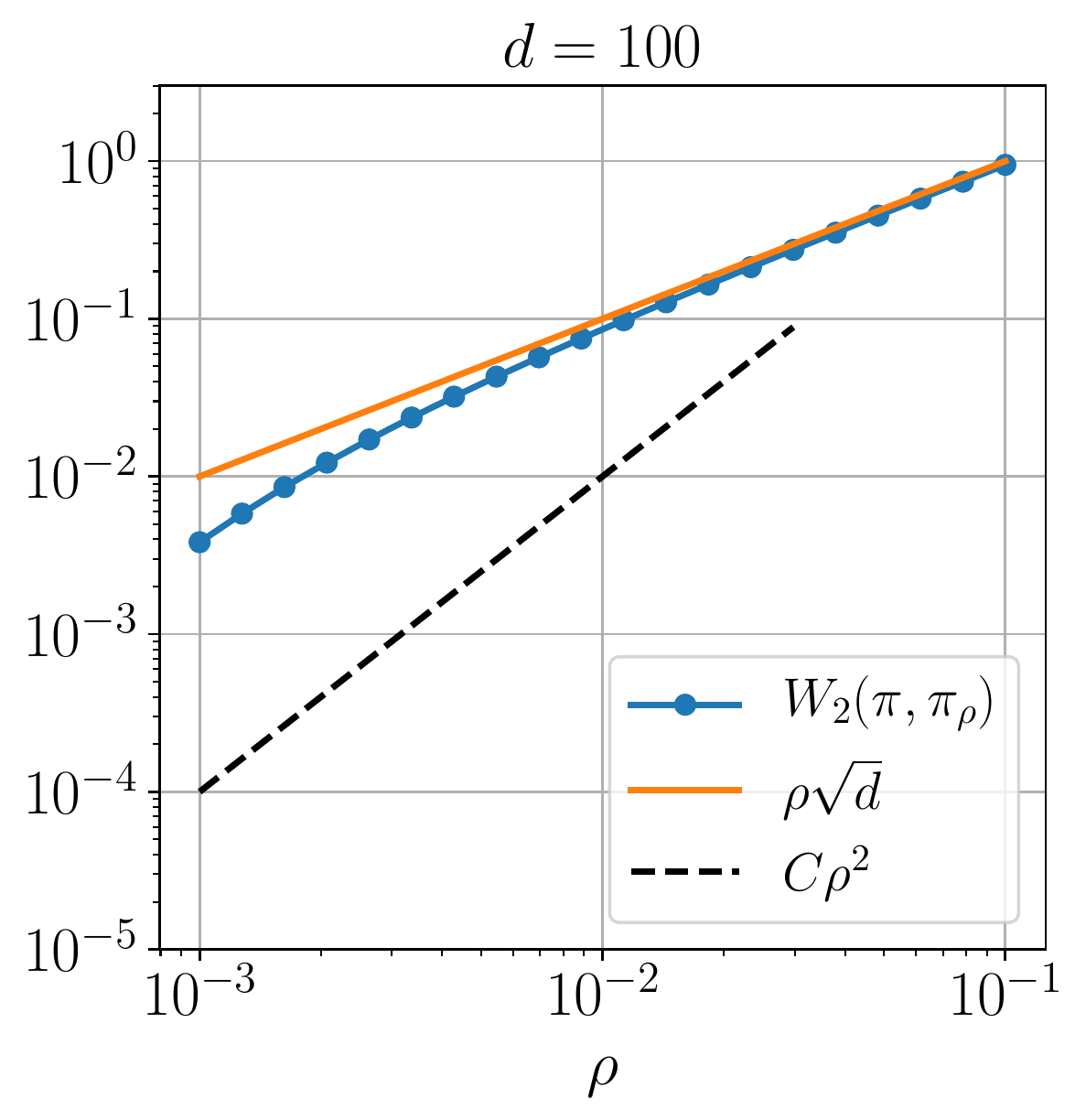}}}
  \mbox{{\includegraphics[scale=0.33]{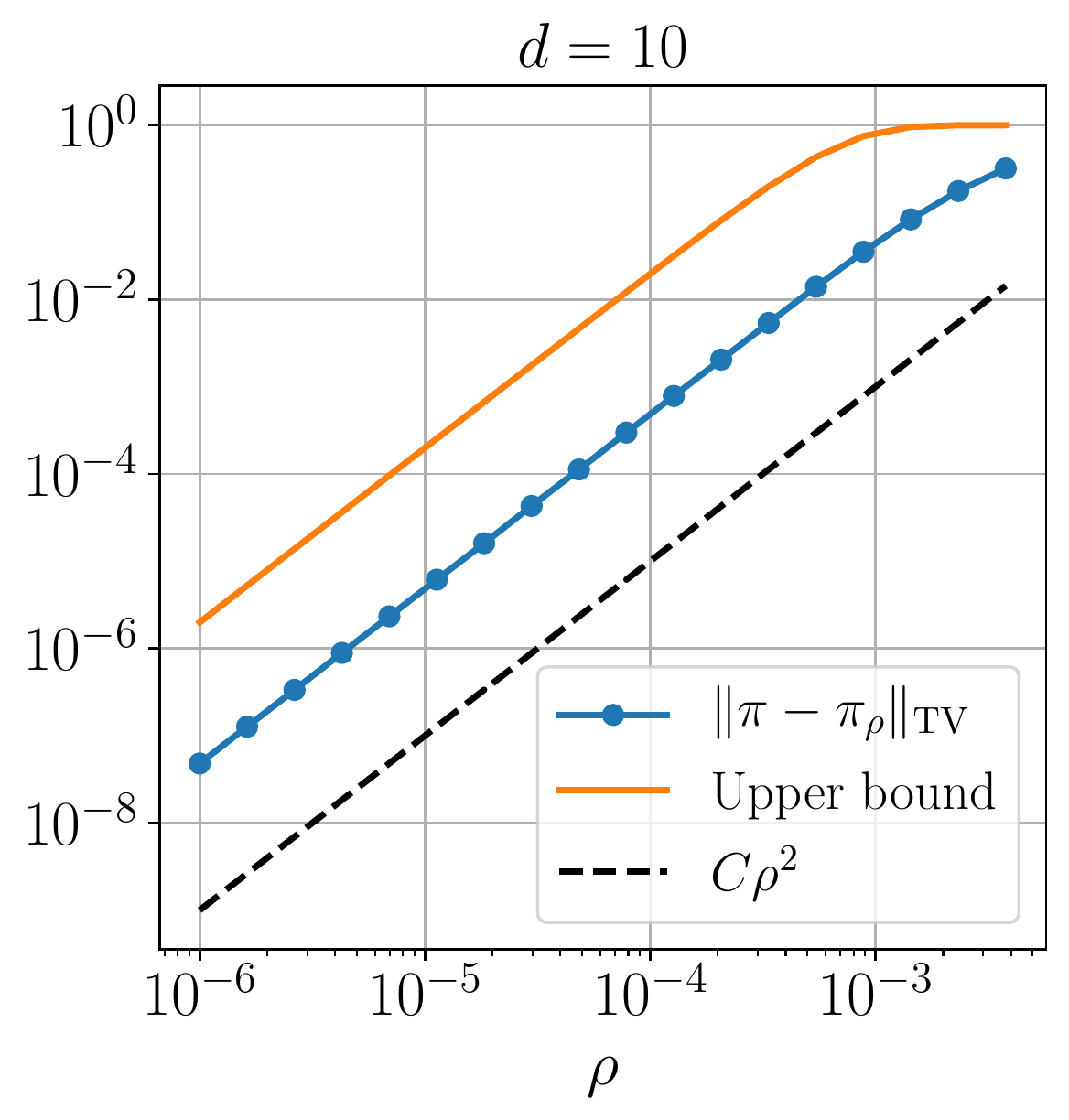}}}
  \mbox{{\includegraphics[scale=0.33]{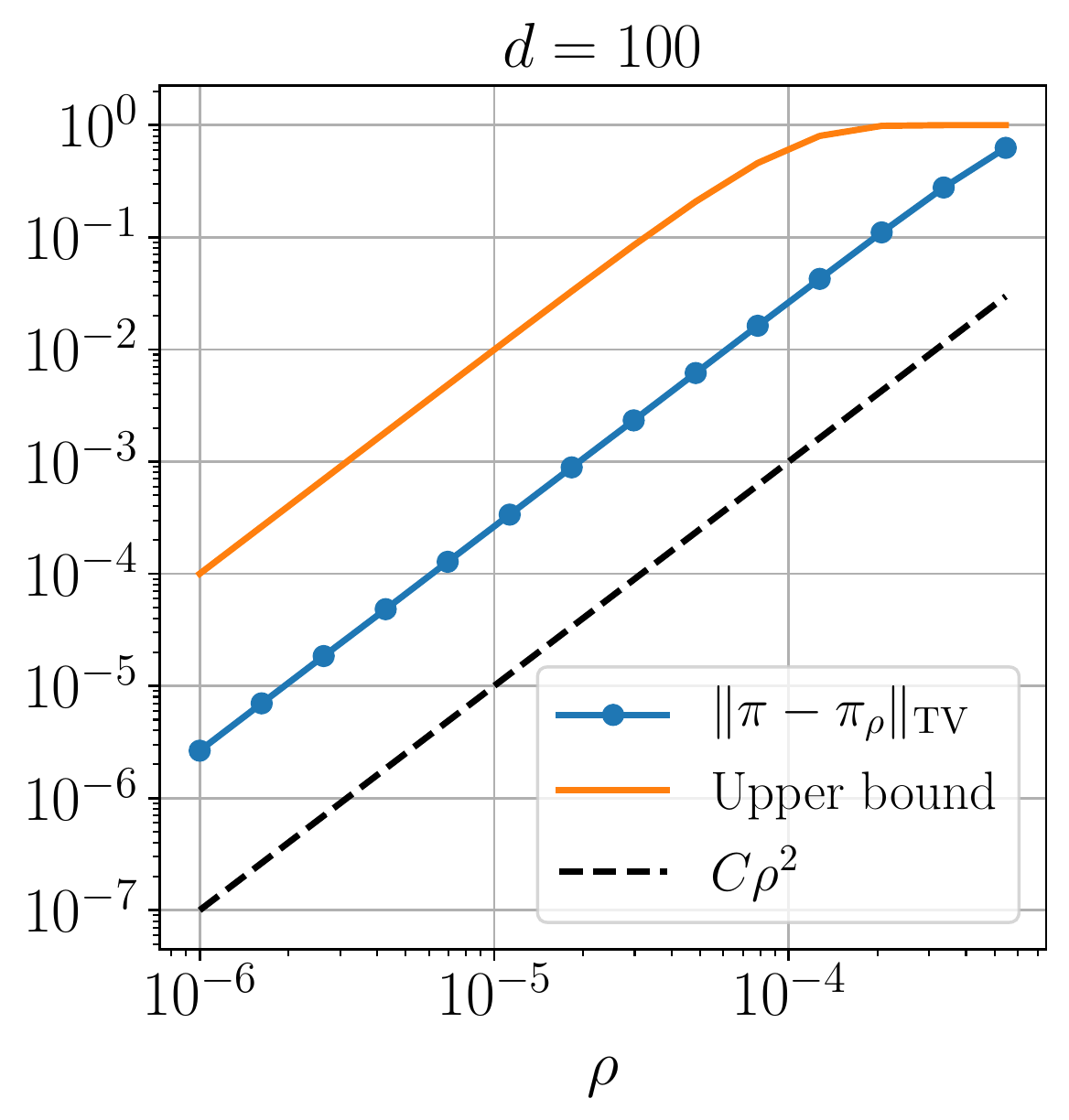}}}
  \caption{For $d \in \{10,100\}$, illustration of the quantitative bounds \eqref{eq:kernel_upp_bound} and \eqref{eq:controlTVnorm_2} associated to 2-Wasserstein and total variation distances, respectively.
  The decay in $\mathcal{O}(\rho^2)$ is shown via the dashed line $C\rho^2$ where $C$ is a constant.}
  \label{fig:gaussian_example_2}
\end{figure}

Figure \ref{fig:gaussian_example_2} shows the behavior of the quantitative bounds derived in Proposition \ref{prop:bound_kernel} and Theorem \ref{theorem:controlTVnorm_2} for $d \in \{10,100\}$.
The Gaussian case allows to compute exactly $W_2(\pi,\pi_{\rho})$ by noting that $W_2^2(\pi,\pi_{\rho}) = \mathrm{Trace}(\B{\Sigma} + \rho^2\B{I}_d - 2\rho\B{\Sigma}^{1/2})$.
On the other hand, $\nr{\pi-\pi_{\rho}}_{\mathrm{TV}}$ has been estimated by using a Monte Carlo approximation.
One can note that the general upper bound on the 2-Wasserstein distance is quite conservative for small $\rho$ since it does not catch the behavior in $\mathcal{O}(\rho^2)$ when $\rho$ is small.
This is essentially due to the fact that this bound only assumes a finite moment property and does not require any regularity assumptions on $\pi$ such as differentiability or strong convexity of its potential.
On the contrary, the bound on the total variation distance, derived under stronger assumptions, manages to achieve a rate of the order $\mathcal{O}(\rho^2)$ for small $\rho$.

\subsection{Sparse linear regression}
\label{example:LASSO} \normalfont

	We study here a generalized version of the least absolute shrinkage and selection operator (lasso) regression problem analyzed by \cite{Park2008}.
  We assume a standard linear regression problem where centered observations $\B{y} \in \mathbb{R}^n$ are related to the unknown parameters $\boldsymbol{\theta} \in \mathbb{R}^{d}$ via the model $\B{y} = \B{X}\boldsymbol{\theta} + \boldsymbol{\varepsilon}$, where $\B{X} \in \mathbb{R}^{n \times d}$ stands for a known standardized design matrix and $\boldsymbol{\varepsilon} \sim \mathcal{N}(\B{0}_n,\sigma^2\B{I}_n)$.
	By considering a generalized Laplacian prior distribution for $\boldsymbol{\theta}$, the target posterior distribution has density for all $\boldsymbol{\theta} \in \mathbb{R}^d$,
	\begin{align}
		\pi(\boldsymbol{\theta}) \triangleq \pi(\boldsymbol{\theta}|\B{y}) \propto \exp\pr{-\dfrac{1}{2\sigma^2}\nr{\B{y}-\B{X}\boldsymbol{\theta}}_2^2 - g(\B{B}\boldsymbol{\theta})}
	\end{align}
  where $g(\B{B}\boldsymbol{\theta}) = \tau\nr{\B{B}\boldsymbol{\theta}}_1$ with $\tau > 0$ and $\B{B} \in \mathbb{R}^{k \times d}$ an arbitrary matrix acting on $\boldsymbol{\theta}$.
	The choice of such a prior may promote a form of sparsity (lasso). 
	For instance, this matrix $\B{B}$ might stand for a $p$-th order difference operator \citep{Bredies2010} which is highly used in signal and image processing problems.
	As an archetypal example, the case $p=1$ leads to the well-known total variation regularization function \citep{Chambolle2010} used to recover piecewise constant signals.

	Note that because of the presence of the matrix $\B{B}$, finding an exact data augmentation leading to an efficient sampling scheme is not possible for the general case $\B{B} \neq \B{I}_d$.
	Instead, an AXDA model makes the posterior sampling task possible.
	Indeed, by regularizing the prior with a Gaussian term, the joint density $\joint$ writes
	\begin{align}
		\joint(\boldsymbol{\theta},\B{z}) \propto \exp\pr{-\dfrac{1}{2\sigma^2}\nr{\B{y}-\B{X}\boldsymbol{\theta}}_2^2 - g(\B{z}) -\dfrac{1}{2\rho^2}\nr{\B{Bx}-\B{z}}_2^2}. \label{eq:lasso_posterior_approx}
	\end{align}
	By resorting to a Gibbs algorithm to sample from \eqref{eq:lasso_posterior_approx}, one can now use a simple data augmentation scheme \citep{Park2008} to sample from the $\B{z}$-conditional.
	On the other hand, sampling from the $\boldsymbol{\theta}$-conditional, which is a multivariate Gaussian distribution, can be undertaken efficiently with state-of-the-art approaches \citep{Papandreou2011,barbos2017clone,Marnissi2018}.

	In this specific case, the potential $g_{\rho}$ associated to the smoothed prior distribution (see \eqref{eq:potential}) has a closed-form expression given for all $\boldsymbol{\theta} \in \mathbb{R}^d$, by
	\begin{align}
		g_{\rho}(\boldsymbol{\theta}) &= \dfrac{k}{2}\log(2\pi\rho^2)-\log\prod_{i=1}^k\displaystyle\int_{\mathbb{R}}\exp\pr{-\tau|z_i|-\dfrac{1}{2\rho^2}(\B{b}_i^T\boldsymbol{\theta}-z_i)^2}\mathrm{d}z_i \nonumber \\
	  &= \dfrac{k}{2}\log(2\pi\rho^2) \\
	  &-\log\prod_{i=1}^k\pr{a(\boldsymbol{\theta})\br{\exp\pr{b(\boldsymbol{\theta})^2}\bbr{1-\mathrm{erf}(b(\boldsymbol{\theta}))} + \exp\pr{c(\boldsymbol{\theta})^2}\bbr{1-\mathrm{erf}(c(\boldsymbol{\theta}))}}} \label{eq:smooth_prior}
	\end{align}
	with $a(\boldsymbol{\theta})=\sqrt{\pi\rho^2/2}\exp\pr{-(\B{b}_i^T\boldsymbol{\theta})^2/(2\rho^2)}$, $b(\boldsymbol{\theta}) = \sqrt{\rho^2/2}(\tau-\B{b}_i^T\boldsymbol{\theta}/\rho^2)$, $c(\boldsymbol{\theta}) = \sqrt{\rho^2/2}(\tau+\B{b}_i^T\boldsymbol{\theta}/\rho^2)$ and $\B{b}_i \in \mathbb{R}^d$ standing for the $i$-th row of $\B{B}$.
	Note that in more general cases where $g_{\rho}$ has no closed form, one can estimate it by a Monte Carlo approximation.
\begin{figure}[!h]
\centering
  \mbox{{\includegraphics[scale=0.38]{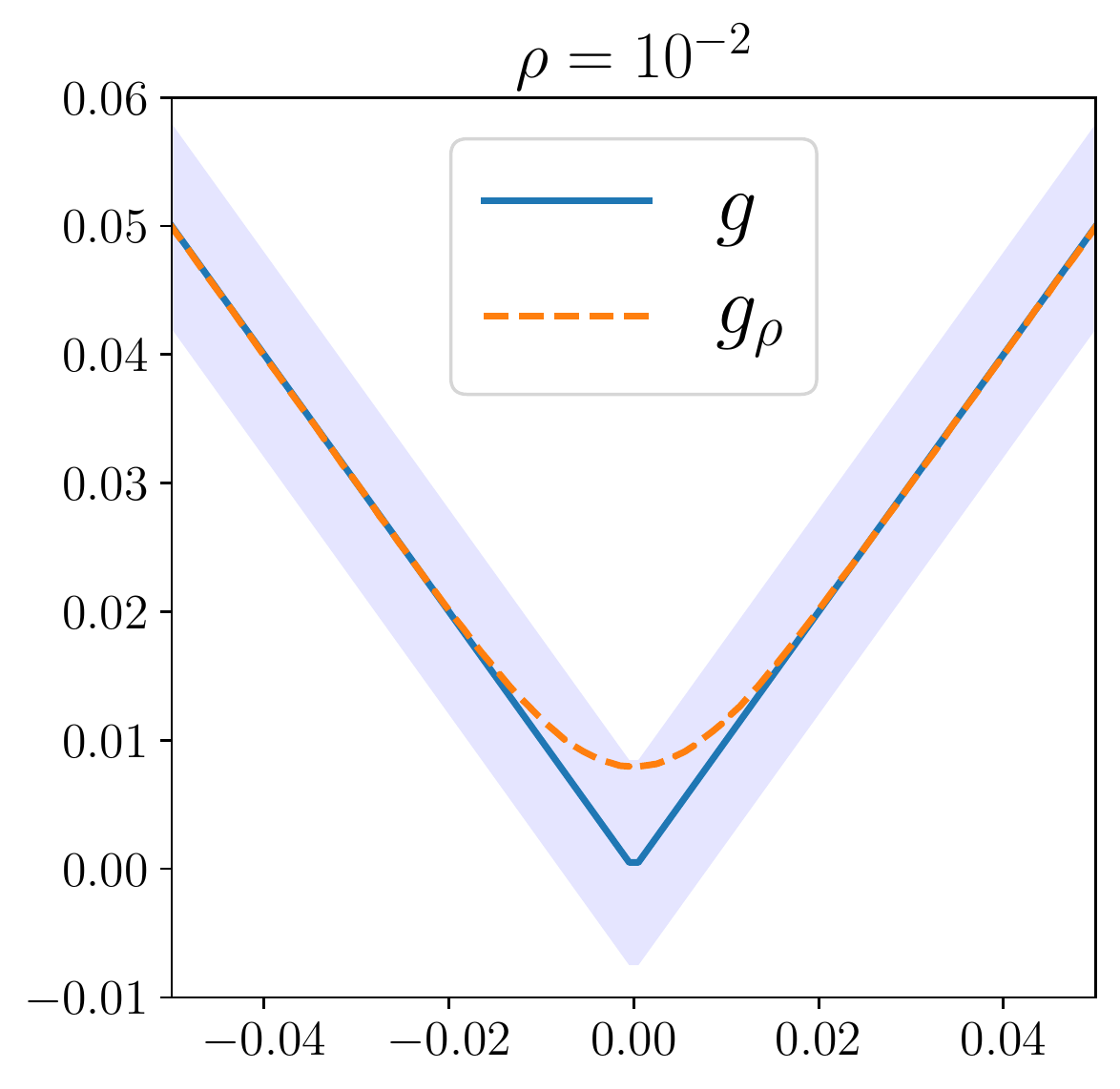}}}
  \mbox{{\includegraphics[scale=0.38]{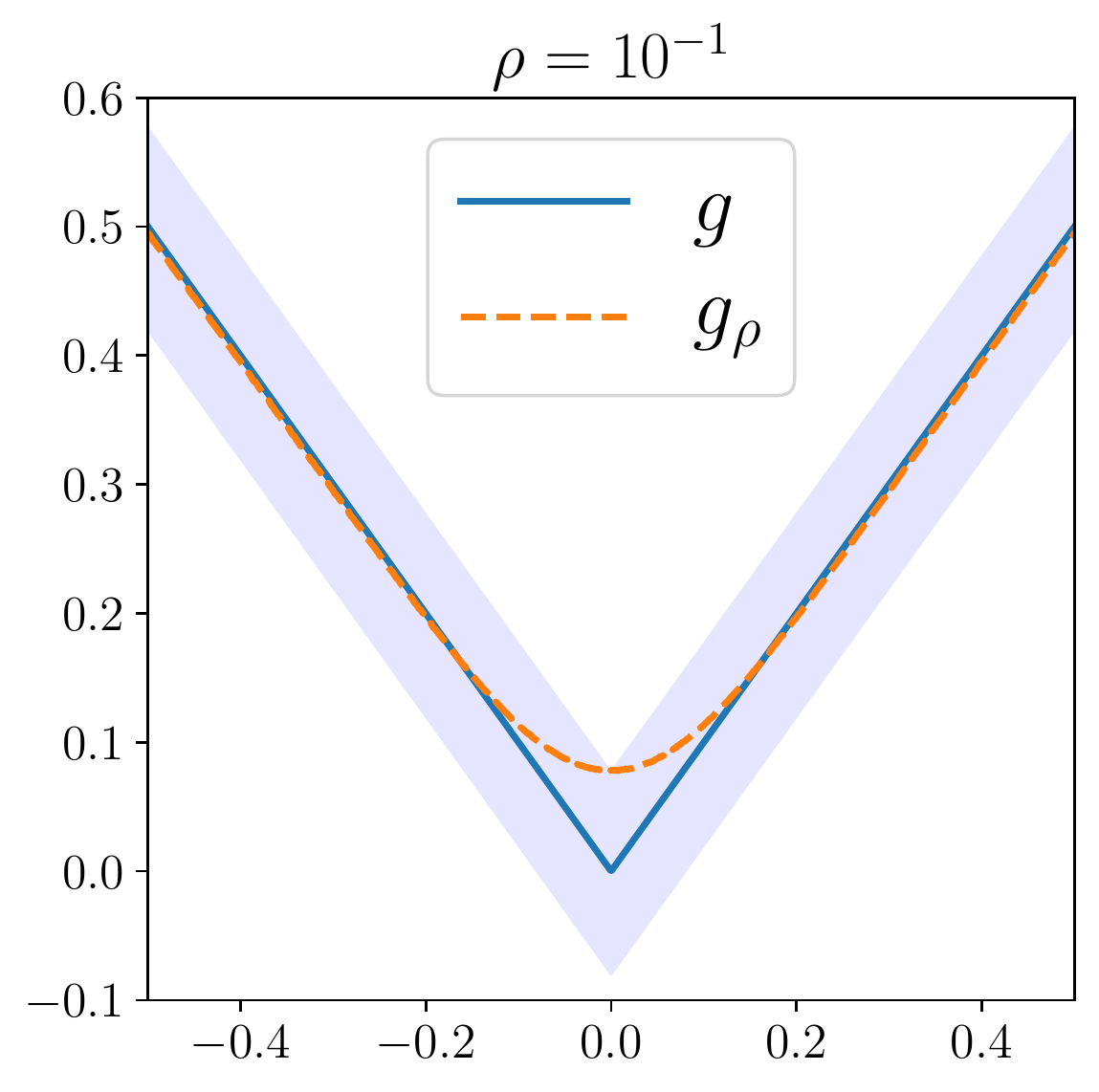}}}
  \mbox{{\includegraphics[scale=0.38]{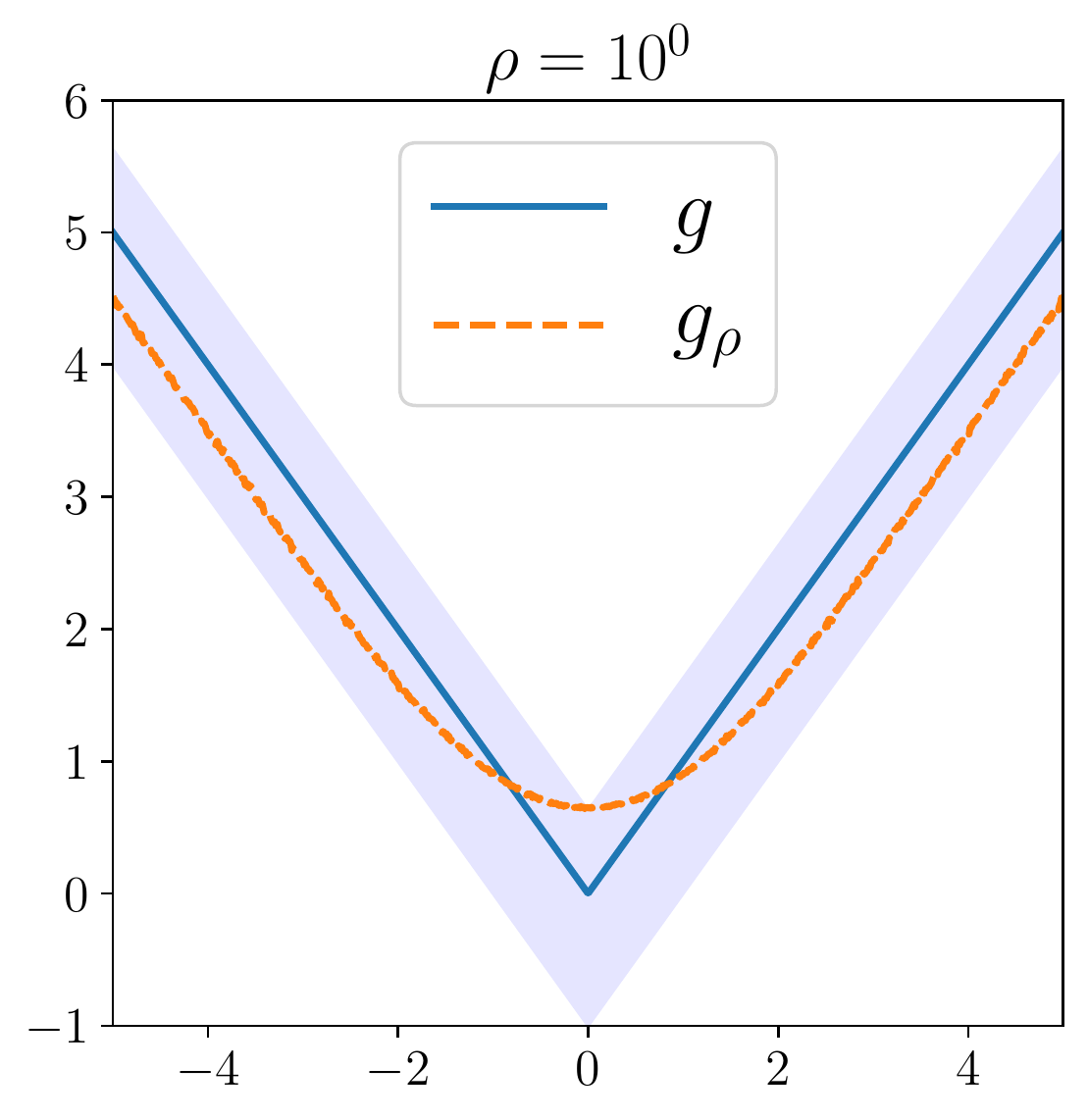}}}
  \mbox{{\includegraphics[scale=0.38]{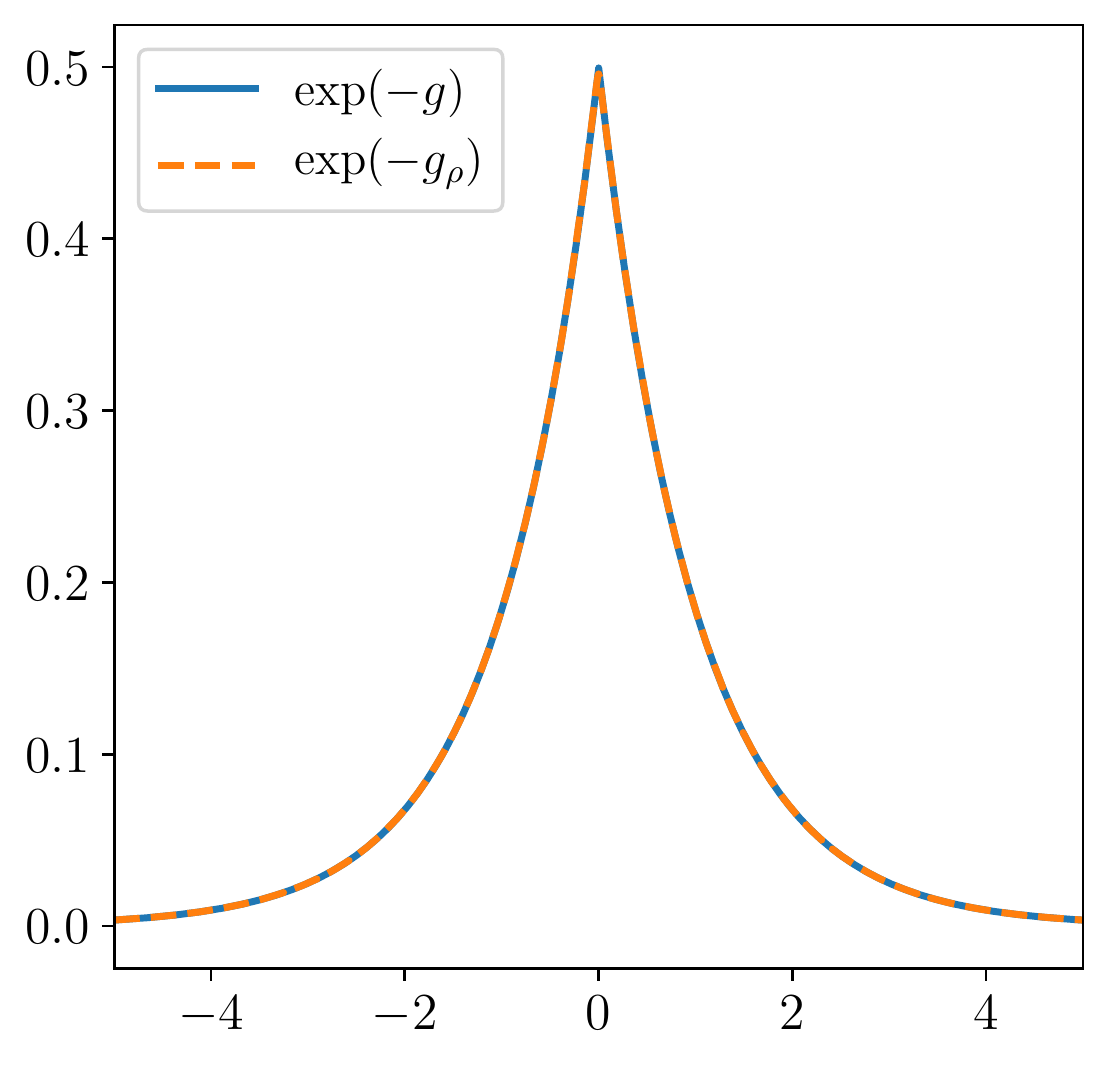}}}
  \mbox{{\includegraphics[scale=0.38]{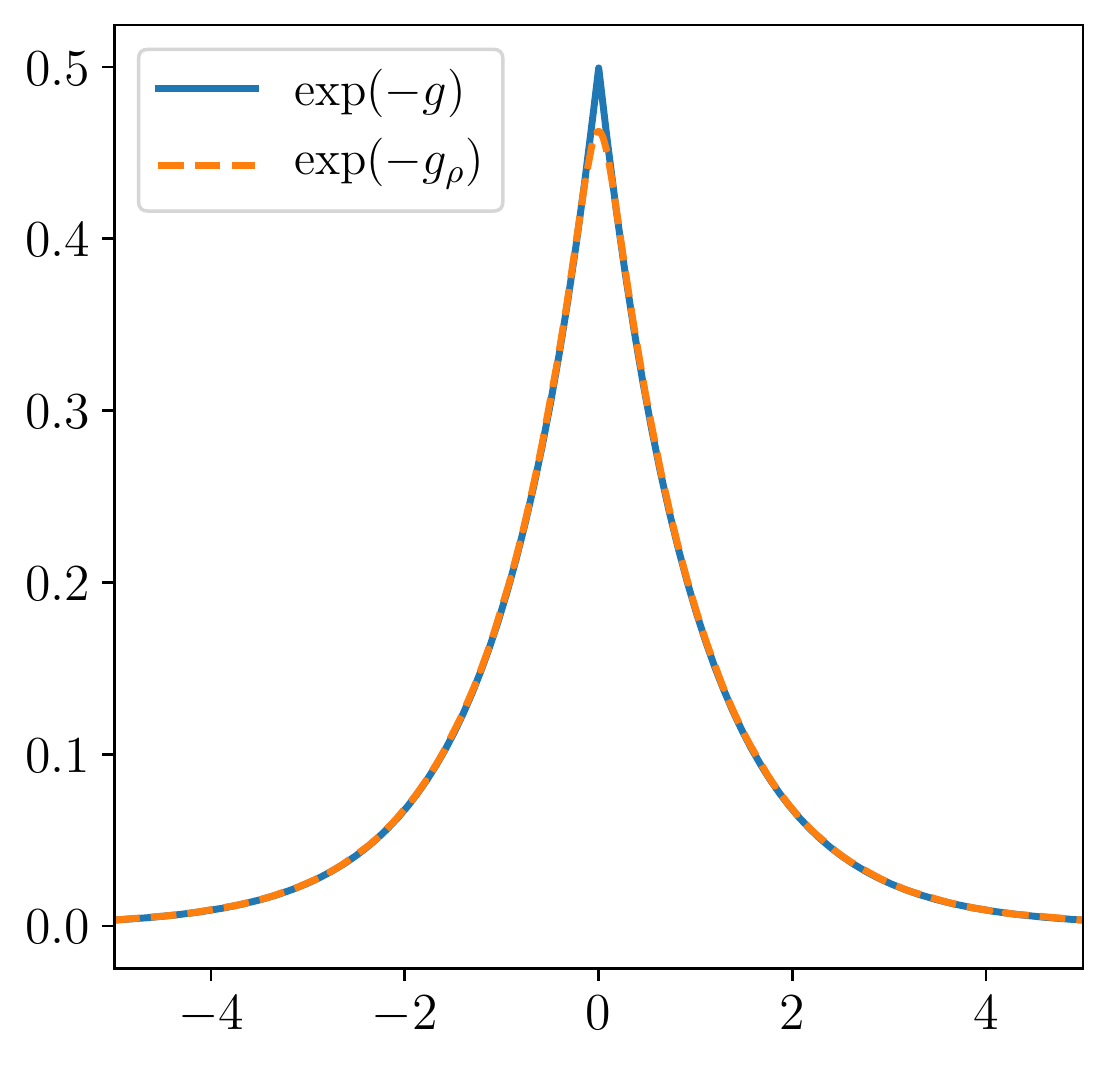}}}
  \mbox{{\includegraphics[scale=0.38]{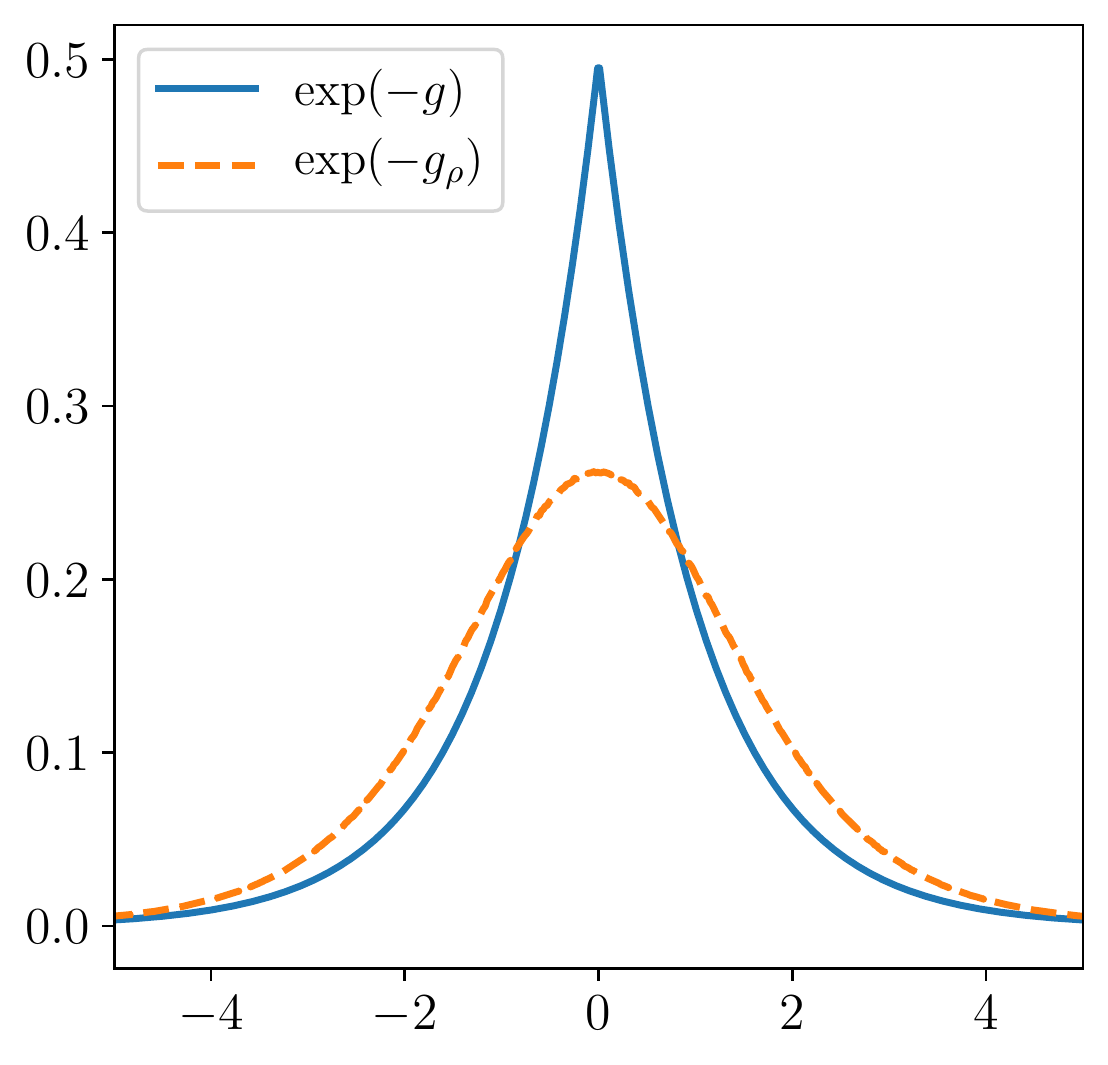}}}
  \mbox{{\includegraphics[scale=0.38]{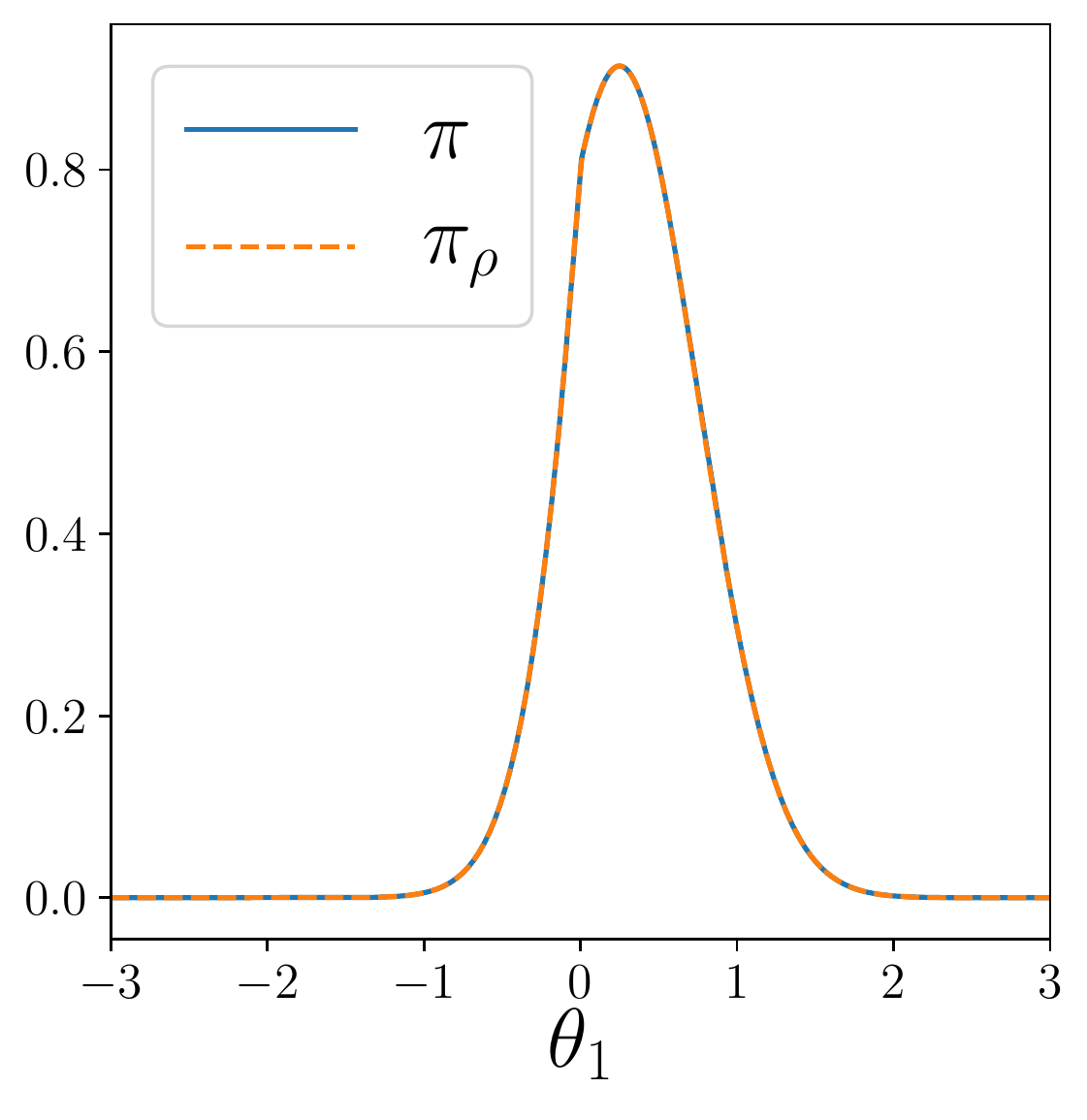}}}
  \mbox{{\includegraphics[scale=0.38]{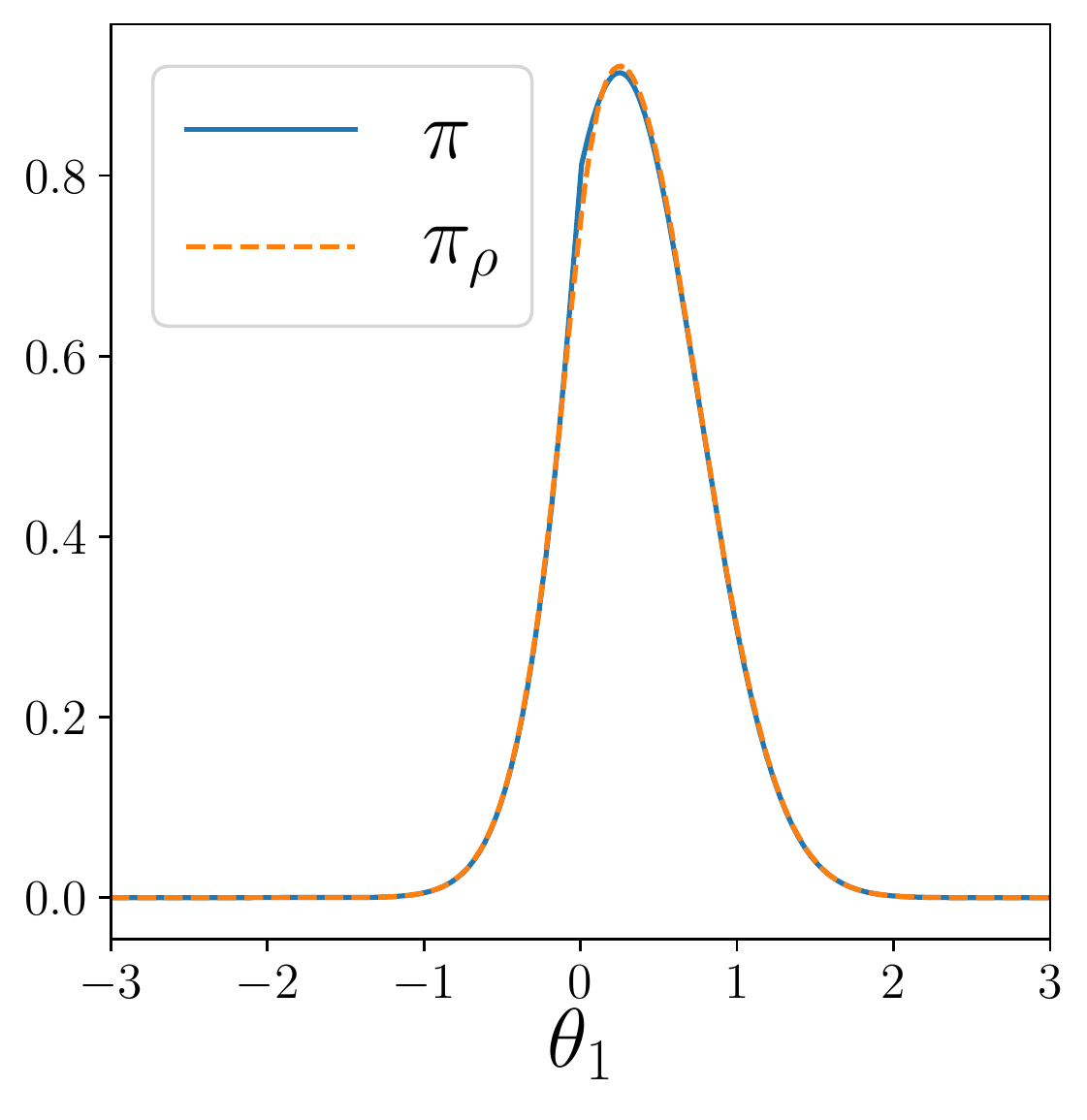}}}
  \mbox{{\includegraphics[scale=0.38]{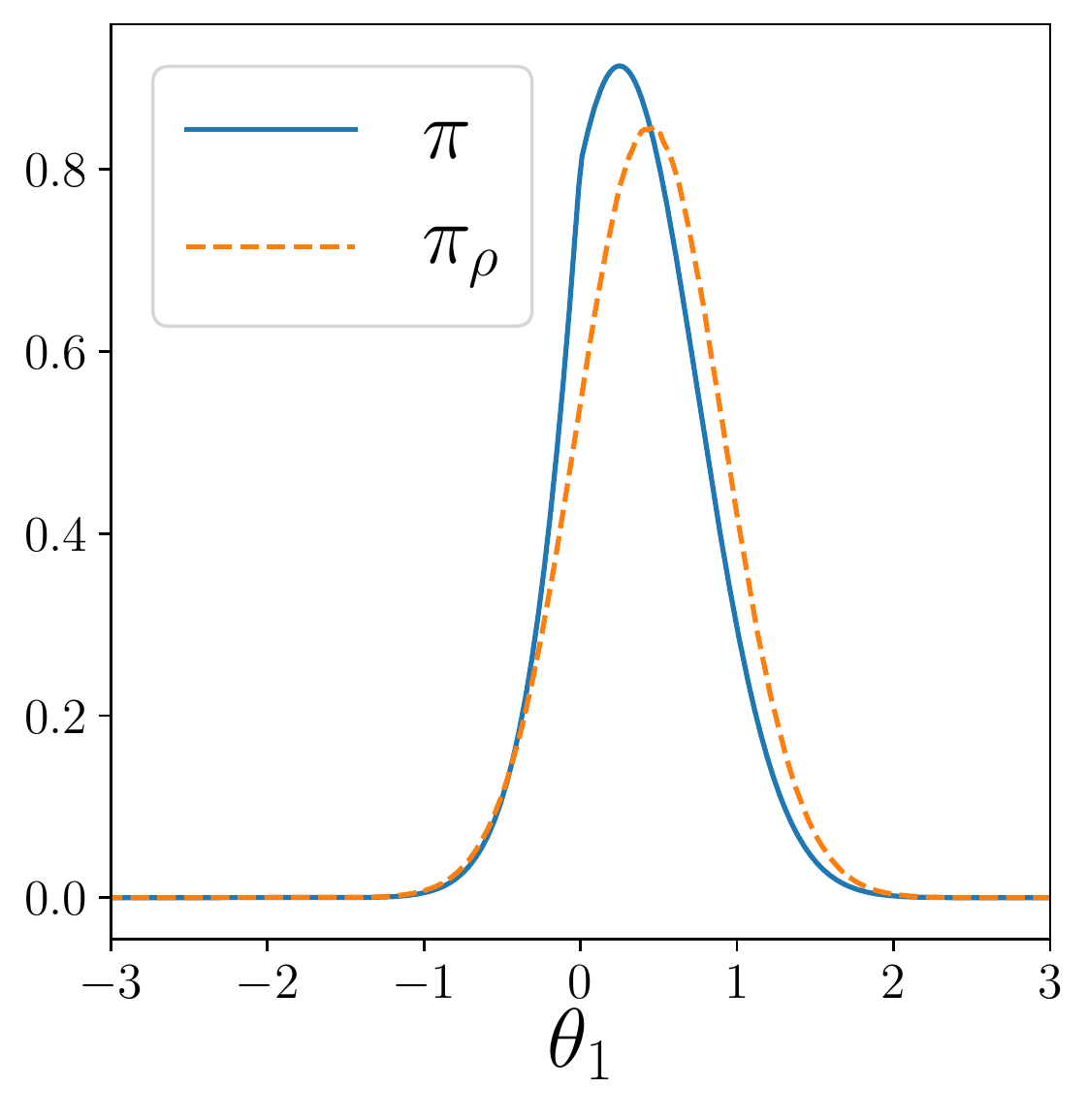}}}
  \caption{From left to right, $\rho=0.01$, $\rho=0.1$ and $\rho=1$. (1st row) Behaviors of $g$ (blue) and $g_{\rho}$ (orange) where the contours of the shaded area correspond to $g + L_{\rho}$ and $g + U_{\rho}$; (2nd row) the corresponding normalized smoothed prior densities proportional to $\exp(-g)$ and $\exp(-g_{\rho})$; (3rd row) posterior densities $\marginal$ w.r.t. $\rho$.}
  \label{fig:prior_potential}
\end{figure}

Figure \ref{fig:prior_potential} shows the behavior of the regularized potential $g_{\rho}$ defined in \eqref{eq:smooth_prior} for several values of the parameter $\rho$ along with the associated smoothed prior and posterior distributions.
For simplicity and pedagogical reasons, the univariate case corresponding to $\boldsymbol{\theta} = \theta_1 \in \mathbb{R}$ and $\B{B} = 1$ has been considered.
The regularization parameter $\tau$ has been set to $\tau = 1$.
The contours of the shaded area correspond to $g + L_{\rho}$ and $g + U_{\rho}$.
The potential $g_{\rho}$ is a smooth approximation of the potential $g$ associated to the initial prior as expected, see Property \emph{iv)} in Proposition \ref{prop:differentiability_prior}.
Note that the inequalities derived in \eqref{eq:bound_prior} are verified.
Although this approximation seems similar to the Moreau-Yosida regularization of a non-smooth potential function \citep{Combettes2011}, the rationale behind this approximation is different.
Indeed, the Moreau-Yosida envelope stands for a particular instance of the infimal convolution between two convex functions (an initial potential and a Gaussian one).
On the other hand, $g_{\rho}$ is the potential associated to a smoothed density obtained by convolution with a Gaussian kernel.
In addition, the third row of Figure \ref{fig:prior_potential} shows the form of the posterior of $\theta_1$ defined in \eqref{eq:lasso_posterior_approx} for $y=1$, $x=2$ and $\sigma=1$ and derived from the smoothed prior distributions shown in Figure \ref{fig:prior_potential}.
For sufficiently small values of $\rho$, the marginal $\marginal$ stands for a quite accurate approximation of the original target $\pi$.

\begin{table}
\caption{Illustration of the bound derived in \eqref{eq:credibility_intervals_control_2} for the marginal posterior $\marginal$ depicted in Section \ref{example:LASSO}. The (1-$\alpha$)-credibility intervals $\mathcal{C}_{\alpha}$ and $\mathcal{C}_{\alpha}^{\rho}$ are the highest posterior density regions associated to each density with $\alpha=0.05$.}
\begin{center}
\begin{tabular}{ccccc}
\bfseries $\rho$ & $\mathcal{C}_{\alpha}$ & $\mathcal{C}^{\rho}_{\alpha}$ & $\int_{\mathcal{C}_{\alpha}^{\rho}}\pi(\theta_1)\mathrm{d}\theta_1$ & $\mathcal{I}_{\alpha}^{\rho}$\\
\hline
$10^{-3}$ & [-0.47,1.24] & [-0.47,1.24] & 0.95 & [0.949,0.951]\\
$10^{-2}$ & idem & [-0.47,1.24] & 0.95 & [0.948,0.952]\\
$10^{-1}$ & idem & [-0.47,1.24] & 0.95 & [0.88,1]\\
$10^0$ & idem & [-0.47,1.37] & 0.96 & [0.34,1]\\
\end{tabular}
\end{center}
\label{table:lasso_credibility}
\end{table}

Table \ref{table:lasso_credibility} illustrates the bounds derived in \eqref{eq:credibility_intervals_control_2} for $\rho\in\left\{1,10^{-1},10^{-2},10^{-3}\right\}$.
For each case, the values of the bounds are summarized in the interval
\begin{align}
\mathcal{I}_{\alpha}^{\rho} = [(1-\alpha)N_{\rho}/D_{-d}(-L_f\rho),\min(1,(1-\alpha)N_{\rho}/D_{-d}(L_f\rho)],
\end{align}
and the real coverage $\int_{\mathcal{C}_{\alpha}^{\rho}}\pi(\theta_1)\mathrm{d}\theta_1$ is also reported.
The (1-$\alpha$)-credibility intervals $\mathcal{C}_{\alpha}$ and $\mathcal{C}_{\alpha}^{\rho}$ have been chosen to be the highest posterior density regions associated to each density with $\alpha=0.05$.
Note that the theoretical coverage interval $\mathcal{I}_{\alpha}^{\rho}$ becomes informative only if $\rho$ is sufficiently small which is not surprising since the assumptions on the potential of $\joint$ are weak.
Indeed, the form of the density (e.g. symmetry or unimodality) is not taken into account in the derived bounds.
Regarding the empirical value of the coverage $\int_{\mathcal{C}_{\alpha}^{\rho}}\pi(\theta_1)\mathrm{d}\theta_1$, we emphasize that the marginal $\marginal$ stands for a conservative approximation of $\pi$ in this example.
Indeed, in each case, the (1-$\alpha$)-credibility interval under $\marginal$ denoted $\mathcal{C}_{\alpha}^{\rho}$ covers at least $100(1-\alpha)$\% of the probability mass under $\pi$.

\subsection{Illustration on an image inpainting problem}
\label{subsec:inpainting}

We illustrate here the correctness of the proposed approach on a multidimensional and non-Gaussian example which classically appears in image processing. 
To this purpose, we consider the observation of a damaged and noisy image $\B{y} \in \mathbb{R}^n$ (represented as a vector by lexicographic ordering) related to the unknown original image $\boldsymbol{\theta} \in \mathbb{R}^d$ by the linear model
\begin{equation}
	\B{y} = \B{H}\boldsymbol{\theta} + \boldsymbol{\varepsilon}, \quad \boldsymbol{\varepsilon} \sim \mathcal{N}(\B{0}_n,\sigma^2\B{I}_n), \label{eq:inpainting_model}
\end{equation}
where $n < d$, $\B{H} \in \mathbb{R}^{n \times d}$ stands for a decimation binary matrix.
The dimension $d$ being typically large (e.g., $10^3 \leq d \leq 10^9$), these problems require scalable inference algorithms.
Since the matrix $\B{H}$ is not invertible, the linear inverse problem \eqref{eq:inpainting_model} is ill-posed.
To cope with this issue, we assign the total variation prior distribution to the unknown parameter $\boldsymbol{\theta}$, leading to the posterior distribution
\begin{equation}
	\pi(\boldsymbol{\theta}  | \B{y}) \propto \exp\pr{-\frac{1}{2\sigma^2}\nr{\B{y} - \B{H}\boldsymbol{\theta}}_2^2 - \tau \sum_{1\leq i \leq d}\nr{(\B{D}\boldsymbol{\theta})_{i}}_2}, \label{eq:posterior}
\end{equation}
where $\tau > 0$ is a regularization parameter, $\B{D}\boldsymbol{\theta} = (\B{D}_1\boldsymbol{\theta},\B{D}_2\boldsymbol{\theta}) \in \mathbb{R}^{2 \times d}$ is the two-dimensional discrete gradient associated to the image $\boldsymbol{\theta}$ and the notation $\B{M}_i$ stands for the $i$-th column of the matrix $\B{M}$, see \cite{Chambolle2010} for more details about the total variation regularization.
The presence of the operator $\B{D}$ and the non-differentiability of the total variation norm rule out the use of common data augmentation schemes and simulation-based algorithms (e.g., Hamiltonian and Langevin Monte Carlo methods).
Possible surrogates are proximal MCMC methods \citep{Pereyra2016B,Durmus2018} which replace the non-differentiable posterior distribution by a smooth approximation based on the proximity operator \citep{Combettes2011} of the total variation norm.
However, the latter does not admit a closed-form expression and iterative routines are commonly used to approximate the latter \citep{Chambolle2004} leading to higher computational costs.

To mitigate these issues, we propose to rely on a particular instance of AXDA by smoothing the total variation prior with a Gaussian term, leading to the approximate joint posterior density
\begin{equation}
	\pi_{\rho}(\boldsymbol{\theta},\B{z} | \B{y}) \propto \exp\pr{-\frac{1}{2\sigma^2}\nr{\B{y} - \B{H}\boldsymbol{\theta}}_2^2 - \tau \sum_{1\leq i \leq d}\nr{\B{Z}_{i}}_2 - \frac{1}{2\rho^2}\nr{\B{Z} - \B{D}\boldsymbol{\theta}}_2^2}, \label{eq:inpainting_approx_model}
\end{equation}
where $\B{Z} = (\B{z}_1,\B{z}_2) \in \mathbb{R}^{2 \times d}$.
By relying on \eqref{eq:inpainting_approx_model}, the inference is now simplified and can be conducted with a Gibbs sampler, see Section 4 in the supplementary material.
Since $\mathrm{ker}(\B{H}) \cap \mathrm{ker}(\B{D}) = \{\B{0}_d\}$, the conditional posterior distribution of $\boldsymbol{\theta}$ is a non-degenerate multivariate Gaussian distribution.
Samples from the latter can be obtained efficiently with the two-dimensional discrete Fourier transform by exploiting the periodic boundary conditions for $\boldsymbol{\theta}$ \citep{Wang2008,Marnissi2018}. 
On the other hand, samples from $\pi_{\rho}(\B{Z}|\boldsymbol{\theta})$ can be drawn efficiently using exact data augmentation, see \cite{kyung2010}.
All the inference details are given in Section 4 of the supplementary material.

\begin{figure}[htb]
\centering
  \mbox{{\includegraphics[scale=0.4]{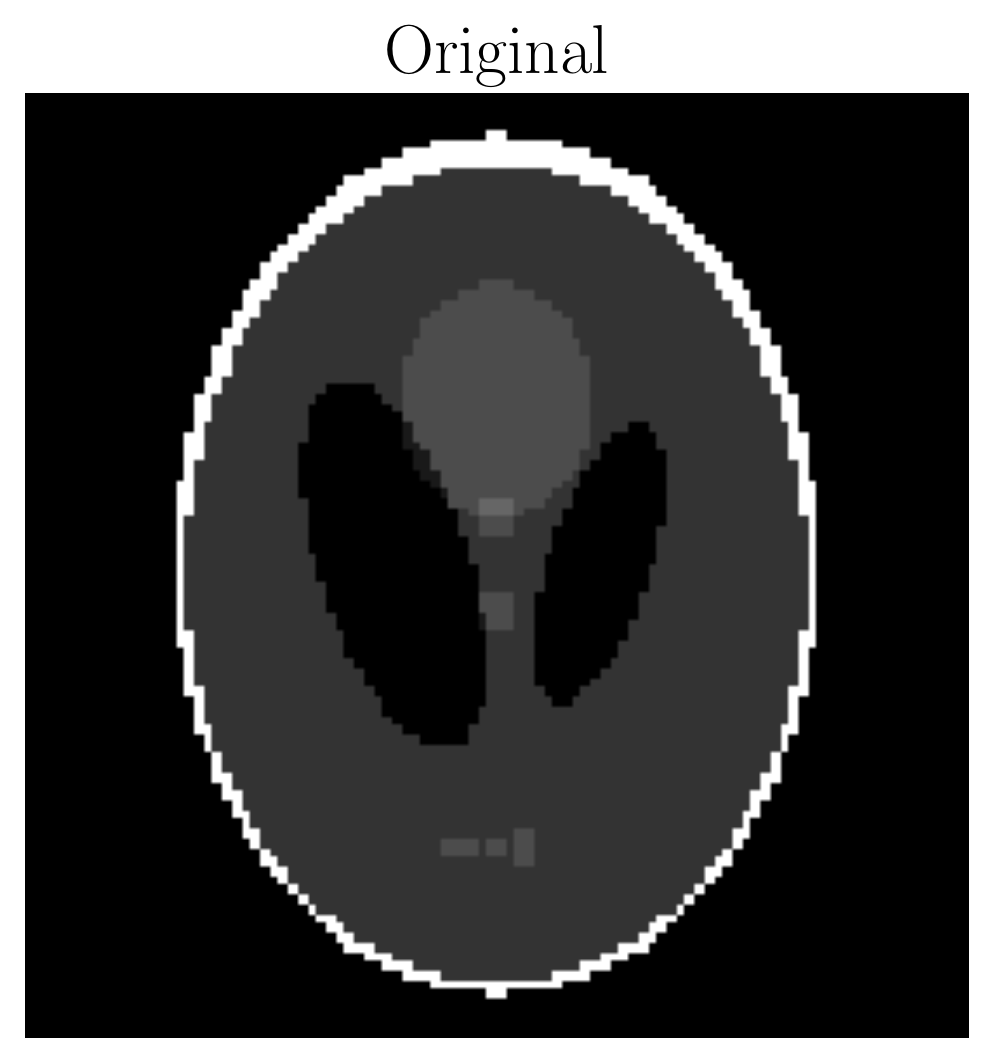}}}
  \mbox{{\includegraphics[scale=0.4]{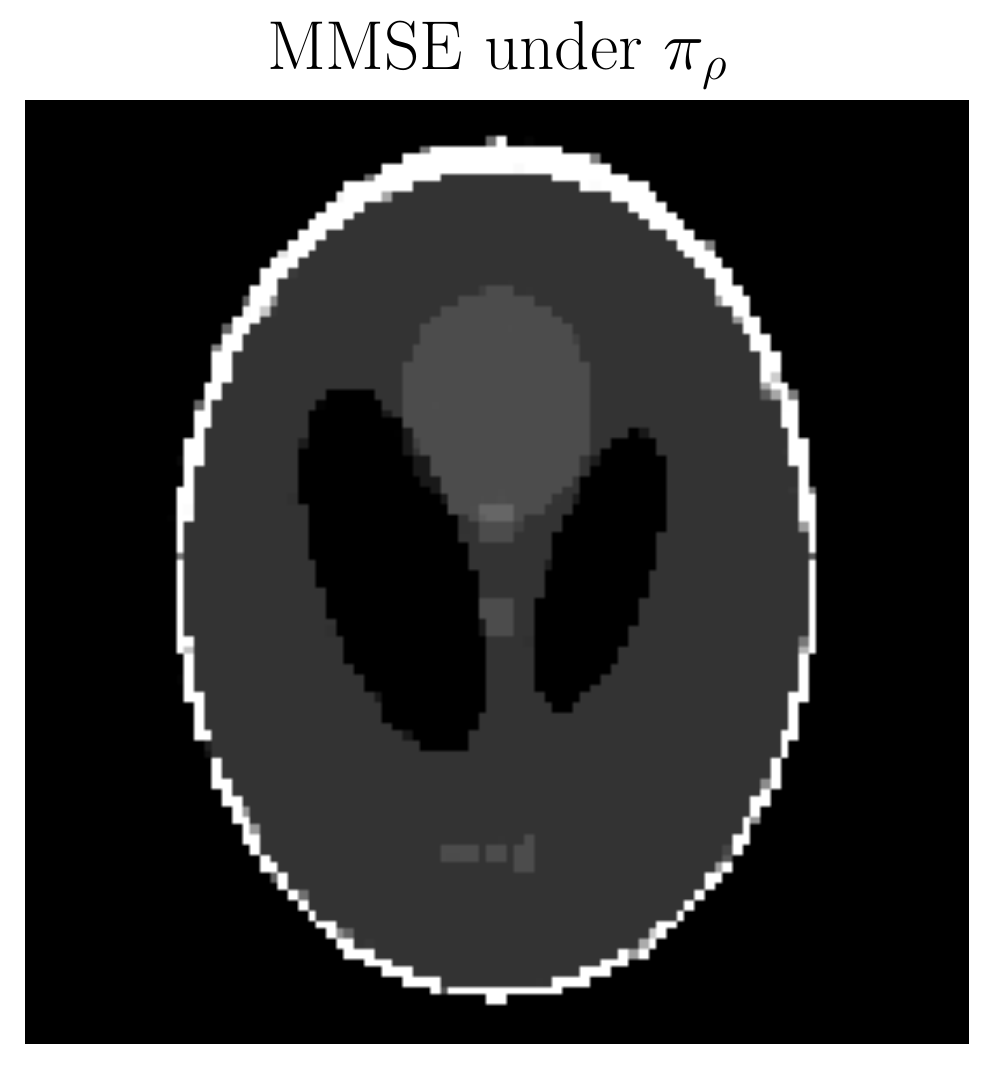}}}
  \mbox{{\includegraphics[scale=0.4]{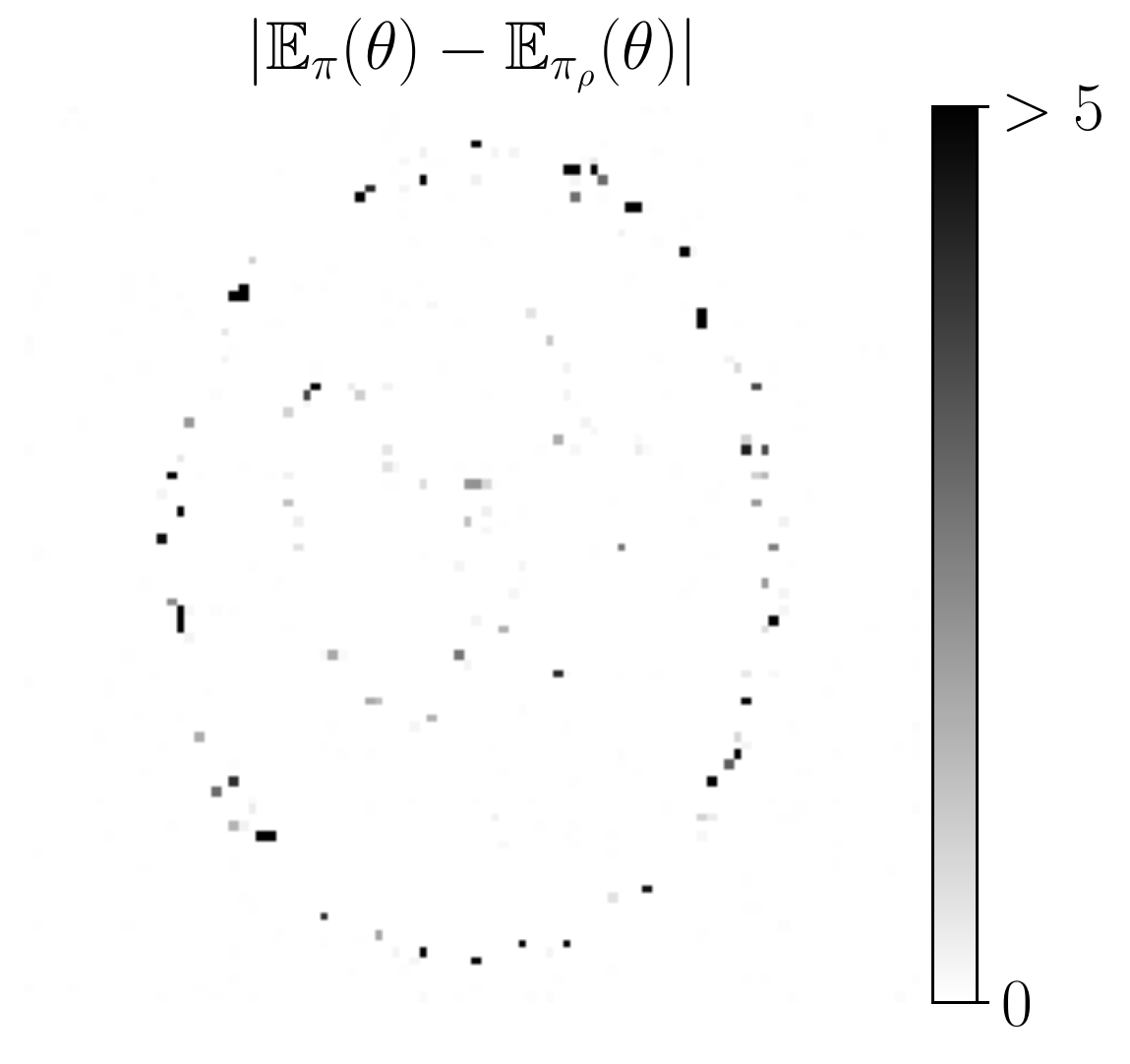}}}
  \caption{From left to right: original image, minimum mean square estimate (MMSE) under $\pi_{\rho}$ and absolute bias between the posterior means under $\pi_{\rho}$ and $\pi$.}
  \label{fig:inpainting_1}
\end{figure}

We illustrate the proposed approximate model $\pi_{\rho}$ by considering the Shepp-Logan phantom magnectic resonance image of size $100 \times 100$ ($d = 10^4$), see Figure \ref{fig:inpainting_1}.
We artifically damaged and added noise to this image to build a noisy observation $\B{y}$ consisting of 90\% randomly selected pixels of the initial image.
The standard deviation of the Gaussian noise and the regularization parameter have been set to $\sigma = 7 \times 10^{-2}$ (corresponding to a SNR of 58dB)  and $\tau = 5$, respectively.
The tolerance parameter has been set to $\rho = 0.1$.

In order to assess the bias of the proposed approach, we implemented the
Moreau-Yosida Metropolis-adjusted Langevin algorithm (MYMALA) of \cite{Durmus2018}, specifically designed to sample exactly from high-dimensional and non-smooth posterior distributions.
For all the MCMC algorithms, the initialization has been set to $\boldsymbol{\theta}^{[0]} = \B{0}_d$.
We generated $10^5$ samples and kept the last $5 \times 10^4$ ones.

Figure \ref{fig:inpainting_1} shows the minimum mean square estimate (MMSE) under $\pi_{\rho}$ along with the original image.
One can denote that the MMSE under $\pi_{\rho}$ is visually similar to the original image and hence coherent with the reconstruction task. 
The relative residual error between the former and the MMSE under $\pi$ is of order 2\%.
The main differences are located on the boundaries of the image, as depicted in the figure on the left which shows the absolute difference between the pixels of the two posterior means $\mathbb{E}_{\pi}(\boldsymbol{\theta})$ and $\mathbb{E}_{\pi_{\rho}}(\boldsymbol{\theta})$.

To emphasize the correctness of the proposed approach beyond the comparision between pointwise estimates, we also paid attention to the comparison between posterior credibility sets induced by both $\pi$ and $\pi_{\rho}$.
To this purpose, we considered the highest posterior density region given by 
\begin{equation}
	\mathcal{C}_{\alpha}^{\star} = \{\boldsymbol{\theta} \in \mathbb{R}^d \ | \ f(\boldsymbol{\theta}) \leq \gamma_{\alpha}\},
\end{equation}
where $\gamma_{\alpha} \in \mathbb{R}$ is such that $\int_{\mathcal{C}_{\alpha}^{\star}}\pi(\boldsymbol{\theta}|\B{y})\mathrm{d}\boldsymbol{\theta} = 1 - \alpha$ and $f$ is the potential function associated to $\pi(\boldsymbol{\theta}|\B{y})$.

\begin{figure}[htb]
\centering
  \mbox{{\includegraphics[scale=0.5]{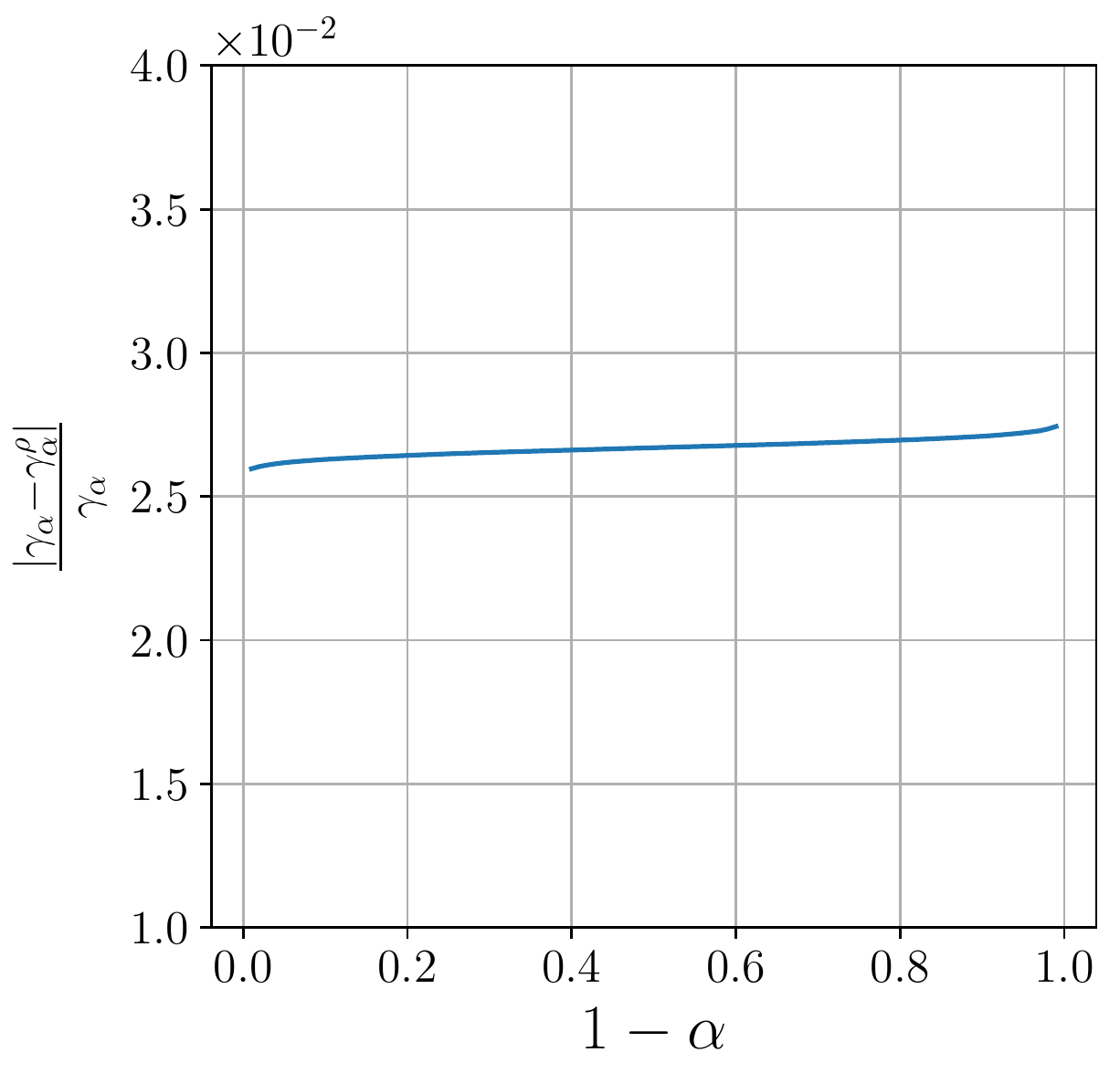}}}
  \mbox{{\includegraphics[scale=0.5]{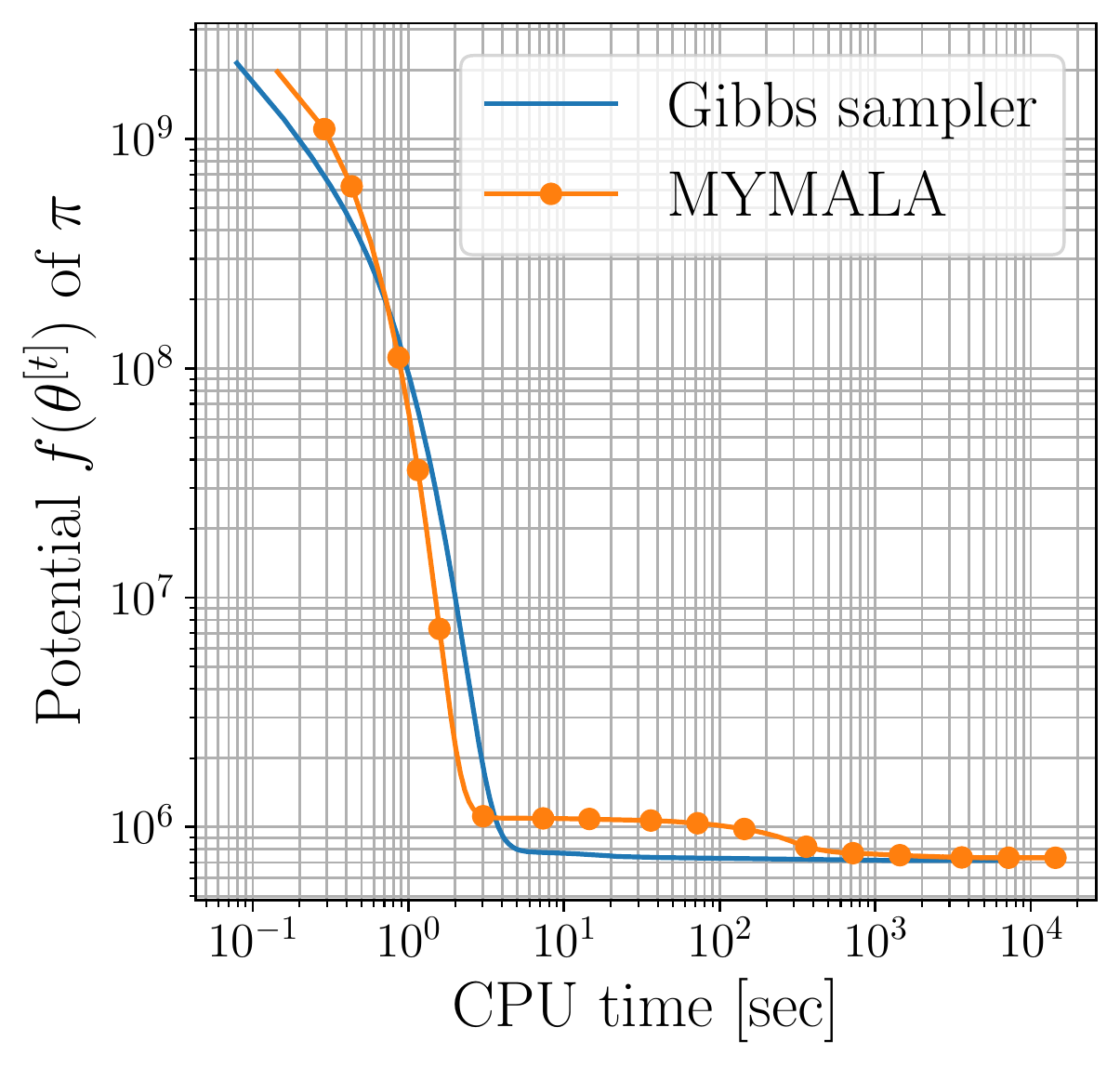}}}
  \caption{(left) relative error between the threshold value estimated with $\pi$ denoted $\gamma_{\alpha}$ and the one estimated with $\pi_{\rho}$ denoted $\gamma_{\alpha}^{\rho}$ and (right) Potential $f = -\log \pi$ w.r.t. the number of iterations $t$ for both MYMALA and the Gibbs sampler targetting $\pi_{\rho}$.}
  \label{fig:inpainting_2}
\end{figure}

Figure \ref{fig:inpainting_2} shows the different values of the scalar summary $\gamma_{\alpha}$ estimated using $\pi$ and the scalar $\gamma_{\alpha}^{\rho}$ estimated using $\pi_{\rho}$ for $\alpha \in [0.01,0.99]$.
Note that the approximation error associated to $\gamma_{\alpha}$ is of order 2.6\% whatever the value of $\alpha$, which supports the use of $\pi_{\rho}$ to conduct Bayesian uncertainty analysis in this problem.
After the burn-in period, the efficiency of the Gibbs algorithm used to sample from $\pi_{\rho}$ has been measured by comparing the effective sample size (ESS) associated to the slowest component of $\boldsymbol{\theta}$ to the one obtained with MYMALA.
We found that the two ESS were roughly similar but the cost per iteration of the Gibbs sampler (0.079 sec/iteration) is almost two times lower than that of MYMALA (0.144 sec/iteration)\footnote{Both algorithms have been implemented in \textsc{Matlab} with the same level of efficiency.}.
In addition, the number of iterations required to reach high-probability regions is much less important for the Gibbs sampler than for MYMALA, showing the interest of AXDA, see Figure \ref{fig:inpainting_2}.

\section{Conclusion}
\label{sec:conclusion}

This paper presented a unifying framework for asymptotically exact data augmentation (AXDA) schemes.
AXDA introduces approximate densities in order to simplify the inference.
By building on existing works which considered special instances of AXDA, we illustrate potential benefits that can be inherited by the proposed framework such as distributed computations, robustness or sophisticated inference schemes from the ABC literature. 
On top of these qualitative properties, we derived a set of theoretical guarantees on the bias involved in the proposed methodology.
The latter encompass a large class of AXDA models and a detailed non-asymptotic analysis has been done for Gaussian smoothing.
These results have been illustrated on several cases that can arise in statistical learning or signal processing showing the broad scope of application of the proposed approach.
In practice,  we emphasize that AXDA models can remarkably improve the inference task in big data and high-dimensional settings.
In summary, at the price of an approximation which comes with theoretical guarantees, AXDA approaches appear to be a general, systematic and efficient way to conduct simple inference in a wide variety of large-scale problems. 
They provide accurate estimates with relevant confidence intervals that are crucial in many applications, in particular when no ground truth is available.

\section*{Acknowledgements}
Part of this work has been supported by the ANR-3IA Artificial and Natural Intelligence Toulouse Institute (ANITI) under grant agreement ANITI ANR-19-PI3A-0004.

\bigskip
\begin{center}
{\large\bf SUPPLEMENTARY MATERIALS}
\end{center}

\begin{description}
\item[Appendices:] The supplementary material includes the proofs of Theorems \ref{theorem:controlTVnorm} and \ref{theorem:controlTVnorm_2}; Propositions \ref{prop:differentiability_prior}, \ref{prop:bound_kernel}, \ref{proposition:Bregman_bias}, \ref{prop:control_prior} and \ref{prop:credibility_intervals_2}; and Corollary \ref{corollary:DLTVnorm} and \ref{theorem:controlTVnorm_generalization}.
It also includes additional details about standard kernels and Bregman divergences, inference details associated to the image inpainting example in Section \ref{subsec:inpainting} and derivations of classical inference algorithms to target AXDA models. (supplementary\_material.pdf, pdf file)

\item[Package for AXDA:] The computer code associated to the illustrations and experiments described in this paper is also available online. 
More precisely, the package “AXDA” contains a \textsc{Python} jupyter notebook to reproduce all the tables and figures of the paper and a \textsc{Matlab}-code associated to the image inpainting example along with a \textsc{README} file. (axda.zip, zipped file)
\end{description}

\newpage
\begin{center}
{\large\bf SUPPLEMENTARY MATERIALS}
\end{center}

\appendix
\renewcommand{\thesection}{\arabic{section}}

\section{Proofs}

\subsection{Proof of Proposition 1}

Property \emph{i)} follows from the fact that $\marginal$ stands for a convolution integral between $\pi$ and $\kappa_{\rho}$, i.e. $\pi_{\rho} = \pi \ast \kappa_{\rho}$.
Therefore, the expectation and variance under $\pi_{\rho}$ are the sum of the expectations and variances of two independent random variables under $\pi$ and $\kappa_{\rho}$ respectively.
Property \emph{ii)} follows directly from \cite[Proposition 8.6]{Folland1999}.
Property \emph{iii)} follows from the fact that log-concavity is preserved by convolution of distributions \cite[Theorem 2.18]{Dharmadhikari1988}.
Finally, Property \emph{iv)} follows from the dominated convergence theorem since $\pi \in L^1$, $\kappa_{\rho} \in \mathcal{C}^{\infty}(\mathbb{R}^d)$ and for all $k\geq 0$, $|\partial^k\kappa_{\rho}|\leq C_k$ \cite[Proposition 8.10]{Folland1999}.

\subsection{Proof of Proposition 2}

The proof can be found in \citet[Lemma 7.1.10]{Ambrosio2008}.
Since it is quite short, we recall it hereafter for completeness.
We have
\begin{align}
  W_p^p(\pi,\pi_{\rho}) &= \min_{\mu}\bbr{\int_{\mathbb{R}^d}\int_{\mathbb{R}^d}\nr{\boldsymbol{\theta} - \B{z}}_2^p\mathrm{d}\mu(\boldsymbol{\theta},\B{z}); \mu \in \Gamma(\pi_{\rho},\pi)} \\
  &\leq \int_{\mathbb{R}^d}\int_{\mathbb{R}^d}\nr{\boldsymbol{\theta}-\B{z}}_2^p\pi_{\rho}(\boldsymbol{\theta},\B{z})\mathrm{d}\boldsymbol{\theta}\mathrm{d}\B{z} \\
  &= \int_{\mathbb{R}^d}\int_{\mathbb{R}^d}\nr{\boldsymbol{\theta}-\B{z}}_2^p\kappa_{\rho}(\B{z},\boldsymbol{\theta})\pi(\B{z})\mathrm{d}\boldsymbol{\theta}\mathrm{d}\B{z} \\
  &= \rho^{-d}\int_{\mathbb{R}^d}\int_{\mathbb{R}^d}\nr{\boldsymbol{\theta}-\B{z}}_2^pK(\rho^{-1}(\boldsymbol{\theta}-\B{z}))\pi(\B{z})\mathrm{d}\boldsymbol{\theta}\mathrm{d}\B{z} \\
  &= \rho^p\int_{\mathbb{R}^d}\nr{\B{u}}_2^pK(\B{u})\mathrm{d}\B{u}\int_{\mathbb{R}^d}\pi(\B{z})\mathrm{d}\B{z} \\
  &=\rho^p\int_{\mathbb{R}^d}\nr{\B{u}}_2^pK(\B{u})\mathrm{d}\B{u}.
\end{align}

\subsection{Proof of Proposition 3}

Let $\boldsymbol{\theta} \in \mathbb{R}^d$.
Since $\pi$ has been assumed to be analytic and twice differentiable with $\B{H}_{\pi}$ being continuous, there exists $\tilde{\boldsymbol{\theta}}$ lying between $\boldsymbol{\theta}$ and $\boldsymbol{\theta}  - \sqrt{\rho}\B{u}$ such that
\begin{align}
  \pi_{\rho}(\boldsymbol{\theta}) &= \int_{\mathbb{R}^d} \pi(\B{z})\kappa_{\rho}(\B{z},\boldsymbol{\theta}) \mathrm{d}\B{z} \\
  &= \dfrac{\displaystyle\int_{\mathbb{R}^d}\pi(\boldsymbol{\theta} - \sqrt{\rho}\B{u}) \exp\pr{-\dfrac{d_{\psi}(\boldsymbol{\theta} - \sqrt{\rho}\B{u},\boldsymbol{\theta})}{\rho}}\mathrm{d}\B{u}}{\displaystyle\int_{\mathbb{R}^d} \exp\pr{-\dfrac{d_{\psi}(\boldsymbol{\theta} - \sqrt{\rho}\B{u},\boldsymbol{\theta})}{\rho}}\mathrm{d}\B{u}} \\
  &= \dfrac{\displaystyle\int_{\mathbb{R}^d}\br{\pi(\boldsymbol{\theta})  - \sqrt{\rho}\nabla\pi(\boldsymbol{\theta})^T\B{u} + \frac{\rho}{2}\B{u}^T\B{H}_{\pi}(\boldsymbol{\tilde{\theta}})\B{u}}\exp\pr{-\dfrac{d_{\psi}(\boldsymbol{\theta} - \sqrt{\rho}\B{u},\boldsymbol{\theta})}{\rho}}\mathrm{d}\B{u}}{\displaystyle\int_{\mathbb{R}^d} \exp\pr{-\dfrac{d_{\psi}(\boldsymbol{\theta} - \sqrt{\rho}\B{u},\boldsymbol{\theta})}{\rho}}\mathrm{d}\B{u}}, 
\end{align}
where $\B{H}_{\pi}$ stands for the Hessian matrix of $\pi$.

It follows that
\begin{align}
  \pi_{\rho}(\boldsymbol{\theta}) &= \pi(\boldsymbol{\theta}) \\
  &- \sqrt{\rho}\nabla \pi(\boldsymbol{\theta})^T\int_{\mathbb{R}^d} \B{u}\frac{\exp\pr{-\dfrac{d_{\psi}(\boldsymbol{\theta} - \sqrt{\rho}\B{u},\boldsymbol{\theta})}{\rho}}\mathrm{d}\B{u}}{\displaystyle\int_{\mathbb{R}^d} \exp\pr{-\dfrac{d_{\psi}(\boldsymbol{\theta} - \sqrt{\rho}\B{u},\boldsymbol{\theta})}{\rho}}\mathrm{d}\B{u}} \label{eq_prop_1}\\
  &+ \frac{\rho}{2} \dfrac{\displaystyle\int_{\mathbb{R}^d}\B{u}^T\B{H}_{\pi}(\tilde{\boldsymbol{\theta}})\B{u}\exp\pr{-\dfrac{d_{\psi}(\boldsymbol{\theta}- \sqrt{\rho}\B{u},\boldsymbol{\theta})}{\rho}}\mathrm{d}\B{u}}{\displaystyle\int_{\mathbb{R}^d} \exp\pr{-\dfrac{d_{\psi}(\boldsymbol{\theta} - \sqrt{\rho}\B{u},\boldsymbol{\theta})}{\rho}}\mathrm{d}\B{u}} \label{eq_prop_2}
\end{align}
We now show that \eqref{eq_prop_1} $= O(\sqrt{\rho})$ and \eqref{eq_prop_2} $= \mathcal{O}(\rho)$.
To this purpose, we use the analyticity and two times differentiability of $d_{\psi}$ w.r.t. to its first argument and the continuity of $\B{H}_{d_{\psi}}$. 
By definition of the Bregman divergence (see Definition 1 in the main paper), $d_{\psi}(\boldsymbol{\theta},\boldsymbol{\theta}) = 0$ and $\nabla_{\B{z}} d_{\psi}(\B{z},\boldsymbol{\theta})\Bigr|_{\substack{\B{z}=\boldsymbol{\theta}}} =\B{0}_d$ so that, for all $\B{u} \in \mathbb{R}^d$,
\begin{equation}
  d_{\psi}(\boldsymbol{\theta} - \sqrt{\rho}\B{u},\boldsymbol{\theta}) =  \frac{\rho}{2}\B{u}^T\B{H}_{d_{\psi}}(\boldsymbol{\theta}')\B{u}, \label{eq_prop_4}
\end{equation}
where $\boldsymbol{\theta}'$ lies between $\boldsymbol{\theta}$ and $\boldsymbol{\theta}  - \sqrt{\rho}\B{u}$.\vspace{1em}

\noindent We first prove \eqref{eq_prop_2} $= \mathcal{O}(\rho)$.
Using \eqref{eq_prop_4}, we can re-write \eqref{eq_prop_2} as 
\begin{align}
  (12) &= \frac{\rho}{2} \dfrac{\displaystyle\int_{\mathbb{R}^d}\B{u}^T\B{H}_{\pi}(\tilde{\boldsymbol{\theta}})\B{u}\exp\pr{-\frac{1}{2}\B{u}^T\B{H}_{d_{\psi}}(\boldsymbol{\theta}')\B{u}}\mathrm{d}\B{u}}{\displaystyle\int_{\mathbb{R}^d} \exp\pr{-\frac{1}{2}\B{u}^T\B{H}_{d_{\psi}}(\boldsymbol{\theta}')\B{u}}\mathrm{d}\B{u}}.
\end{align}
Since $\lim_{\rho \rightarrow 0} \boldsymbol{\theta}' = \boldsymbol{\theta}$ and $\lim_{\rho \rightarrow 0} \tilde{\boldsymbol{\theta}} = \boldsymbol{\theta}$, we will use the dominated convergence theorem using that
\begin{align}
  &\lim_{\rho \rightarrow 0} \B{u}^T\B{H}_{\pi}(\tilde{\boldsymbol{\theta}})\B{u}\exp\pr{-\frac{1}{2}\B{u}^T\B{H}_{d_{\psi}}(\boldsymbol{\theta}')\B{u}} = \B{u}^T\B{H}_{\pi}(\boldsymbol{\theta})\B{u}\exp\pr{-\frac{1}{2}\B{u}^T\B{H}_{d_{\psi}}(\boldsymbol{\theta})\B{u}}.
\end{align}
In addition, since $d_{\psi}$ is strictly convex w.r.t. its first argument, $\B{H}_{d_{\psi}}$ is a symmetric and positive-definite matrix.
By using that $\nr{\B{H}_{\pi}} \leq C < \infty$ and $\nr{\B{H}_{d_{\psi}}} \geq c > 0$, we have:
\begin{equation}
  \left |\B{u}^T\B{H}_{\pi}(\tilde{\boldsymbol{\theta}})\B{u}\exp\pr{-\frac{1}{2}\B{u}^T\B{H}_{d_{\psi}}(\boldsymbol{\theta}')\B{u}} \right| \leq C \nr{\B{u}}_2^2 \exp\pr{-\frac{c}{2}\B{u}^T\B{u}},
\end{equation}
which is integrable on $\mathbb{R}^d$.
From the dominated convergence theorem, it follows that 
\begin{equation}
\displaystyle\int_{\mathbb{R}^d}\B{u}^T\B{H}_{\pi}(\tilde{\boldsymbol{\theta}})\B{u}\exp\pr{-\frac{1}{2}\B{u}^T\B{H}_{d_{\psi}}(\boldsymbol{\theta}')\B{u}}\mathrm{d}\B{u} = \displaystyle\int_{\mathbb{R}^d}\B{u}^T\B{H}_{\pi}(\boldsymbol{\theta})\B{u}\exp\pr{-\frac{1}{2}\B{u}^T\B{H}_{d_{\psi}}(\boldsymbol{\theta})\B{u}}\mathrm{d}\B{u} + o(1).
\end{equation}
Similarly,
\begin{equation}
\displaystyle\int_{\mathbb{R}^d}\exp\pr{-\frac{1}{2}\B{u}^T\B{H}_{d_{\psi}}(\boldsymbol{\theta}')\B{u}}\mathrm{d}\B{u} = \displaystyle\int_{\mathbb{R}^d}\exp\pr{-\frac{1}{2}\B{u}^T\B{H}_{d_{\psi}}(\boldsymbol{\theta})\B{u}}\mathrm{d}\B{u} + o(1).
\end{equation}
Hence,   
\begin{align}
  (12) &= \frac{\rho}{2} \dfrac{\displaystyle\int_{\mathbb{R}^d}\B{u}^T\B{H}_{\pi}(\boldsymbol{\theta})\B{u}\exp\pr{-\frac{1}{2}\B{u}^T\B{H}_{d_{\psi}}(\boldsymbol{\theta})\B{u}}\mathrm{d}\B{u}}{\displaystyle\int_{\mathbb{R}^d} \exp\pr{-\frac{1}{2}\B{u}^T\B{H}_{d_{\psi}}(\boldsymbol{\theta})\B{u}}\mathrm{d}\B{u}} + o(\rho) \\
  &= \frac{\rho}{2} \mathrm{Trace}\pr{\B{H}_{\pi}(\boldsymbol{\theta})\B{H}_{d_{\psi}}(\boldsymbol{\theta})^{-1}} + o(\rho).
\end{align}

\noindent We now prove \eqref{eq_prop_1} $= \mathcal{O}(\sqrt{\rho})$.
Using \eqref{eq_prop_4}, it follows that
\begin{align}
  (11) &= - \sqrt{\rho}\nabla \pi(\boldsymbol{\theta})^T\dfrac{\displaystyle\int_{\mathbb{R}^d} \B{u}\exp\pr{-\frac{1}{2}\B{u}^T\B{H}_{d_{\psi}}(\boldsymbol{\theta}')\B{u}}\mathrm{d}\B{u}}{\displaystyle\int_{\mathbb{R}^d}\exp\pr{-\frac{1}{2}\B{u}^T\B{H}_{d_{\psi}}(\boldsymbol{\theta}')\B{u}}\mathrm{d}\B{u}}.
\end{align}
Again, since $\nr{H_{d_{\psi}}}$ has been assumed to be lower bounded, it follows from the dominated convergence theorem that
\begin{align}
  (11) &= - \sqrt{\rho}\nabla \pi(\boldsymbol{\theta})^T\dfrac{\displaystyle\int_{\mathbb{R}^d} \B{u}\exp\pr{-\frac{1}{2}\B{u}^T\B{H}_{d_{\psi}}(\boldsymbol{\theta})\B{u}}\mathrm{d}\B{u}}{\displaystyle\int_{\mathbb{R}^d} \exp\pr{-\frac{1}{2}\B{u}^T\B{H}_{d_{\psi}}(\boldsymbol{\theta})\B{u}}\mathrm{d}\B{u}} + o(\sqrt{\rho}) \\
  &= - \sqrt{\rho}\nabla \pi(\boldsymbol{\theta})^T\dfrac{\displaystyle\int_{\mathbb{R}^d} \B{u}\exp\pr{-\frac{1}{2}\B{u}^T\B{H}_{d_{\psi}}(\boldsymbol{\theta})\B{u}}\mathrm{d}\B{u}}{\displaystyle\int_{\mathbb{R}^d} \exp\pr{-\frac{1}{2}\B{u}^T\B{H}_{d_{\psi}}(\boldsymbol{\theta})\B{u}}\mathrm{d}\B{u}} + o(\sqrt{\rho}) \\
  &= o(\sqrt{\rho}) = \mathcal{O}(\sqrt{\rho}).
\end{align}

\subsection{Proof of Theorem 1}
\label{proof:theorem_2}

We are interested in controlling w.r.t. $\rho$ the quantity $\nr{\marginal-\pi}_{\mathrm{TV}}$.
To this purpose, let assume that $f$ satisfies $(A_1)$ in the main paper.
In the following, we will assume for simplicity reasons that $\pi$ stands for a pdf associated to the random variable $\boldsymbol{\theta}$.
The case when $\pi$ is a likelihood is treated right after.
Under this convention, it follows
\begin{align}
  \nr{\marginal-\pi}_{\mathrm{TV}} &= \dfrac{1}{2}\int_{\mathbb{R}^d}\left|\marginal(\boldsymbol{\theta})-\pi(\boldsymbol{\theta})\right|\mathrm{d}\boldsymbol{\theta}\nonumber\\
  &= \dfrac{1}{2}\int_{\mathbb{R}^d}\pi(\boldsymbol{\theta})\left|\dfrac{C_{\pi}}{C_{\marginal\text{}}}\mathcal{K}(\boldsymbol{\theta})-1\right|\mathrm{d}\boldsymbol{\theta}, \label{eq:proofTV_1}
\end{align}
where $C_{\pi}$ and $C_{\marginal\text{}}$ are the normalizing constants associated to $\pi$ and $\marginal$, respectively, and
\begin{align}
  \mathcal{K}(\boldsymbol{\theta}) &= \dfrac{\marginal(\boldsymbol{\theta})C_{\pi_{\rho}}}{\pi(\boldsymbol{\theta})C_{\pi}} \\
  &= \int_{\mathbb{R}^d}\exp\pr{f(\boldsymbol{\theta}) - f(\B{z}) - \dfrac{1}{2\rho^2}\nr{\boldsymbol{\theta} - \B{z}}_2^2}\mathrm{d}\B{z}.\label{eq:proofTV_I}
\end{align}
Note that 
\begin{equation}
  \int_{\mathbb{R}^d}\mathcal{K}(\boldsymbol{\theta})\pi(\boldsymbol{\theta})\mathrm{d}\boldsymbol{\theta} = \dfrac{C_{\pi_{\rho}}}{C_{\pi}} \label{eq:propI}.
\end{equation}
Since $f$ is assumed to be $L_f$-Lipschitz, we have
\begin{align}
  \mathcal{K}(\boldsymbol{\theta}) \leq \int_{\mathbb{R}^d}\exp\pr{L_f\nr{\boldsymbol{\theta}-\B{z}}_2 - \dfrac{1}{2\rho^2}\nr{\boldsymbol{\theta}-\B{z}}^2_2}\mathrm{d}\B{z}.
\end{align}
We make the change of variables $\B{u} = \B{z}-\boldsymbol{\theta}$, which leads to
\begin{align}
  \mathcal{K}(\boldsymbol{\theta}) &\leq \int_{\mathbb{R}^d}\exp\pr{L_f\nr{\B{u}}_2 - \dfrac{1}{2\rho^2}\nr{\B{u}}^2_2}\mathrm{d}\B{u}.
\end{align}
Then, with another change of variables $t = \nr{\B{u}}_2$, it follows 
\begin{align}
  \mathcal{K}(\boldsymbol{\theta}) \leq \dfrac{2\pi^{d/2}}{\Gamma\pr{\dfrac{d}{2}}}\int_0^{\infty}t^{d-1}\exp\pr{L_ft - \dfrac{1}{2\rho^2}t^2}\mathrm{d}t.
\end{align}
This integral admits a closed-form expression \cite[Formula 3.462 1.]{Gradshteyn2015} by introducing the special parabolic cylinder function $D_{-d}$ defined for all $d>0$ and $z \in \mathbb{R}$ by
\begin{align}
D_{-d}(z) = \dfrac{\exp(-z^2/4)}{\Gamma(d)}\int_0^{+\infty}e^{-xz - x^2/2}x^{d-1}\mathrm{d}x.
\end{align}
Then,
\begin{align}
  \mathcal{K}(\boldsymbol{\theta}) \leq A(\rho), \label{eq:bound_I_A}
\end{align}
where
\begin{align}
  A(\rho) = \dfrac{2\pi^{d/2}\rho^d\Gamma(d)\exp\pr{\dfrac{L_f^2\rho^2}{4}}}{\Gamma\pr{\dfrac{d}{2}}}D_{-d}\pr{-L_f\rho}. \label{eq:A}
\end{align}
Then, with \eqref{eq:propI} and \eqref{eq:bound_I_A}, we also have 
\begin{align}
  \dfrac{C_{\pi}}{C_{\marginal\text{}}} \geq \dfrac{1}{A(\rho)}.
\end{align}
We now use the triangle inequality in \eqref{eq:proofTV_1} which leads to
\begin{align}
  \nr{\marginal-\pi}_{\mathrm{TV}} &\leq \dfrac{1}{2}\pr{\int_{\mathbb{R}^d}\left|\dfrac{C_{\pi}}{C_{\marginal\text{}}}\mathcal{K}(\boldsymbol{\theta})-\dfrac{1}{A(\rho)}\mathcal{K}(\boldsymbol{\theta})\right|\pi(\boldsymbol{\theta})\mathrm{d}\boldsymbol{\theta} + \int_{\mathbb{R}^d}\left|\dfrac{1}{A(\rho)}\mathcal{K}(\boldsymbol{\theta})-1\right|\pi(\boldsymbol{\theta})\mathrm{d}\boldsymbol{\theta}}\nonumber\\
  &= \dfrac{1}{2}\pr{\int_{\mathbb{R}^d}\pr{\dfrac{C_{\pi}}{C_{\marginal\text{}}}\mathcal{K}(\boldsymbol{\theta})-\dfrac{1}{A(\rho)}\mathcal{K}(\boldsymbol{\theta})}\pi(\boldsymbol{\theta})\mathrm{d}\boldsymbol{\theta} + \int_{\mathbb{R}^d}\pr{1-\dfrac{1}{A(\rho)}\mathcal{K}(\boldsymbol{\theta})}\pi(\boldsymbol{\theta})\mathrm{d}\boldsymbol{\theta}}.\label{eq:proofTV_2}
\end{align}
The first term in this upper bound writes
\begin{align}
\int_{\mathbb{R}^d}\pr{\dfrac{C_{\pi}}{C_{\marginal\text{}}}-\dfrac{1}{A(\rho)}}\mathcal{K}(\boldsymbol{\theta})\pi(\boldsymbol{\theta})\mathrm{d}\boldsymbol{\theta} &= 1 -\dfrac{1}{A(\rho)}\int_{\mathbb{R}^d}\mathcal{K}(\boldsymbol{\theta})\pi(\boldsymbol{\theta})\mathrm{d}\boldsymbol{\theta} \nonumber\\
&= \int_{\mathbb{R}^d}\pr{1 -\dfrac{1}{A(\rho)}\mathcal{K}(\boldsymbol{\theta})}\pi(\boldsymbol{\theta})\mathrm{d}\boldsymbol{\theta}.
\end{align}

This allows us to bound \eqref{eq:proofTV_2}, that is
\begin{align}
  \nr{\marginal-\pi}_{\mathrm{TV}} &\leq \int_{\mathbb{R}^d}\pr{1 - \dfrac{1}{A(\rho)}\mathcal{K}(\boldsymbol{\theta})}\pi(\boldsymbol{\theta})\mathrm{d}\boldsymbol{\theta}. \label{eq:proofTV_3}
\end{align}

Using one more time the $L_f$-Lipschitz assumption on $f$, we have for all $\boldsymbol{\theta},\B{z}$,
\begin{align}
&-(f(\B{z})-f(\boldsymbol{\theta})) \geq -|f(\B{z})-f(\boldsymbol{\theta})| \geq -L_f \nr{\boldsymbol{\theta}-\B{z}}_2, \\
 \text{so that }&\mathcal{K}(\boldsymbol{\theta}) \geq \int_{\mathbb{R}^d}\exp\pr{-L_f\nr{\boldsymbol{\theta}-\B{z}}_2 - \dfrac{1}{2\rho^2}\nr{\boldsymbol{\theta} - \B{z}}_2^2}\mathrm{d}\B{z}.
\end{align}

With the same changes of variables as above, it follows
\begin{align}
    \mathcal{K}(\boldsymbol{\theta}) \geq B(\rho), \label{eq:bound_I_B}
  \end{align}
where
\begin{align}
  B(\rho) = \dfrac{2\alpha_{\mathrm{Vol}}\pi^{d/2}\rho^d\Gamma(d)\exp\pr{\dfrac{L_f^2\rho^2}{4}}}{\Gamma\pr{\dfrac{d}{2}}}D_{-d}\pr{L_f\rho}.\label{eq:B}
\end{align}

Then we have $1 - \dfrac{1}{A(\rho)}\mathcal{K}(\boldsymbol{\theta}) \leq 1 - \dfrac{B(\rho)}{A(\rho)}$ which combined with \eqref{eq:proofTV_3} yields
\begin{align}
  \nr{\marginal-\pi}_{\mathrm{TV}} \leq 1 - \dfrac{D_{-d}\pr{L_f\rho}}{D_{-d}\pr{-L_f\rho}}.
\end{align}

\textbf{Note:} When $\pi = \pi(\B{y}|\boldsymbol{\theta})$ is a likelihood, \eqref{eq:proofTV_1} becomes
\begin{align}
  \nr{\marginal-\pi}_{\mathrm{TV}} &= \dfrac{1}{2}\int_{\mathbb{R}^n}\pi(\B{y}|\boldsymbol{\theta})\left|\dfrac{C_{\pi}}{C_{\marginal\text{}}}\mathcal{K}(\B{y};\boldsymbol{\theta})-1\right|\mathrm{d}\B{y}.
\end{align}
Since $L_f$ is assumed to be independent of $\B{y}$, the same type of proof can be followed in this case and yields the same quantitative bound.

\subsection{Proof of Corollary 1}
\label{proof:coro_1}

The parabolic cylinder function when $d>0$ has the following expression \cite[Formula 9.241 2.]{Gradshteyn2015}
\begin{align}
  D_{-d}(z) = \dfrac{\exp(-z^2/4)}{\Gamma(d)}\int_0^{+\infty}e^{-xz - x^2/2}x^{d-1}\mathrm{d}x.
\end{align}
In the limiting case when $z\rightarrow0$, a first order Taylor expansion of $e^{-xz}$ gives
\begin{align}
  D_{-d}(z) &= \dfrac{\exp(-z^2/4)}{\Gamma(d)}\int_0^{+\infty}e^{- x^2/2}x^{d-1}(1-xz + o(z))\mathrm{d}x \nonumber \\
  &= \dfrac{\exp(-z^2/4)}{\Gamma(d)}\pr{\int_0^{+\infty}e^{- x^2/2}x^{d-1}\mathrm{d}x - z\int_0^{+\infty}e^{- x^2/2}x^{d}\mathrm{d}x + o(z)} \nonumber \\
  &= \dfrac{\exp(-z^2/4)}{\Gamma(d)}\pr{\Gamma\pr{\dfrac{d}{2}}2^{d/2-1} - z\Gamma\pr{\dfrac{d+1}{2}}2^{d/2-1/2} + o(z)}, \label{eq:DL_D}
\end{align}
recording that $\int_0^{+\infty}e^{- x^2/2}x^{d}\mathrm{d}x = \Gamma((d+1)/2)2^{d/2-1/2}$ \citep[Formula 3.383 11.]{Gradshteyn2015}.
Using \eqref{eq:DL_D} for $z = \pm \rho L_f$ yields
\begin{align}
1 - \dfrac{D_{-d}(L_f\rho)}{D_{-d}(-L_f\rho)} &= 1-\dfrac{\dfrac{\exp(-(\rho L_f)^2/4)}{\Gamma(d)}\pr{\Gamma\pr{\dfrac{d}{2}}2^{d/2-1} - \rho L_f\Gamma\pr{\dfrac{d+1}{2}}2^{d/2-1/2} + o(\rho)}}{\dfrac{\exp(-(\rho L_f)^2/4)}{\Gamma(d)}\pr{\Gamma\pr{\dfrac{d}{2}}2^{d/2-1} + \rho L_f\Gamma\pr{\dfrac{d+1}{2}}2^{d/2-1/2} + o(\rho)}} \nonumber \\
&= 1-\dfrac{\Gamma\pr{\dfrac{d}{2}}2^{d/2-1} - \rho L_f\Gamma\pr{\dfrac{d+1}{2}}2^{d/2-1/2} + o(\rho)}{\Gamma\pr{\dfrac{d}{2}}2^{d/2-1}\pr{1 + \rho \dfrac{L_f\Gamma\pr{\dfrac{d+1}{2}}\sqrt{2}}{\Gamma\pr{\dfrac{d}{2}}} + o(\rho)}} \nonumber \\
&= 1-\pr{1 - \rho \dfrac{L_f\Gamma\pr{\dfrac{d+1}{2}}\sqrt{2}}{\Gamma\pr{\dfrac{d}{2}}} + o(\rho)}\pr{1 - \rho \dfrac{L_f\Gamma\pr{\dfrac{d+1}{2}}\sqrt{2}}{\Gamma\pr{\dfrac{d}{2}}} + o(\rho)} \nonumber \\
&= \dfrac{2\sqrt{2}\Gamma\pr{\dfrac{d+1}{2}}}{\Gamma\pr{\dfrac{d}{2}}}L_f\rho + o(\rho). \label{eq:equivalent_bound}
\end{align}

\subsection{Dependence of \eqref{eq:equivalent_bound} with respect to the dimension}
The gamma function $\Gamma$ can be expressed for all $z > 0$ as $\Gamma(z) = \int_0^{+\infty}x^{z-1}e^{-x}\mathrm{d}x$.
When $z$ is large, Stirling-like approximations give the following equivalent for $\Gamma(z+1/2)$ and $\Gamma(z)$:
\begin{align}
  \Gamma(z+1/2) &\underset{z\rightarrow+\infty}\sim \sqrt{2\pi}z^ze^{-z} \\
  \Gamma(z) &\underset{z\rightarrow+\infty}\sim \sqrt{2\pi}z^{z-1/2}e^{-z}.
\end{align}
So that when $d$ is large
\begin{align}
  \dfrac{2\sqrt{2}\Gamma\pr{\dfrac{d+1}{2}}}{\Gamma\pr{\dfrac{d}{2}}}L_f\rho &\underset{d\rightarrow+\infty}\sim \dfrac{2\sqrt{2}\sqrt{2\pi}(d/2)^{d/2}e^{-d/2}}{\sqrt{2\pi}(d/2)^{d/2-1/2}e^{-d/2}}L_f\rho \nonumber \\
  &\underset{d\rightarrow+\infty}\sim 2\sqrt{2}(d/2)^{1/2}L_f\rho \nonumber \\
  &\underset{d\rightarrow+\infty}\sim 2L_f\rho d^{1/2}.
\end{align}

\subsection{Proof of Theorem 2}

We now prove another bound on the TV distance when $f$ satisfies $(A_2),(A_3)$ and $(A_4)$ in the main paper.
The beginning of the proof follows the same lines as in the proof of Theorem 1 above when $\pi$ stands for a pdf associated to $\boldsymbol{\theta}$.
Hence, we have from \eqref{eq:proofTV_1} that
\begin{align}
  \nr{\marginal-\pi}_{\mathrm{TV}} &= \dfrac{1}{2}\int_{\mathbb{R}^d} \pi(\boldsymbol{\theta})\left|1 - \mathcal{K}(\boldsymbol{\theta})\dfrac{C_{\pi}}{C_{\marginal\text{}}}\right|\mathrm{d}\boldsymbol{\theta}. \label{eq:proof_lemma_2_0}
\end{align}
We now use the convexity of $f$ to write for all $\boldsymbol{\theta} \in \mathbb{R}^d,\B{z} \in\mathbb{R}^{d}$,
\begin{align}
  f(\boldsymbol{\theta}) - f(\B{z}) \leq \nabla f(\boldsymbol{\theta})^T(\boldsymbol{\theta}-\B{z}). \label{eq:proof_lemma_2_1}
\end{align}
  By using \eqref{eq:proof_lemma_2_1} and \eqref{eq:proofTV_I}, it follows that
  \begin{align}
      \mathcal{K}(\boldsymbol{\theta}) &\leq \int_{\mathbb{R}^d}\exp\pr{\nabla f(\boldsymbol{\theta})^T(\boldsymbol{\theta}-\B{z}) - \dfrac{1}{2\rho^2}\nr{\boldsymbol{\theta}-\B{z}}_2^2}\mathrm{d}\B{z} \nonumber\\
      &= \exp\pr{\dfrac{\rho^2}{2}\nr{\nabla f(\boldsymbol{\theta})}_2^2}\int_{\mathbb{R}^{d}}\exp\pr{ - \dfrac{1}{2\rho^2}\nr{\B{z} - \boldsymbol{\theta}-\rho^2\nabla f(\boldsymbol{\theta})}_2^2}\mathrm{d}\B{z} \nonumber \\
      &= \exp\pr{\dfrac{\rho^2}{2}\nr{\nabla f(\boldsymbol{\theta})}_2^2}(2\pi\rho^2)^{d /2} = B_1(\boldsymbol{\theta}).\label{eq:proof_lemma_2_2}
  \end{align}

  By using again the convexity of $f$, we also have for all $\boldsymbol{\theta} \in \mathbb{R}^d,\B{z} \in \mathbb{R}^{d}$,
  \begin{align}
      f(\boldsymbol{\theta}) - f(\B{z}) \geq \nabla f(\B{z})^T(\boldsymbol{\theta}-\B{z}). \label{eq:proof_lemma_2_3}
  \end{align}
  Then, \eqref{eq:proof_lemma_2_3} leads to 
  \begin{align}
      \mathcal{K}(\boldsymbol{\theta}) &\geq \int_{\mathbb{R}^{d}}\exp\pr{\nabla f(\B{z})^T(\boldsymbol{\theta}-\B{z}) - \dfrac{1}{2\rho^2}\nr{\boldsymbol{\theta}-\B{z}}_2^2}\mathrm{d}\B{z} \nonumber\\
      &= \int_{\mathbb{R}^{d}}\exp\pr{\nabla f(\boldsymbol{\theta})^T(\boldsymbol{\theta}-\B{z}) - \dfrac{1}{2\rho^2}\nr{\boldsymbol{\theta}-\B{z}}_2^2}\nonumber \\
      &\hspace{1.7cm}\times \exp\pr{ - (\nabla f(\boldsymbol{\theta})-\nabla f(\B{z}))^T(\boldsymbol{\theta}-\B{z})}\mathrm{d}\B{z}.
  \end{align}
  We now use $(A_2)$ in the main paper which leads to
  \begin{align}
      \mathcal{K}(\boldsymbol{\theta}) &\geq \int_{\mathbb{R}^{d}}\exp\pr{\nabla f(\boldsymbol{\theta})^T(\boldsymbol{\theta}-\B{z}) - \pr{\dfrac{1+2\rho^2M_f}{2\rho^2}}\nr{\boldsymbol{\theta}-\B{z}}_2^2}\mathrm{d}\B{z} \nonumber\\
      &= \exp\pr{\dfrac{\rho^2}{2(1+2\rho^2M_f)}\nr{\nabla f(\boldsymbol{\theta})}_2^2}\pr{\dfrac{2\pi\rho^2}{1+2\rho^2M_f}}^{d /2} = B_2(\boldsymbol{\theta}). \label{eq:proof_lemma_2_4}
  \end{align}
  
  We now apply the triangle inequality in \eqref{eq:proof_lemma_2_0} which yields
 \begin{align}
  \nr{\marginal-\pi}_{\mathrm{TV}} &\leq \dfrac{1}{2}\int_{\mathbb{R}^d} \left|\dfrac{C_{\pi}}{C_{\marginal\text{}}}-\dfrac{1}{B_1(\boldsymbol{\theta})}\right|\mathcal{K}(\boldsymbol{\theta})\pi(\boldsymbol{\theta})\mathrm{d}\boldsymbol{\theta} + \dfrac{1}{2}\int_{\mathbb{R}^d}\left|\dfrac{\mathcal{K}(\boldsymbol{\theta})}{B_1(\boldsymbol{\theta})}-1\right|\pi(\boldsymbol{\theta})\mathrm{d}\boldsymbol{\theta} \nonumber\\
  &= \dfrac{1}{2}\int_{\mathbb{R}^d} \left|\dfrac{C_{\pi}}{C_{\marginal\text{}}}-\dfrac{1}{B_1(\boldsymbol{\theta})}\right|\mathcal{K}(\boldsymbol{\theta})\pi(\boldsymbol{\theta})\mathrm{d}\boldsymbol{\theta} + \dfrac{1}{2}\int_{\mathbb{R}^d}\pr{1-\dfrac{\mathcal{K}(\boldsymbol{\theta})}{B_1(\boldsymbol{\theta})}}\pi(\boldsymbol{\theta})\mathrm{d}\boldsymbol{\theta}.  \label{eq:proof_lemma_2_5} 
\end{align}

The absolute value in the first term of \eqref{eq:proof_lemma_2_5} can be removed by noting that
\begin{align}
    \dfrac{C_{\pi}}{C_{\pi_{\rho}}} &= \dfrac{\displaystyle\int_{\mathbb{R}^d}\exp(-f(\boldsymbol{\theta}))\mathrm{d}\boldsymbol{\theta}}{\displaystyle\int_{\mathbb{R}^d}\exp\pr{-f(\B{z})}\displaystyle\int_{\mathbb{R}^d} \exp\pr{- \dfrac{1}{2\rho^2}\nr{\B{z}-\boldsymbol{\theta}}_2^2}\mathrm{d}\boldsymbol{\theta}\mathrm{d}\B{z}} \nonumber \\
    &\geq \dfrac{\displaystyle\int_{\mathbb{R}^d}\exp(-f(\boldsymbol{\theta}))\mathrm{d}\boldsymbol{\theta}}{\displaystyle\int_{\mathbb{R}^d}\exp\pr{-f(\B{z})}\displaystyle\int_{\mathbb{R}^d} \exp\pr{- \dfrac{1}{2\rho^2}\nr{\B{z}-\boldsymbol{\theta}}_2^2}\mathrm{d}\boldsymbol{\theta}\mathrm{d}\B{z}} \nonumber \\
    &= \pr{2\pi\rho^2}^{-d / 2} \nonumber \\
    &= \dfrac{\exp\pr{\dfrac{\rho^2}{2}\nr{\nabla f(\boldsymbol{\theta})}^2}}{B_1(\boldsymbol{\theta})} \nonumber \\
    &\geq \dfrac{1}{B_1(\boldsymbol{\theta})}.
\end{align}

Then \eqref{eq:proof_lemma_2_5} becomes
\begin{align}
  &\nr{\marginal-\pi}_{\mathrm{TV}} 
  \leq \int_{\mathbb{R}^d}\pr{1-\dfrac{\mathcal{K}(\boldsymbol{\theta})}{B_1(\boldsymbol{\theta})}}\pi(\boldsymbol{\theta})\mathrm{d}\boldsymbol{\theta} \nonumber \\
  &\leq \int_{\mathbb{R}^d}\pr{1-\dfrac{B_2(\boldsymbol{\theta})}{B_1(\boldsymbol{\theta})}}\pi(\boldsymbol{\theta})\mathrm{d}\boldsymbol{\theta}.
  \label{eq:proof_lemma_2_6} 
\end{align}
 We now use the fact that $-\exp(-u) \leq u - 1$ for all $u\geq 0$ which yields
 \begin{align}
  \nr{\marginal-\pi}_{\mathrm{TV}} &\leq 1+\pr{1+2\rho^2M_f}^{-d /2}\int_{\mathbb{R}^d}\pr{\dfrac{\rho^4 M_f\nr{\nabla f(\boldsymbol{\theta})}^2}{1+2\rho^2M_f}-1}\pi(\boldsymbol{\theta})\mathrm{d}\boldsymbol{\theta} \nonumber \\
  \text{(with $(A_3)$) }&= 1 - \pr{1+2\rho^2M_f}^{-d /2}\pr{1 - \dfrac{ \rho^4 M_f \mathsf{M}_f}{1+2\rho^2M_f}}.
  \label{eq:proof_lemma_2_7} 
\end{align}

The result in Corollary 3 in the main paper comes from a straightforward Taylor expansion of \eqref{eq:proof_lemma_2_7}.

\subsection{Proof of Corollary 3}

Equation \eqref{eq:proofTV_I} becomes
\begin{align}
  \mathcal{K}(\boldsymbol{\theta}) = \prod_{j=1}^J\int_{\mathbb{R}^d}\exp\pr{f_j(\boldsymbol{\theta}) - f_j(\B{z}_j) - \dfrac{1}{2\rho_j^2}\nr{\boldsymbol{\theta}-\B{z}_j}^2_2}\mathrm{d}\B{z}_j = \prod_{j=1}^J\mathcal{K}_j(\boldsymbol{\theta}). \label{eq:proofTV2_I}
\end{align}
Bounding each term in \eqref{eq:proofTV2_I} and following the proof of Theorem 1 detailed above completes the proof.

\subsection{Proof of Proposition 4}

By using \eqref{eq:bound_I_A} and \eqref{eq:bound_I_B} we have for all $\boldsymbol{\theta} \in \mathbb{R}^d$,
\begin{align}
  B(\rho) & \leq \int_{\mathbb{R}^d}\exp\pr{f(\boldsymbol{\theta}) - f(\B{z}) - \dfrac{1}{2\rho^2}\nr{\boldsymbol{\theta}-\B{z}}^2_2}\mathrm{d}\B{z} &&\leq A(\rho) \nonumber\\
  B(\rho)\exp(-f(\boldsymbol{\theta})) & \leq \int_{\mathbb{R}^d}\exp\pr{- f(\B{z}) - \dfrac{1}{2\rho^2}\nr{\boldsymbol{\theta}-\B{z}}^2_2}\mathrm{d}\B{z} &&\leq A(\rho)\exp(-f(\boldsymbol{\theta})) \nonumber\\
-\log A(\rho) + f(\boldsymbol{\theta}) & \leq -\log\int_{\mathbb{R}^d}\exp\pr{- f(\B{z}) - \dfrac{1}{2\rho^2}\nr{\boldsymbol{\theta}-\B{z}}^2_2}\mathrm{d}\B{z} &&\leq -\log B(\rho) + f(\boldsymbol{\theta})\nonumber
\end{align}
So that
\begin{align}
   -\log A(\rho) + \dfrac{d}{2}\log(2\pi\rho^2) \leq f_{\rho}(\boldsymbol{\theta}) - f(\boldsymbol{\theta}) \leq -\log B(\rho) + \dfrac{d}{2}\log(2\pi\rho^2).
\end{align}
The result of Proposition 4 follows from the definition of $A(\rho)$ and $B(\rho)$.

\subsection{Proof of Proposition 5}

By using \eqref{eq:bound_I_A} and \eqref{eq:bound_I_B} it follows, for all $\boldsymbol{\theta} \in \mathbb{R}^d$,
\begin{align}
B(\rho) &\leq \mathcal{K}(\boldsymbol{\theta}) \leq A(\rho) \\
B(\rho)C_{\pi}\pi(\boldsymbol{\theta}) &\leq \mathcal{K}(\boldsymbol{\theta})C_{\pi}\pi(\boldsymbol{\theta})\leq A(\rho)C_{\pi}\pi(\boldsymbol{\theta}).
\end{align}
Using \eqref{eq:proofTV_I} yields
\begin{align}
B(\rho)\pi(\boldsymbol{\theta}) &\leq \pi_{\rho}(\boldsymbol{\theta})\dfrac{C_{\pi_{\rho}}}{C_{\pi}} \leq A(\rho)\pi(\boldsymbol{\theta}) \\
B(\rho)\pi(\boldsymbol{\theta}) &\leq \pi_{\rho}(\boldsymbol{\theta})(2\pi\rho^2)^{d/2} \leq A(\rho)\pi(\boldsymbol{\theta}).
\end{align}
Using \eqref{eq:A} and \eqref{eq:B} gives
\begin{align}
\dfrac{N_{\rho}}{D_{-d}(-L_f\rho)}\marginal(\boldsymbol{\theta}) \leq \pi(\boldsymbol{\theta})\leq \dfrac{N_{\rho}}{D_{-d}(L_f\rho)}\marginal(\boldsymbol{\theta}), \label{eq:cred_1}
\end{align}
where the constant $N_{\rho}$ has been defined in (39) in the main paper.

Let $\mathcal{C}_{\alpha}^{\rho}$ an arbitrary $(1-\alpha)$-credibility region under $\marginal$.
By integrating \eqref{eq:cred_1} on $\mathcal{C}_{\alpha}^{\rho}$, 
\begin{align}
\dfrac{N_{\rho}}{D_{-d}(-L_f\rho)}(1-\alpha) \leq \int_{\mathcal{C}_{\alpha}^{\rho}}\pi(\boldsymbol{\theta})\mathrm{d}\boldsymbol{\theta}\leq \dfrac{N_{\rho}}{D_{-d}(L_f\rho)}(1-\alpha). \label{eq:bound_coverage}
\end{align}
Since $\mathcal{C}^{\rho}_{\alpha} \subseteq \mathbb{R}^d$ and $\int_{\mathbb{R}^d}\pi(\boldsymbol{\theta})\mathrm{d}\boldsymbol{\theta} = 1$, the upper bound in \eqref{eq:bound_coverage} can be replaced by \\$\min\bbr{1,\dfrac{N_{\rho}}{D_{-d}(L_f\rho)}(1-\alpha)}$.

\subsection{Lipschitz loss functions - Dependence w.r.t. the number of observations}

Combining Corollary 4 in the main paper with \eqref{eq:equivalent_bound}, we have:
\begin{align}
\nr{\pi - \pi_{\rho}}_{\mathrm{TV}} &\leq 1 - \prod_{i=1}^n\pr{1 - \dfrac{2\sqrt{2}\Gamma\pr{\dfrac{d+1}{2}}}{\Gamma\pr{\dfrac{d}{2}}}L_{f_i}\rho + o(\rho)} \\
&= \dfrac{2\sqrt{2}\Gamma\pr{\dfrac{d+1}{2}}}{\Gamma\pr{\dfrac{d}{2}}}\rho \sum_{i=1}^nL_{f_i} + o(\rho) \\
&\leq \dfrac{2\sqrt{2}\Gamma\pr{\dfrac{d+1}{2}}\displaystyle\max_{i \in [n]}L_{f_i}}{\Gamma\pr{\dfrac{d}{2}}} n \rho + o(\rho).
\end{align}

\newpage

\section{On the choice of \texorpdfstring{$\kappa_{\rho}$}{the smoothing function}}

\begin{figure}[h]
\centering
  \mbox{{\includegraphics[scale=0.5]{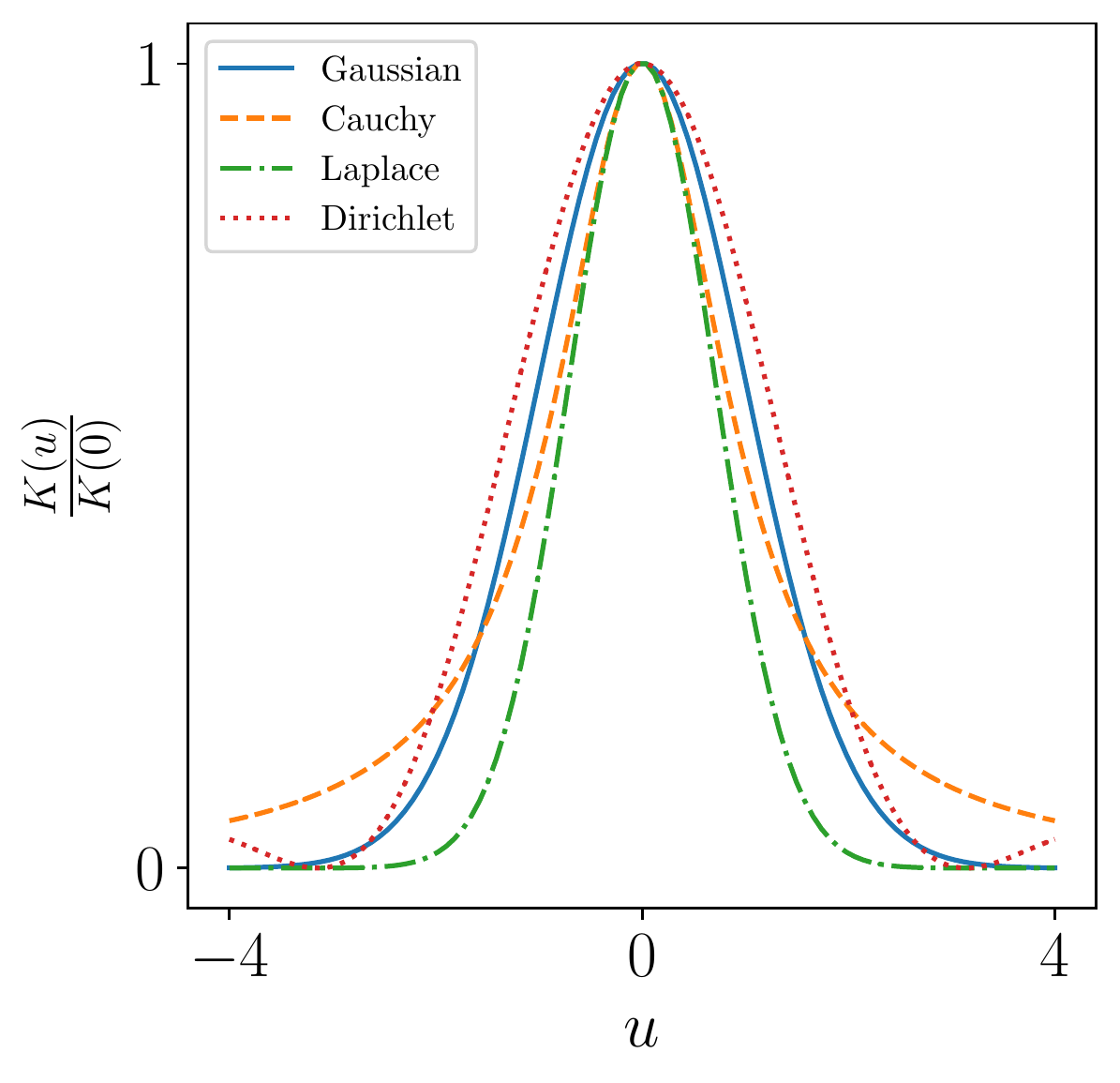}}}
  \mbox{{\includegraphics[scale=0.5]{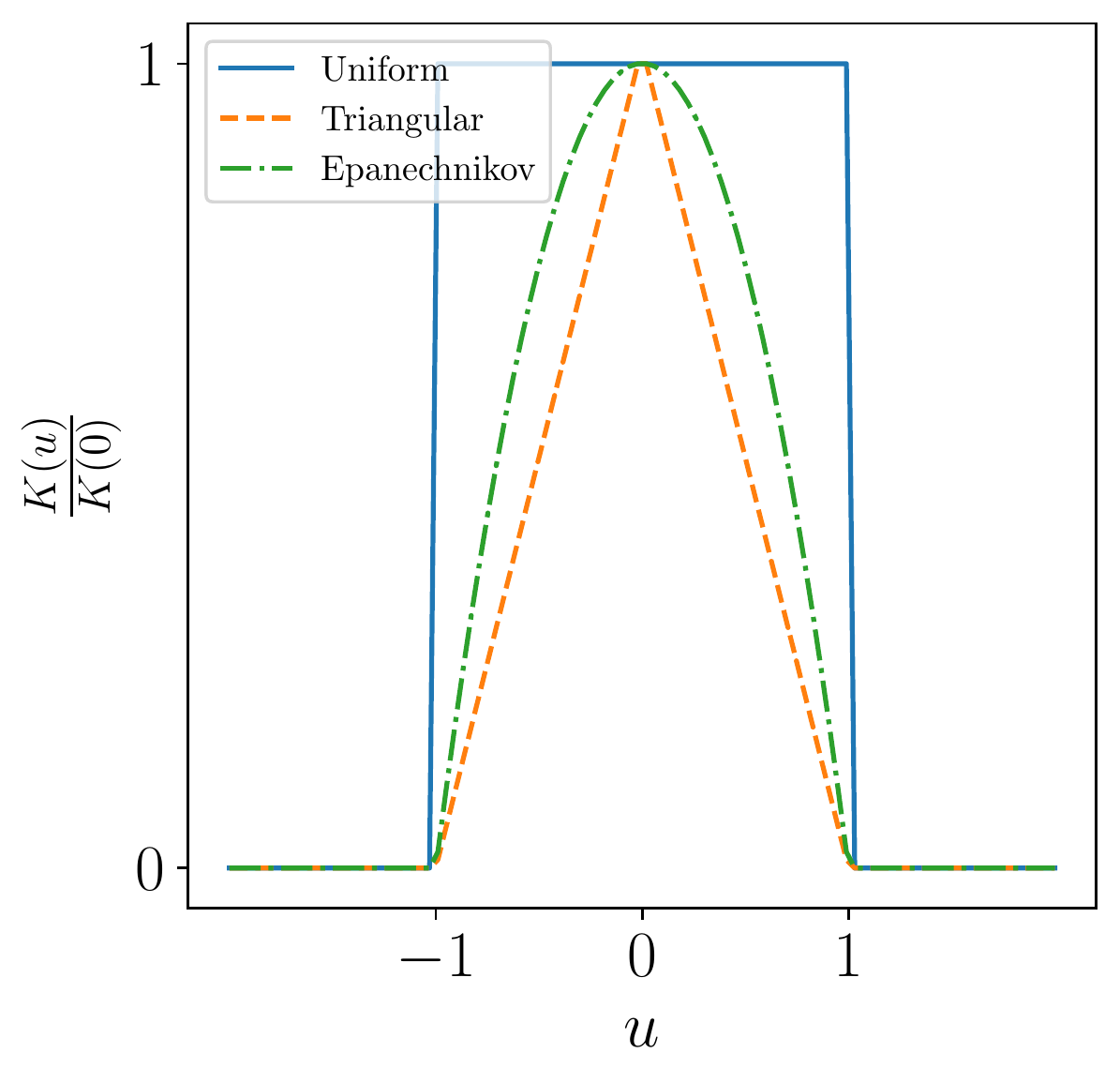}}}
  \caption{(left) Normalized non compactly-supported kernels; (right) normalized compactly-supported kernels detailed in Table \ref{table:kernels}.}
  \label{fig:kernels}
\end{figure}

Section 2.2 of the main document provides two distinct ways to choose $\kappa_{\rho}$. We can first consider a kernel $K$ that is a positive function such that $\displaystyle\int_{\mathbb{R}^d}K(\B{u})\mathrm{d}\B{u} = 1$ and $K(-\B{u}) = K(\B{u})$, for all $\B{u} \in \mathbb{R}^d$.
Based on the latter, we  define for all $\B{z},\boldsymbol{\theta} \in \Theta$, $\kappa_{\rho}(\B{z},\boldsymbol{\theta}) \propto_{\B{z}} \rho^{-d}K(\rho^{-1}(\boldsymbol{\theta}-\B{z}))$ \citep{Dang2012}.
Table \ref{table:kernels} lists some classical examples of symmetric kernels $K(\cdot)$ which are not necessarily compactly supported.
For sake of simplicity, we only define univariate versions of them but they can obviously be generalized in higher dimension.
Figure \ref{fig:kernels} illustrates these kernels.
Some of them have for instance been used in ABC approaches \citep{Sisson2018}.

\begin{table}[h]
\caption{Examples of classical kernels $K$ that can be used to define an appropriate density $\kappa_{\rho}$ verifying Property 1 in the main paper.}
\begin{center}
\begin{tabular}{ccc}
name & support & $K(u)$\\
\hline
Gaussian & $\mathbb{R}$ & $\frac{1}{\sqrt{2\pi}}\exp\pr{-u^2/2}$\\ 
Cauchy & $\mathbb{R}$ & $\frac{1}{\pi(1 + u^2)}$\\ 
Laplace & $\mathbb{R}$ & $\frac{1}{2}\exp\pr{-|u|}$\\ 
Dirichlet & $\mathbb{R}$ & $\frac{\sin^2(u)}{\pi u^2}$\\
Uniform & $[-1,1]$ & $\frac{1}{2}\mathbbm{1}_{|u| \leq 1}$\\ 
Triangular & $[-1,1]$ & $(1-|u|)\mathbbm{1}_{|u| \leq 1}$\\ 
Epanechnikov & $[-1,1]$ & $\frac{3}{4}(1-u^2)\mathbbm{1}_{|u| \leq 1}$\\ 
\end{tabular}
\end{center}
\label{table:kernels}
\end{table} 

Another natural choice consists in resorting to a divergence function $\phi$ such that $\kappa_{\rho}(\B{z},\boldsymbol{\theta}) \propto_{\B{z}} \exp(-\rho^{-1}\phi(\B{z},\boldsymbol{\theta}))$. Particular instances of these functions are the family of the Bregman divergences, which are ubiquitous tools in signal processing, machine learning and optimization (see Definition 1 of the main document). Table \ref{table:kernels_divergences} recalls classical examples of Bregman divergences that can be used to define $\phi$. Again, only univariate examples of such potentials are provided but they can be easily extended to the multivariate case.

\begin{table}[h]
\caption{Examples of divergence functions $\phi$ that can be used to define an appropriate density $\kappa_{\rho}$ verifying Property 1 in the main paper.}
\begin{center}
\begin{tabular}{ccc}
name & $\Theta$ & $\phi(z,\theta)$\\
\hline
Squared loss & $\mathbb{R}$ & $(z-\theta)^2$\\ 
Absolute loss & $\mathbb{R}$ & $|z-\theta|$\\ 
Logistic loss & $[0,1]$ & $z\log\pr{\frac{z}{\theta}} + (1-z)\log\pr{\frac{1-z}{1-\theta}}$\\ 
Itakura-Saito divergence & $\mathbb{R}_+$ & $\frac{z}{\theta} - \log\pr{\frac{z}{\theta}} - 1$\\ 
Kullback-Leibler divergence & $[0,1]$ & $z\log\pr{\frac{z}{\theta}}$\\
\end{tabular}
\end{center}
\label{table:kernels_divergences}
\end{table} 

Note that, reciprocally, tight connections have been already drawn between Bregman divergences and regular exponential family distributions. Indeed, when the function $\kappa_{\rho}$ belongs to the exponential family, its associated potential function defined by $-\log \kappa_{\rho}$ can be expressed as a Bregman divergence up to an additional term \citep[Theorem 4]{Banerjee2005}
\begin{equation}
  -\log \kappa_{\rho}(\B{z},\boldsymbol{\theta}) = d_{\psi}(\B{z},\mathbb{E}(\B{z}|\boldsymbol{\theta})) - \log b_{\psi}(\B{z}).
\end{equation}

\section{Illustration for Lipschitz loss functions used in statistical learning}
\normalfont

\begin{figure}[!h]
\centering
  \mbox{{\includegraphics[scale=0.35]{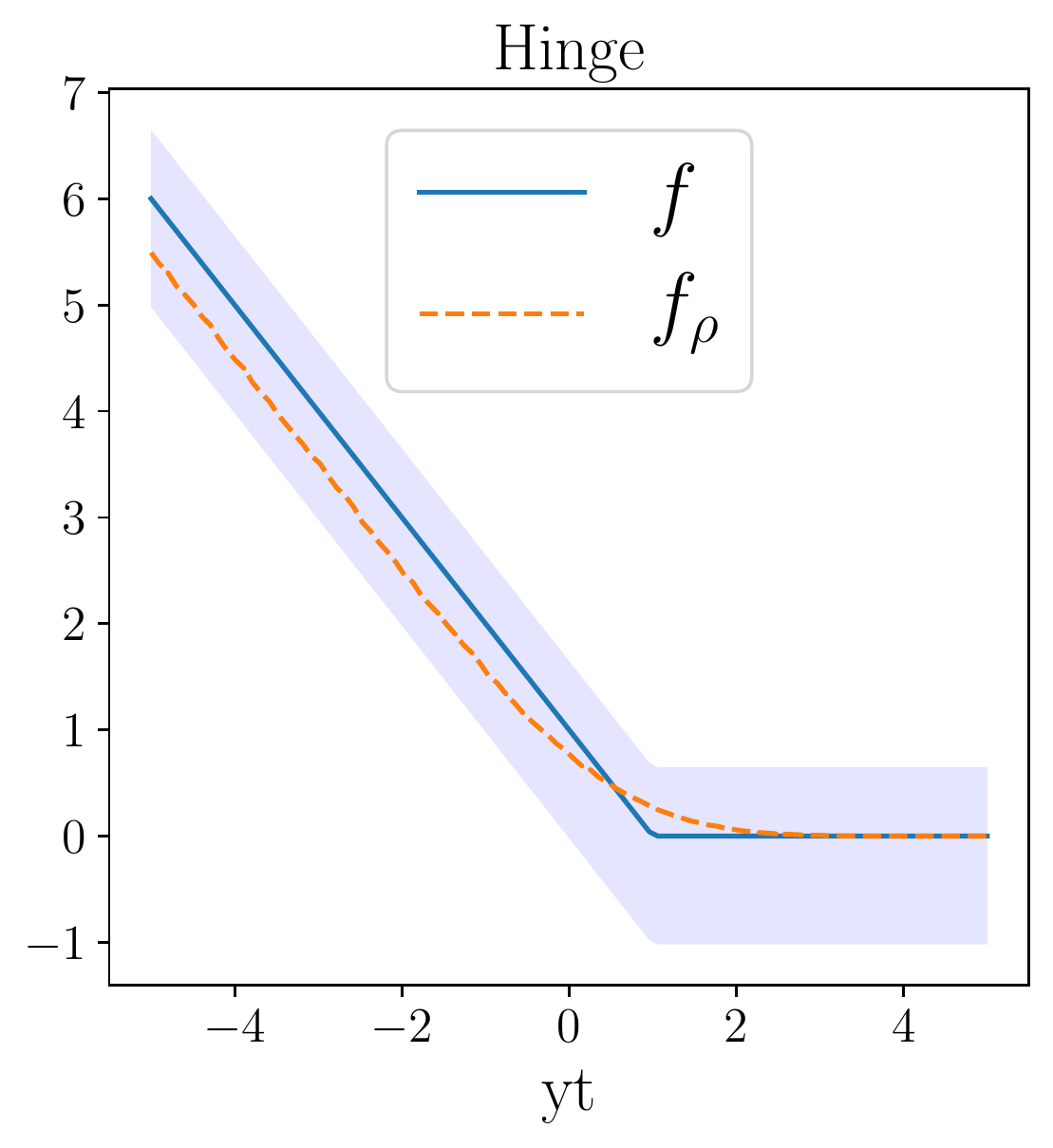}}}
  \mbox{{\includegraphics[scale=0.35]{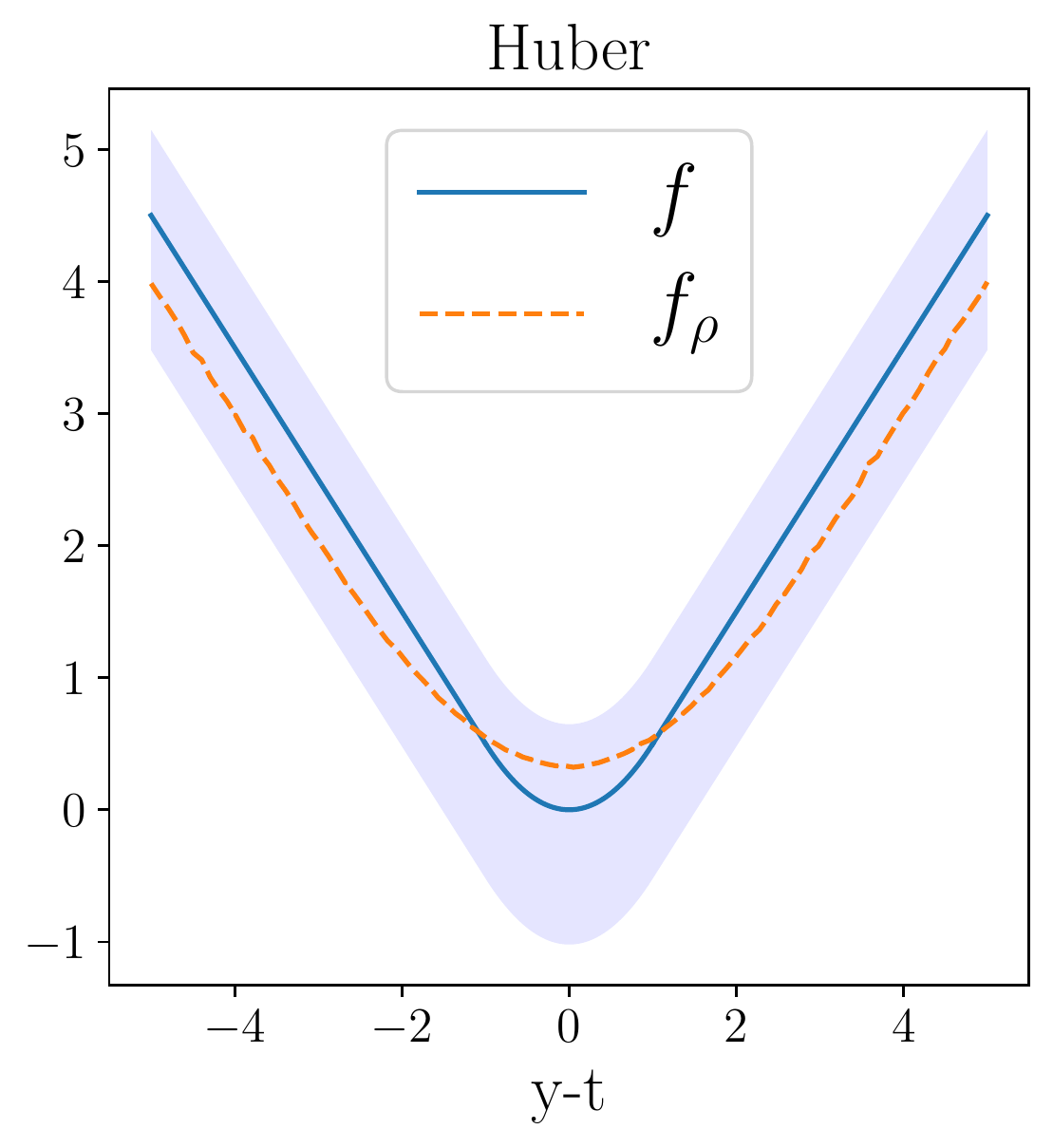}}}
  \mbox{{\includegraphics[scale=0.35]{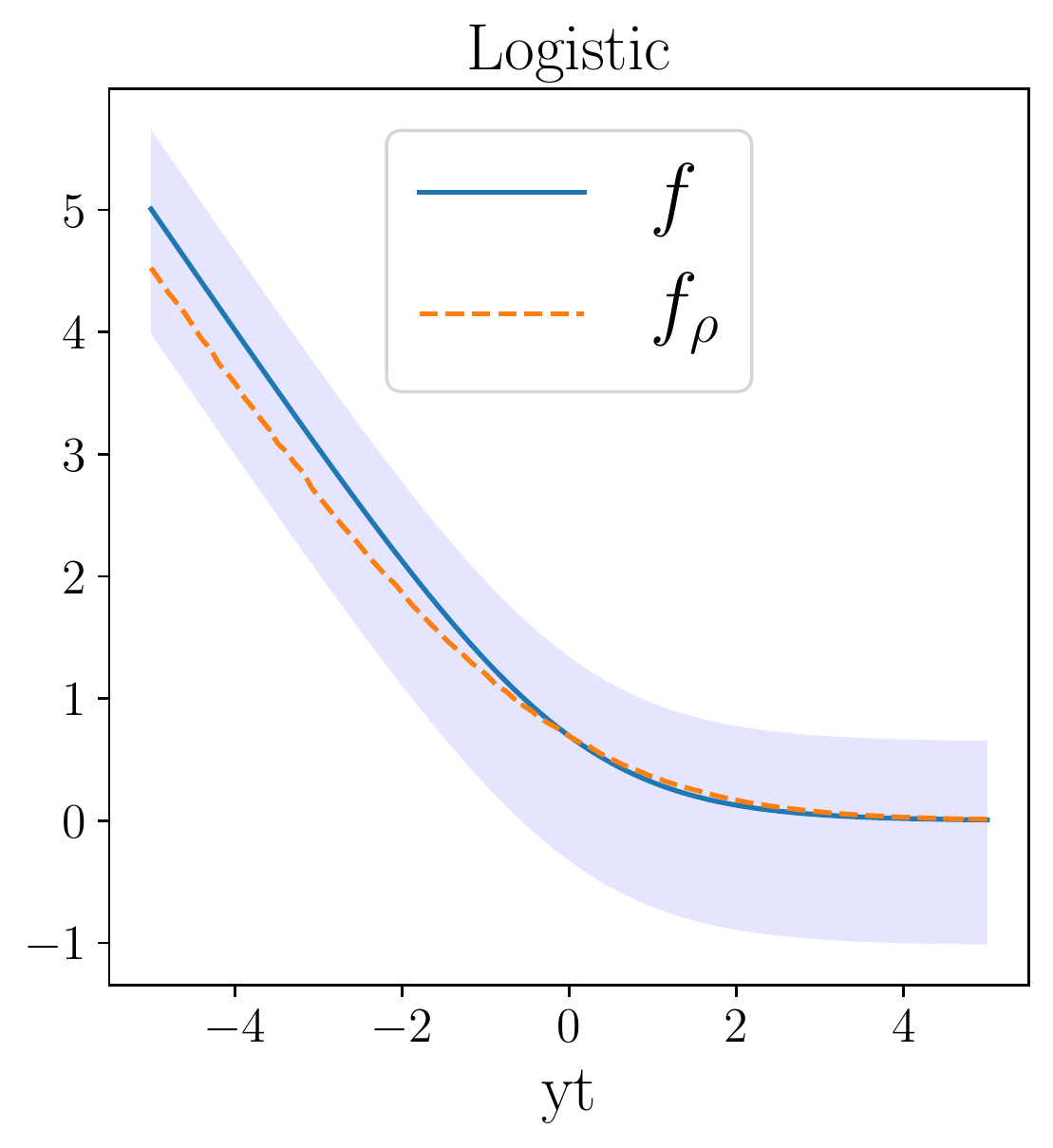}}}
  \mbox{{\includegraphics[scale=0.35]{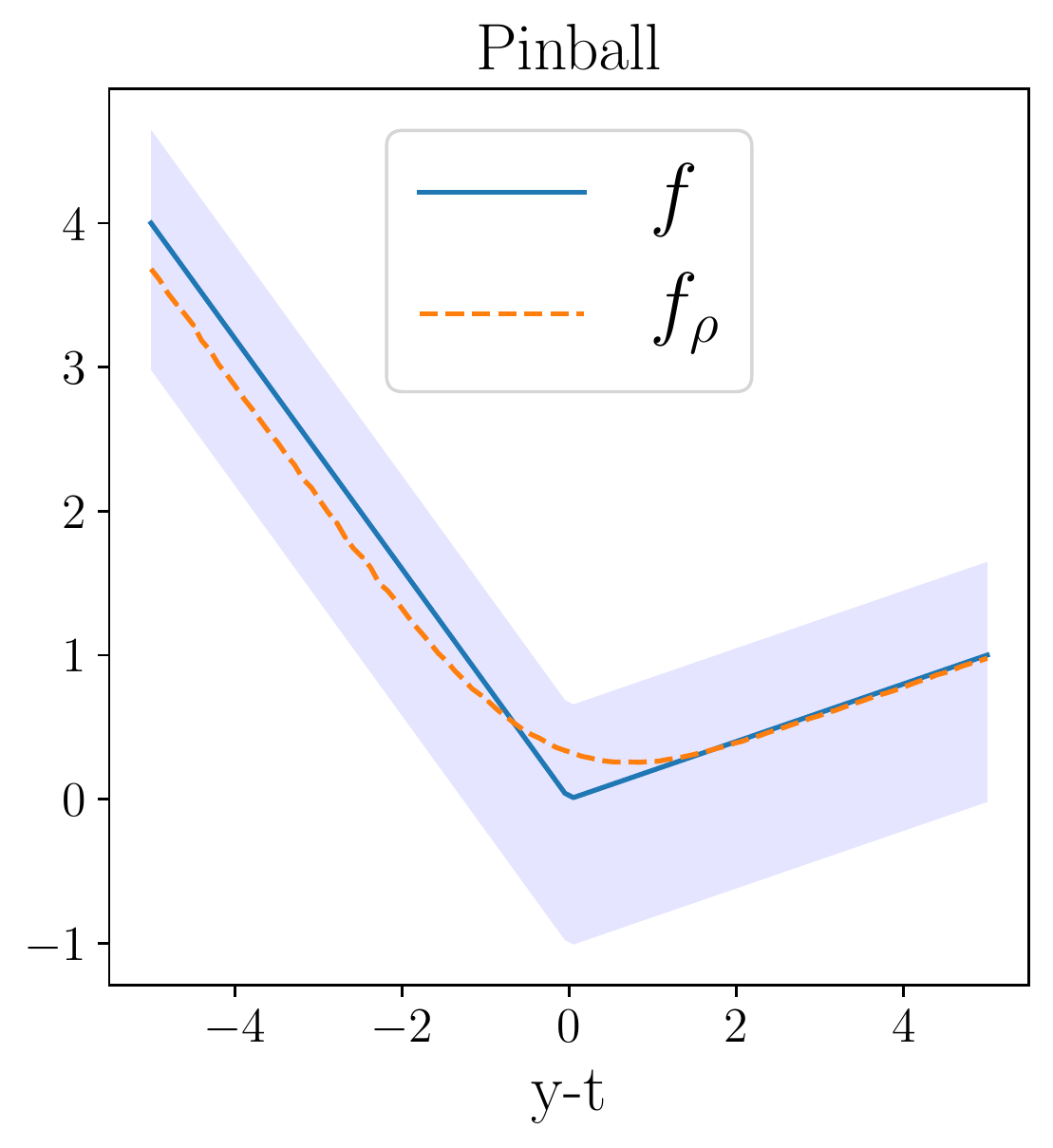}}}
  \caption{Loss functions of Table \ref{table:Lipschitz_loss} along with their associated regularized loss $f_{\rho}$ with $\rho=1$ estimated with a Monte Carlo approximation. The Huber and pinball losses have been plotted with $\delta=1$ and $\tau = 0.2$, respectively. The contours of the shaded area correspond to $f + L_{\rho}$ and $f + U_{\rho}$.}
  \label{fig:example5_loss}
\end{figure}

Some of the results of Section 4.3 in the main paper assume that the potential function $f$ associated to $\pi$ is Lipschitz.
Interestingly, such Lipschitz functions are used in standard statistical learning problems to evaluate the discrepancy between observations and model outputs \citep{vandeGeer2016}.
\begin{table}
\caption{Lipschitz loss functions $f$ used in standard statistical learning problems.
Their domain of definition is denoted $\mathcal{D}_f$ and $y$ stands for an observation.
The notation ``reg.'' stands for regression.}
\begin{center}
\begin{tabular}{c c c p{8cm}}
name & problem & $\mathcal{D}_f$ & \multicolumn{1}{c}{$f(y;t)$} \\
\hline
hinge & SVM & $\{-1,1\} \times \mathbb{R}$ & $\max\pr{0, 1 - yt}$\\
\multirow{ 1}{*}{Huber} & \multirow{ 1}{*}{robust reg.} & \multirow{ 1}{*}{$\mathbb{R} \times \mathbb{R}$} & $\left\{
                \begin{array}{l}
                  (y-t)^2/(2\delta) \ \text{if} \ |y-t|\leq\delta\\
                  |y-t|-\delta/2 \ \text{otherwise, where} \ \delta > 0
                \end{array}
              \right.$\\ 
logistic & logistic reg. & $\{-1,1\} \times \mathbb{R}$ & $\log(1+\exp(-yt))$\\
\multirow{ 1}{*}{pinball} & \multirow{ 1}{*}{quantile reg.} & \multirow{ 1}{*}{$\mathbb{R} \times \mathbb{R}$} & \small $\tau \max(0,t - y) + (1 - \tau)\max(0,y - t)$, $\tau \in (0,1)$\\
\end{tabular}
\end{center}
\label{table:Lipschitz_loss}
\end{table}
Table \ref{table:Lipschitz_loss} lists some of them along with their definition and associated statistical problems.
Note that the absolute loss stands for a particular instance of the pinball loss with $\tau = 0.5$.
Figure \ref{fig:example5_loss} illustrates the form of these losses and associated regularized potentials $f_{\rho}$ with $\rho=1$ obtained via a Monte Carlo approximation. 

Without loss of generality, these problems consider a likelihood function that can be written as in (32) in the main paper with
\begin{align}
  f_j(y_j;\boldsymbol{\theta}) = f(y_j;\B{x}_j^T\boldsymbol{\theta}), \label{eq:example5_func}
\end{align}
where for $j \in [n]$, $\B{x}_j$ is the feature vector associated with observation $y_j$; $f$ is one of the loss functions in Table \ref{table:Lipschitz_loss} and $\boldsymbol{\theta} \in \mathbb{R}^d$ is the parameter to infer.
Since all the loss functions listed in Table \ref{table:Lipschitz_loss} are Lipschitz continuous w.r.t. their second argument $t$ with Lipschitz constant equal to 1, the potential $f_j$ in \eqref{eq:example5_func} is also Lipschitz with constant $L_{f_j} = \nr{\B{x}_j}_2$.
Motivated by the robustness properties inherited by AXDA, see Section 3.3 in the main paper, we consider the smoothing of the likelihood contribution associated to each observation $f_j$ with a Gaussian kernel.
The results of Corollary 3 in the main paper can then be applied to $\pi$ defined in (32) in the main paper.

In practice, to illustrate the behavior of the upper bound in Corollary 3 w.r.t. the number of observations, we fixed the dimension $d$ and considered several values of $n$ ranging from 1 to $10^4$.
For each $n$, we randomly generated sets of features $\left\{\B{x}_{j}\right\}_{j \in [n]}$ and we normalized the columns of the matrix $\B{X}^T=\left[\B{x}_1,\ldots,\B{x}_n\right]^T$ such that each entry is a random number between $0$ and $1$.
The latter operation is classical in machine learning and is also called feature scaling.

\begin{figure}[htb]
\centering
  \mbox{{\includegraphics[scale=0.5]{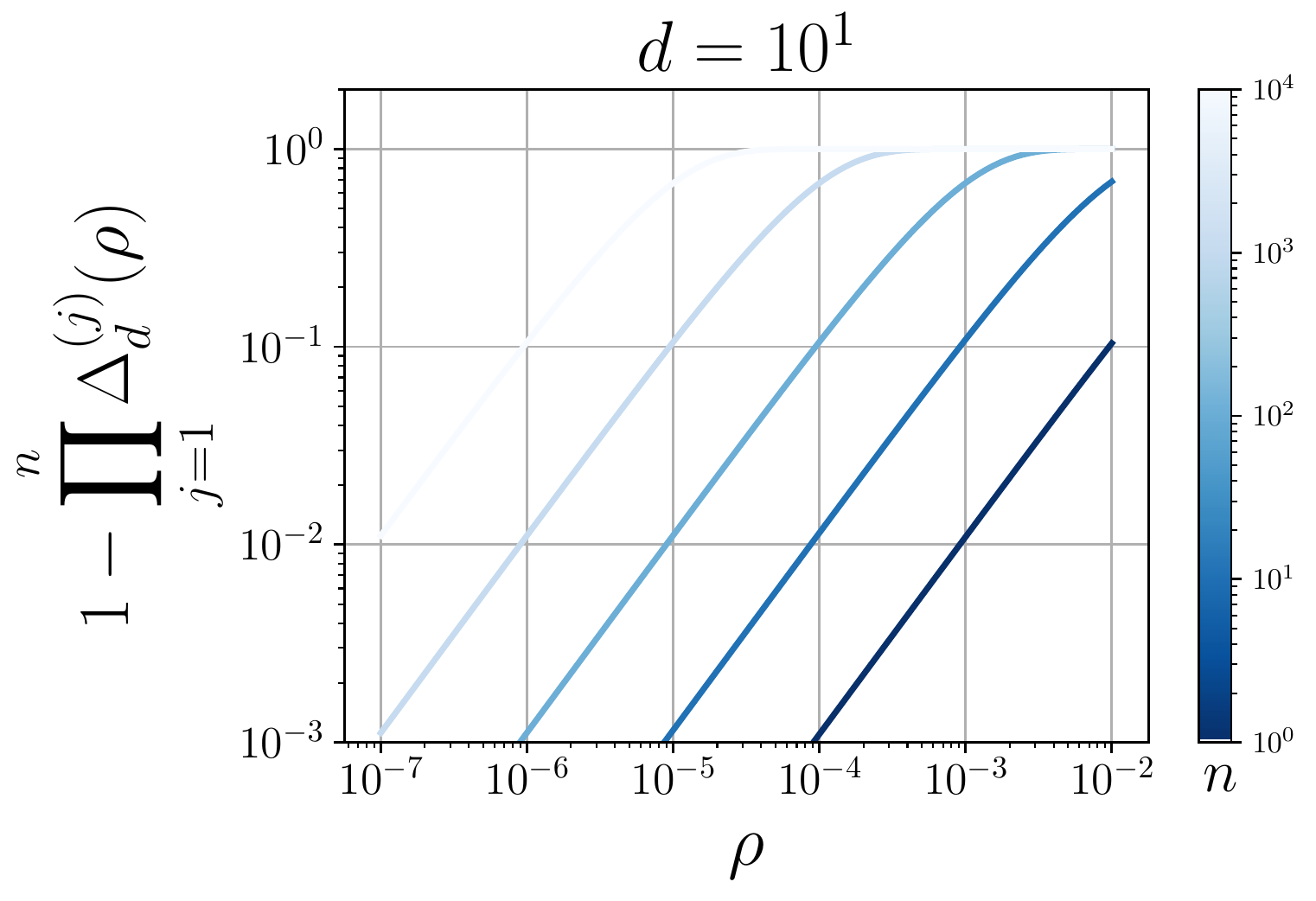}}}
  \mbox{{\includegraphics[scale=0.5]{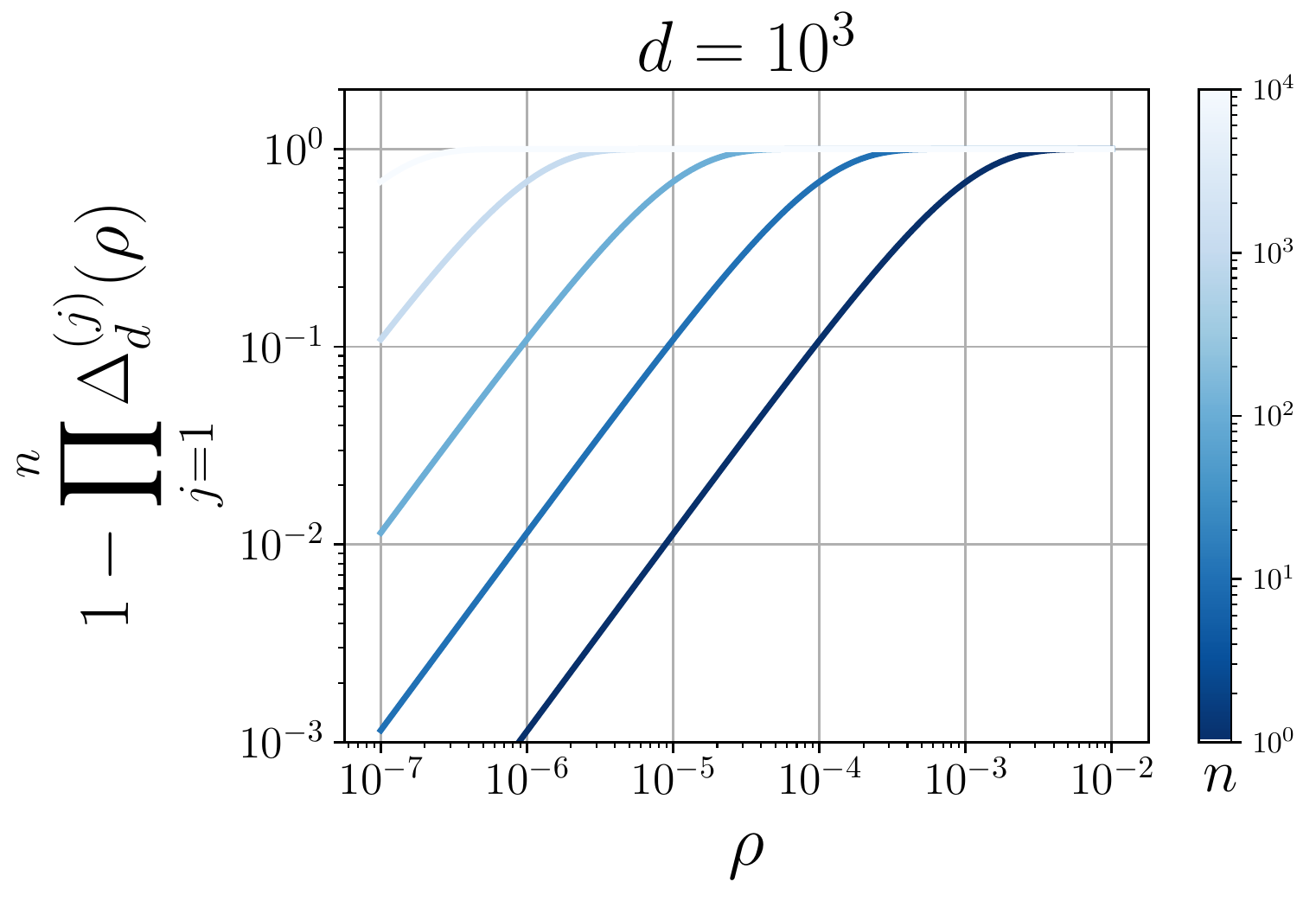}}}
  \caption{Behavior of the upper bound in Corollary 3 w.r.t. $\rho$ and $n$ for several values of the dimension $d$. The notation $\Delta^{(j)}_d(\rho)$ has been defined in Corollary 3.}
  \label{fig:example5}
\end{figure}

Figure \ref{fig:example5} shows the behavior of the upper bound in Corollary 3 for two values of the dimension $d=10$ and $d=10^3$.
As expected, the bound becomes less informative for a fixed value of $\rho$ as the number of likelihood approximations increases with the size of the dataset $n$.
Nonetheless, the effect of $n$ on the bound is not highly prohibitive.
In both cases $d=10$ and $d=10^3$, $\rho$ and $n$ appear to be complementary variables: increasing the value of the latter and decreasing the value of the former by the same factor roughly gives the same bound value.
Actually, one can show that the dependence of the bound when $\rho$ is small is of the order $\mathcal{O}(n\rho)$ for a fixed dimension $d$, see the supplementary material.
Obviously, one can limit this dependence on $n$ by splitting \textit{blocks} of observations in minibatches instead of splitting each observation.
This splitting strategy has for instance been considered by \cite{Rendell2018}.

\section{Inference details for the image inpainting example}

In this section, we detail the steps of the Gibbs sampler used to sample from the posterior distribution $\pi_{\rho}(\boldsymbol{\theta},\B{z}|\B{y})$ in Section 5.4 in the main paper.

\subsection{Sampling the auxiliary vector}

The conditional distribution associated to the auxiliary variable $\B{Z} = (\B{z}_1,\B{z}_2)$ writes
\begin{align}
\pi_{\rho}(\B{Z}|\boldsymbol{\theta}) \propto &\exp\pr{- \tau\sum_{1\leq i \leq d} \nr{\B{Z}_i}_2 - \dfrac{1}{2\rho^2}\nr{\B{Z} - \B{D} \boldsymbol{\theta}}_2^2}. \label{eq:inpainting_cond_z0}
\end{align}

This conditional distribution can be sampled exactly by using data augmentation.
Indeed, one can re-write the distribution involving the non-differentiable potential $\nr{\cdot}_2$ as a mixture of normal and gamma distributions \cite[Section 3.1]{kyung2010}.
Hence, sampling from \eqref{eq:inpainting_cond_z0} can be performed with the following two steps
\begin{align*}
  &\text{Draw } \dfrac{1}{\gamma_i} \sim \mathrm{InverseGaussian}\pr{\dfrac{\tau}{\nr{\B{Z}_{i}}_2},\tau^2} \forall i \in [d], \text{ if $\nr{\B{Z}_{i}}_2 > 0$} \\
  &\text{Draw } \dfrac{1}{\gamma_i} \sim \mathrm{InverseGaussian}\pr{\dfrac{3}{2},\dfrac{\tau^2}{2}} \forall i \in [d], \text{ if $\nr{\B{Z}_{i}}_2 = 0$} \\
  &\text{Draw } \B{z}_{1,i} \sim \mathcal{N}\pr{\dfrac{\gamma_i (D_1\boldsymbol{\theta})_i}{\rho^2 + \gamma_i},\dfrac{\rho^2\gamma_i}{\rho^2 + \gamma_i}} \forall i \in [d], \\
  &\text{Draw } \B{z}_{2,i} \sim \mathcal{N}\pr{\dfrac{\gamma_i (D_2\boldsymbol{\theta})_i}{\rho^2 + \gamma_i},\dfrac{\rho^2\gamma_i}{\rho^2 + \gamma_i}} \forall i \in [d].
\end{align*}

Note that all these sampling steps can be performed efficiently by ``vectorizing'' them.

\subsection{Sampling the parameter of interest}

The conditional distribution associated to the image to recover $\boldsymbol{\theta}$ writes
\begin{align}
\pi_{\rho}(\boldsymbol{\theta}|\B{Z},\B{y}) \propto &\exp\pr{- \dfrac{1}{2\rho^2}\nr{\B{Z}-\B{D} \boldsymbol{\theta}}_2^2 - \dfrac{1}{2\sigma^2}\nr{\B{y} - \B{H}\boldsymbol{\theta}}_2^2}. \label{eq:inpainting_cond_theta}
\end{align}

The distribution \eqref{eq:inpainting_cond_theta} is a non-degenerate Gaussian distribution $\mathcal{N}(\boldsymbol{\mu}_{\boldsymbol{\theta}},\B{\Sigma}_{\boldsymbol{\theta}})$ where 
\begin{align}
   \B{\Sigma}_{\boldsymbol{\theta}} &= \pr{\rho^{-2}\B{D}^T\B{D} + \sigma^{-2}\B{H}^T\B{H}}^{-1} \\
   \boldsymbol{\mu}_{\boldsymbol{\theta}} &= \B{\Sigma}_{\boldsymbol{\theta}}\pr{\sigma^{-2}\B{H}^T\B{y} + \rho^{-2}\B{D}^T\B{Z}}. 
\end{align}

Sampling from this multivariate distribution can be done efficiently in $O(d \log d)$ floating point operations by resorting to the two-dimensional discrete Fourier transform.
Indeed, under periodic boundary conditions for $\boldsymbol{\theta}$, the matrix $\B{D}^T\B{D}$ is a block circulant matrix and hence diagonalizable in the Fourier domain.
On the other hand, $\B{H}^T\B{H}$ stands for a diagonal matrix with some zeros on the diagonal corresponding to the missing pixels.
Since these two matrices cannot be diagonalized in the same domain, we use the auxiliary variable method of \cite{Marnissi2018} to decouple them.
Let $\eta\nr{\B{H}^T\B{H}}_{\mathrm{S}} < \sigma^2$ where $\nr{\B{M}}_{\mathrm{S}}$ is the spectral norm of the matrix $\B{M}$.
Then, we have the following two-step sampling scheme
\begin{align*}
  &\text{Draw } \B{v} \sim \mathcal{N}\pr{\pr{\dfrac{\B{I}_d}{\eta} - \dfrac{\B{H}^T\B{H}}{\sigma^2}}\boldsymbol{\theta},\dfrac{\B{I}_d}{\eta} - \dfrac{\B{H}^T\B{H}}{\sigma^2}}, \\
  &\text{Draw } \boldsymbol{\theta} \sim \mathcal{N}\pr{\boldsymbol{\mu}_{\boldsymbol{\theta}},\B{\Sigma}_{\boldsymbol{\theta}}},
\end{align*}
where 
\begin{align*}
  &\B{\Sigma}_{\boldsymbol{\theta}} = \pr{\dfrac{\B{I}_d}{\eta} - \dfrac{\B{D}^T\B{D}}{\rho^2}}^{-1},\\
  &\boldsymbol{\mu}_{\boldsymbol{\theta}} = \B{\Sigma}_{\boldsymbol{\theta}}\pr{\B{v} + \dfrac{\B{H}^T}{\sigma^2}\B{y} + \dfrac{\B{D}^T}{\rho^2}\B{Z}}.
\end{align*}

\section{Inference algorithms based on AXDA}
\label{sec:inference}

Motivated by the issues detailed in Section 2, the good expected properties reviewed in Section 3 as well as the theoretical results shown in Section 4 in the main paper, this section shows that AXDA may allow to derive more efficient and distributed inference algorithms ranging from simulation to optimization-based methods. To do so, the potential benefits of AXDA in comparison with  direct inference from $\pi$ are presented and discussed.
MCMC and VB methods based on AXDA models are detailed to explore the distribution of the parameters to infer. 
Optimization-based approaches such as ADMM and the EM-algorithm are also derived if maximum likelihood (ML) or maximum a posteriori (MAP) estimates based on the approximate density $\marginal$ are desired.

From now on, we assume that $\Theta=\mathbb{R}^d$ and we consider a target density with the general form
\begin{align}
  \pi(\boldsymbol{\theta}) \propto \prod_{j=1}^J \pi_j(\B{A}_j\boldsymbol{\theta}) \propto \prod_{j=1}^J \displaystyle\exp\pr{-f_j(\B{A}_j\boldsymbol{\theta})}. \label{eq:target_density_2}
\end{align}
Based on this target density, the augmented density $\joint$ is assumed to take the form
\begin{align}
  \joint(\boldsymbol{\theta},\B{z}_{1:J}) \propto \prod_{j=1}^J \pi_j(\B{z}_j)\kappa_{\rho}(\B{z}_j,\B{A}_j\boldsymbol{\theta}). \label{eq:split_density_2}
\end{align}
This writing permits to highlight the benefits of using the augmented density $\joint$ instead of $\pi$ for each of the different inference approaches detailed in the sequel.

\subsection{Monte Carlo sampling from AXDA}
\label{subsec:simulation}

A standard way to sample from the joint density $\joint$ is to consider a Gibbs sampler as in Algorithm \ref{algo:Gibbs}.
This algorithm can make the inference tractable, simpler and/or faster by targetting $\joint$ instead of $\pi$.
First, by splitting the initial potential $\sum_j f_j$, $\joint$ admits simpler and lower-dimensional conditional posterior distributions, each of them possibly defined by a single potential $f_j$. Within a Gibbs algorithm, these conditional posteriors yield simpler sampling steps, which may embed efficient dedicated state-of-the-art sampling methods.
Second, given the current iterate $\boldsymbol{\theta}^{[t]}$, sampling each auxiliary variable $\B{z}_j^{[t]}$ can be performed in an independent and parallel manner for a faster inference.
This is of particular interest in big data settings where datasets are stored on multiple kernels or machines \citep{Rendell2018}.
In addition, \cite{Vono2019} experimentally showed that considering AXDA-based models can even improve the convergence properties of classical MCMC methods such as Langevin Monte Carlo by embedding them.
A detailed description of additional benefits of AXDA simulation-based methods and their illustration on image processing and machine learning problems can be found in \cite{Rendell2018} and \cite{Vono2019}.

\renewcommand{\baselinestretch}{0.8}
\begin{algorithm}
    \caption{Gibbs sampler}
    \label{algo:Gibbs}
     \SetKwInOut{Input}{Input}
     \Input{Functions $f_j$, tolerance parameter $\rho$, initialization $\B{z}^{[0]}$ and total nb. of iterations $T_{\mathrm{MC}}$}
   \For{$t \leftarrow 1$ \KwTo $T_{\mathrm{MC}}$}{%
   \textit{\% Drawing the variable of interest $\boldsymbol{\theta}$} \\
   $\boldsymbol{\theta}^{[t]} \sim  \joint(\boldsymbol{\theta}|\B{z}^{[t-1]}) = \displaystyle\prod_{j=1}^J \kappa_{\rho}(\B{z}_j^{[t-1]},\B{A}_j\boldsymbol{\theta})$\;
   \textit{\% Drawing the splitting variables $\B{z}_j$} \\
   \For{$j\gets1$ \KwTo $J$}{
   $\B{z}_j^{[t]} \sim  \joint(\B{z}_j|\boldsymbol{\theta}^{[t]}) = \pi_j(\B{z}_j)\kappa_{\rho}(\B{z}_j,\B{A}_j\boldsymbol{\theta}^{[t]})$\;
   }
   }
   \SetKwInOut{Output}{Output}
     \Output{Collection of samples $\bbr{\boldsymbol{\theta}^{[t]}}_{t = 1}^{T_{\mathrm{MC}}}$ asymptotically distributed according to $\joint$.}
\end{algorithm}
\renewcommand{\baselinestretch}{1.3}

\begin{example}\normalfont
  We consider in this example the penalized logistic regression problem.
  We assume that $n$ binary responses $\B{y} \in \{-1,1\}^n$ are observed and correspond to conditionally independent Bernoulli random variables with probability of success $\sigma(\B{x}_j^T\boldsymbol{\theta})$.
  The function $\sigma$ is the sigmoid function, $\B{x}_{j} \in \mathbb{R}^d$ stands for the feature vector associated to observation $y_j$ and $\boldsymbol{\theta} \in \mathbb{R}^d$ are the unknown regression coefficients to infer.
  We consider a zero-mean Gaussian prior distribution on $\boldsymbol{\theta}$ with precision $2\tau$, that is $g(\boldsymbol{\theta}) = \tau\nr{\boldsymbol{\theta}}_2^2$.
  The target $\pi$ then stands for the posterior distribution of the unknown regression coefficients $\boldsymbol{\theta}$
  \begin{align}
    \pi(\boldsymbol{\theta}|\B{y}) \propto \displaystyle\exp\pr{-g(\boldsymbol{\theta})-\sum_{j=1}^n\log\br{1+\exp\pr{y_j\B{x}_j^T\boldsymbol{\theta}}}}. \label{eq:target_density_logReg}
  \end{align}
  By denoting $f_{(n+1)} = g$ and for all $j \in [n]$ with $J = n$, $f_j(u) = \log\br{1+\exp\pr{y_{j}u}}$, the posterior distribution in \eqref{eq:target_density_logReg} has the form \eqref{eq:target_density_2} with $J=n+1$.
  Following the work of \cite{Polson2013}, one can derive a promising DA scheme from $\pi$ based on the Polya-Gamma distribution.
  Hence, a Gibbs sampler can be used to sample from each conditional distribution as detailed by \cite{Polson2013}.
  However, this Gibbs sampler scales poorly in high-dimensional settings as pointed out by \cite{Durmus2016}. 
  The AXDA alternative is Algorithm \ref{algo:Gibbs} with a quadratic potential and resulting conditional distributions
  \begin{align}
    \pi_{\rho}(z_j|\boldsymbol{\theta},y_{j}) &\propto \displaystyle\exp\pr{-\log\br{1+\exp\pr{y_jz_j}} - \dfrac{1}{\rho^2}(z_j - \B{x}_j^T\boldsymbol{\theta})^2} \quad \forall j \in [n] \label{eq:cond_z}\\
    \pi_{\rho}(\boldsymbol{\theta}|\B{z}_{1:n}) &\propto \displaystyle\exp\pr{-\tau\nr{\boldsymbol{\theta}}_2^2- \sum_{j=1}^n \dfrac{1}{\rho^2}(z_j - \B{x}_j^T\boldsymbol{\theta})^2}. \label{eq:cond11}
  \end{align}
  Thanks to this splitting scheme, the inference is simpler, might be distributed, and sampling from these conditional distributions can be done exactly and efficiently.
  Indeed, since \eqref{eq:cond_z} is univariate and log-concave, one can use adaptive rejection sampling \citep{Gilks1992} while sampling the variable of interest $\boldsymbol{\theta}$ from \eqref{eq:cond11} boils down to high-dimensional Gaussian sampling and efficient methods can be applied.
\end{example}

\subsection{Variational Bayes inference from AXDA}
\label{subsec:VB}
AXDA can also be a major asset when conducting variational Bayes (VB) inference, providing important benefits such as simplicity and parallelization.
VB methods \citep{Bishop2000,Opper2001} circumvent the direct sampling from a target density such as $\joint$ by defining an approximation of the latter denoted $\tilde{\pi}_{\rho}$.
The best approximation is found by minimizing the Kullback-Leibler (KL) divergence between $\tilde{\pi}_{\rho}$ and $\joint$ restricted to a set of tractable candidates $\tilde{\pi}_{\rho}$.
Depending on this set, a lot of VB approximation methods exist in the literature, see \cite{Bishop2006} and \cite{Pereyra2016A} for reviews.
In this section, we will consider the widely-used mean-field approximation method where the approximate density $\tilde{\joint\text{}}$ is chosen among the set of conditionally independent (w.r.t. $\rho$) densities, that is $\tilde{\pi}_{\rho}(\boldsymbol{\theta},\B{z}_{1:J}) = \tilde{\pi}_{\rho}(\boldsymbol{\theta})\prod_{j=1}^J\tilde{\pi}_{\rho}(\B{z}_j)$.
Under this constraint, the optimal choice of the VB approximation is given by
\begin{align}
  &\log \tilde{\pi}_{\rho}(\boldsymbol{\theta}) = \sum_{j=1}^J\mathbb{E}_{\tilde{\pi}_{\rho}(\B{z}_j)}\log\kappa_{\rho}(\B{z}_j,\B{A}_j\boldsymbol{\theta}) \\
  &\log \tilde{\pi}_{\rho}(\B{z}_j) = -f_j(\B{z}_j) + \mathbb{E}_{\tilde{\pi}_{\rho}(\boldsymbol{\theta})}\log\kappa_{\rho}(\B{z}_j,\B{A}_j\boldsymbol{\theta}) \label{eq:VB_z}.
\end{align}
The above VB-marginals require to compute expectations under each marginal distribution which are often functions of moments under each marginal.
Similarly to what has been encountered for Gibbs sampling in the previous section, deriving a VB approach based on $\joint$, instead of $\pi$, yields important benefits for parallel and possibly easier computations.
Indeed, the VB-marginal in \eqref{eq:VB_z} shows again that each potential $f_j$ contributes independently given $\boldsymbol{\theta}$.
The updates of expectations under \eqref{eq:VB_z} are thereby simplified since (i) the VB-marginals \eqref{eq:VB_z} are simpler than those obtained from a mean-field approximation of $\pi$ and (ii) the moments under the latter can be computed in parallel or distributed.
\begin{example*}\normalfont
  \cite{Jaakkola2000} considered a local VB algorithm for the penalized logistic regression problem.
  Instead of using a local VB approach and finding bounds on each individual function $f_j$, the use of $\joint$ instead of $\pi$ permits to consider directly a global VB approach such as the mean-field approximation.
  In addition, similarly to Algorithm \ref{algo:Gibbs}, the updates of \eqref{eq:VB_z} and the associated expectations can be computed in parallel and efficiently by using state-of-the-art existing methods.
  For instance, the expectations $\mathbb{E}_{\tilde{\pi}_{\rho}(\B{z}_j)}$ can be approximated efficiently using rejection sampling.
\end{example*}

\subsection{Optimizing AXDA meets quadratic penalty methods}
\label{subsec:optimization}

Computing the MAP or ML estimate under the AXDA model \eqref{eq:split_density_2} boils down to solve the optimization problem
\begin{align}
  \min_{\boldsymbol{\theta},\B{z}_{1:J}} \sum_{j=1}^Jf_j(\B{z}_j) - \log \kappa_{\rho}(\B{z}_j,\B{A}_j\boldsymbol{\theta}). \label{eq:quadPenalty_optim_problem}
\end{align}
If $\kappa_{\rho}$ stands for a Gaussian kernel, the problem \eqref{eq:quadPenalty_optim_problem} can be viewed as a quadratically penalized formulation of the initial problem $\min_{\boldsymbol{\theta}} \sum_{j=1}^Jf_j(\boldsymbol{\theta})$, see \citet[Section 17.1]{NoceWrig06}.
As expected, the solution of \eqref{eq:quadPenalty_optim_problem} stands for an approximate solution w.r.t. the initial optimization problem.
The associated algorithm is depicted in Algorithm \ref{algo:ADMM}.
Regarding this algorithm, one can clearly see the benefit of using a variable splitting approach as in AXDA: the initial potential is split into $J$ individual potentials with no operator acting on $\boldsymbol{\theta}$.
Therefore, the corresponding minimization problems are simpler (e.g., associated proximity operators may become available) and can be handled in parallel.

We eventually point out that the benefits of Algorithm \ref{algo:ADMM} highlighted previously are also shared with the ADMM \citep{Boyd2011}.
Instead of solving the approximate optimization problem \eqref{eq:quadPenalty_optim_problem} which encodes the splitting operation with a quadratic regularization term, the latter builds on Lagrangian duality in order to provide an exact solution to the initial minimization problem $\min f_j$.

\renewcommand{\baselinestretch}{0.8}
\begin{algorithm}%
    \caption{Quadratic penalty minimization}
    \label{algo:ADMM}
     \SetKwInOut{Input}{Input}
     \Input{Functions $f_j$, penalty parameter $\rho$, $t \leftarrow 0$ and $\B{z}_{1:J}^{[0]}$}
   \While{stopping criterion not satisfied}{%
   \textit{\% Minimization w.r.t. $\boldsymbol{\theta}$} \\
   $\boldsymbol{\theta}^{[t]} \in \underset{\boldsymbol{\theta}}{\arg \min}\ \displaystyle\sum_{j=1}^J \dfrac{1}{2\rho^2}\nr{\B{A}_j\boldsymbol{\theta}-\B{z}_j^{[t-1]}}_2^2$\;
   \For{$j\gets1$ \KwTo $J$}{
   \textit{\% Minimization w.r.t. $\B{z}_j$}\\
   $\B{z}_j^{[t]} \in \underset{\B{z}_j}{\arg \min}\ f_j(\B{z}_j) + \dfrac{1}{2\rho^2}\nr{\B{A}_j\boldsymbol{\theta}^{[t]}-\B{z}_j}_2^2$\;
   }
   \textit{\% Updating iterations counter} \\
   $ t \leftarrow t + 1$ \;
   }
   \SetKwInOut{Output}{Output}
     \Output{MAP or ML estimate depending on the considered problem.}
\end{algorithm}
\renewcommand{\baselinestretch}{1.3}

\begin{example*}
\label{example:logistic_reg_inference} \normalfont

Computing directly the MAP estimate with classical forward-backward algorithms (e.g., the fast iterative shrinkage-thresholding algorithm (FISTA) \citep{Beck2009}) associated to $\pi$ is challenging if the observations are distributed among multiple nodes.
In addition, proximity operators associated to $f_j$ are generally not available in closed-form because of the operators $\B{A}_j$.
Algorithm \ref{algo:ADMM} permits to tackle these issues by splitting the initial objective function.
In particular, the minimization w.r.t. $\B{z}_j$ for $j \in [J]$ corresponds to an unidimensional $l_2$ regularized logistic regression problem that can be dealt with gradient-based methods with few iterations. 
The minimization w.r.t. $\boldsymbol{\theta}$ boils down to the solving of a linear system where efficient solvers can be applied.
Note that such a splitting scheme avoids the use of stochastic gradient methods.
\end{example*}

\subsection{Expectation-maximization for AXDA}
\label{subsec:EM}

An EM algorithm under the augmented density $\joint(\boldsymbol{\theta},\B{z})$ will target the MAP or ML estimator, see Algorithm \ref{algo:EM}.
If the expectations in the E-step cannot be evaluated, one can use a Monte Carlo approximation to approximate them \citep{Wei1990}.
The benefits of using the augmented density $\joint$ instead of $\pi$ are threefolds.
Firstly, as pointed out in Section 2 in the main paper, exact DA schemes based on $\pi$ cannot be derived in general cases and corresponding EM algorithms cannot be implemented.
Instead, considering $\joint$ gives a quite systematic way of introducing latent variables in the original statistical model.
Secondly, the expectations involved in the E-step of Algorithm \ref{algo:EM} can be simpler to derive than the expectation under $\pi$.
Indeed, the latter involves the whole potential $\sum_jf_j$ while the former involves regularized parts $f_j - \log \kappa_{\rho}$ of this potential separately.
Finally, conditionally on $\boldsymbol{\theta}^{[t]}$, the random variables $\B{z}_j$ are independent.
Thus, each expectation involved in the E-step can be computed in parallel.

\renewcommand{\baselinestretch}{0.8}
\begin{algorithm}
    \caption{EM}
    \label{algo:EM}
     \SetKwInOut{Input}{Input}
     \Input{Functions $f_j$, penalty parameter $\rho$, $t \leftarrow 0$ and $\boldsymbol{\theta}^{[0]}$}
   \While{stopping criterion not satisfied}{%
   \textit{\% E-step}\\
   Define $Q(\boldsymbol{\theta}|\boldsymbol{\theta}^{[t]}) = \displaystyle\sum_{j=1}^J\mathbb{E}_{\joint(\B{z}_j|\boldsymbol{\theta}^{[t]})}\pr{-f_j(\B{z}_j) +\log \kappa_{\rho}(\B{z}_j,\B{A}_j\boldsymbol{\theta})}$\;
   \textit{\% M-step}\\;
   Compute $\boldsymbol{\theta}^{[t+1]} =  \underset{\boldsymbol{\theta}}{\arg \max}\ Q(\boldsymbol{\theta}|\boldsymbol{\theta}^{[t]})$\;
   \textit{\% Updating iterations counter} \\
   $ t \leftarrow t + 1$ \;
   }
   \SetKwInOut{Output}{Output}
     \Output{MAP or ML estimate depending on the considered problem.}
\end{algorithm}
\renewcommand{\baselinestretch}{1.3}

\begin{example*}\normalfont
  Again, following the work of \cite{Polson2013}, if the potential $g$ of the prior distribution is quadratic or corresponds to a sparsity-promoting $\ell_p$-penalization ($0<p\leq1$), a simple EM-algorithm can be derived as detailed by \cite{Scott2013}.
  However, although this EM algorithm can be generalized to an online version, it does not scale to distributed and high-dimensional problems.
  On the other hand, the E-step of Algorithm \ref{algo:EM} can be processed in parallel by computing the $J$ expectations on individual nodes: thanks to the AXDA approach, the algorithm is therefore suited to distributed and high-dimensional scenarios.
\end{example*}

\newpage
\renewcommand{\baselinestretch}{0.5}
\bibliographystyle{asa}
\renewcommand{\baselinestretch}{0.5}
\bibliography{biblio}

\end{document}